\documentclass[3p]{elsarticle}

\usepackage{lineno, hyperref}
\modulolinenumbers[5]

\journal{ }
\usepackage{amsmath}

\usepackage[cmintegrals]{newtxmath}

\usepackage{bbding}
\usepackage{makecell}

\usepackage{threeparttable}

\usepackage{amsmath,amssymb,amsfonts}
\usepackage{algorithmic}
\usepackage{graphicx}
\usepackage{textcomp}
\usepackage{caption}
\usepackage{float}
\usepackage{enumitem}

\usepackage{subfigure}
\usepackage{hyperref}
\usepackage{xcolor}
\usepackage{url}
\usepackage{array}
\usepackage{seqsplit}
\usepackage{multirow}
\usepackage{listings}
\graphicspath{{figures/}}

\newtheorem{definition}{Definition}

\mathchardef\mhyphen="2D

\newcommand{\env}{\mathcal{Z}}
\newcommand{\adv}{\mathcal{A}}

\newcommand{\comm}{\mathsf{C}}

\newcommand{\shard}{\mathsf{shard}}

\newcommand{\hash}{\mathsf{H}}
\newcommand{\tx}{\mathsf{tx}}

\newcommand{\logs}{\mathsf{LOG}}
\newcommand{\lists}{\mathsf{list}}
\newcommand{\chains}{\mathsf{chain}}

\newcommand{\newnodes}{\mathsf{newnodes}}

\newcommand{\biga}{\mathsf{A}}
\newcommand{\bigb}{\mathsf{B}}
\newcommand{\bigc}{\mathsf{C}}

\newcommand{\ns}{\textsc{Ns}}
\newcommand{\na}{\textsc{Na}}
\newcommand{\er}{\textsc{Er}}
\newcommand{\sr}{\textsc{Sr}}
\newcommand{\isc}{\textsc{Isc}}
\newcommand{\cstp}{\textsc{Cstp}}
\newcommand{\mm}{\textsc{Mm}}

\usepackage{tabularx}
\usepackage{amssymb}
\usepackage{pifont}
\newcommand{\cmark}{\textcolor{blue}{\text{\ding{51}}}}%
\newcommand{\xmark}{\textcolor{red}{\text{\ding{55}}}}%
\usepackage{makecell}
\usepackage{array}
\newcolumntype{x}[1]{>{\centering\arraybackslash\hspace{0pt}}p{#1}}
\usepackage{booktabs}
\usepackage{tikz}

\usepackage{ragged2e}

\newenvironment{component}[1][htb]
{
	\begin{algorithm}[#1]%
	}{\end{algorithm}}

\usepackage[algoruled,linesnumbered,vlined,noend]{algorithm2e}

\definecolor{mypink1}{rgb}{0.858, 0.188, 0.478}
\definecolor{mypink2}{RGB}{219, 48, 122}
\definecolor{mypink3}{cmyk}{0, 0.7808, 0.4429, 0.1412}
\definecolor{mygray}{gray}{0.6}
\definecolor{myorange}{RGB}{255,97,3}
\definecolor{myblue}{RGB}{51,161,201}

\usepackage{subfigure}

\hypersetup{
	colorlinks=true,
	linkcolor=blue,
	filecolor=magenta,      
	urlcolor=cyan,
}

\begin{document}

\begin{frontmatter}

\title{Building Blocks of Sharding Blockchain Systems: Concepts, Approaches, and Open Problems}


\author[1]{Yizhong~Liu}
\ead{liuyizhong@buaa.edu.cn}
\author[1]{Jianwei~Liu}
\ead{liujianwei@buaa.edu.cn}
\author[2]{Marcos Antonio~Vaz Salles}
\ead{vmarcos@di.ku.dk}
\author[1]{Zongyang~Zhang\corref{aa}}
\ead{zongyangzhang@buaa.edu.cn}
\author[1]{Tong~Li}
\ead{leetong@buaa.edu.cn}
\author[1]{Bin~Hu}
\ead{hubin0205@buaa.edu.cn}
\author[2]{Fritz~Henglein}
\ead{henglein@di.ku.dk}
\author[3]{Rongxing~Lu}
\ead{rlu1@unb.ca}

\address[1]{School of Cyber Science and Technology, Beihang University, Beijing, China}
\address[2]{Department of Computer Science, University of Copenhagen, Copenhagen, Denmark}
\address[3]{Faculty of Computer Science, University of New Brunswick, Fredericton, Canada}
\cortext[aa]{Corresponding author}

\begin{abstract}
Sharding is the prevalent approach to breaking the trilemma of simultaneously achieving decentralization, security, and scalability in traditional blockchain systems, which are implemented as replicated state machines relying on atomic broadcast for consensus on an immutable chain of valid transactions. Sharding is to be understood broadly as techniques for dynamically partitioning nodes in a blockchain system into subsets (shards) that perform storage, communication, and computation tasks without fine-grained synchronization with each other.
Despite much recent research on sharding blockchains, much remains to be explored in the design space of these systems. Towards that aim, we conduct a systematic analysis of existing sharding blockchain systems and derive a conceptual decomposition of their architecture into functional components and the underlying assumptions about system models and attackers they are built on.  
The functional components identified are node selection, epoch randomness, node assignment, intra-shard consensus, cross-shard transaction processing, shard reconfiguration, and motivation mechanism. We describe interfaces, functionality, and properties of each component and show how they compose into a sharding blockchain system.
For each component, we systematically review existing approaches, identify potential and open problems, and propose future research directions. We focus on potential security attacks and performance problems, including system throughput and latency concerns such as confirmation delays.
We believe our modular architectural decomposition and in-depth analysis of each component, based on a comprehensive literature study, provides a systematic basis for conceptualizing state-of-the-art sharding blockchain systems, proving or improving security and performance properties of components, and developing new sharding blockchain system designs.

\end{abstract}


\end{frontmatter}

\section{Introduction}
A traditional blockchain system is a peer-to-peer (P2P) distributed system with decentralized governance. It implements a replicated state machine that relies on an atomic broadcast (total event order consensus) protocol for consensus on an immutable chain (sequence) of valid transactions.  
A transaction can be any record of data but is usually a digitally signed statement that expresses a transfer of ownership of a digital resource (asset). In the seminal and paradigmatic Bitcoin system \cite{N08}\footnote{Where a blockchain is called \emph{chain of blocks}; neither ``blockchain'' nor ``block-chain'' occur in the paper.}, a transaction is a cryptographically signed transfer of a built-in synthetic currency, Bitcoin, by and to anonymous parties identified only by public keys they themselves have generated.  Any node can join and leave the P2P network at any time, without authentication.  The nodes receive transactions submitted by clients, collect them into blocks of valid transactions, and compete with other nodes to have their block extend the currently longest chain of such blocks.  The node operators are incentivized to do this by receiving a fee, in Bitcoin, whenever their block is the consensual continuation to the currently longest chain.\footnote{This is a simplified description with technical infelicities.}

An idealized blockchain system strives for the combination of decentralized control (no single party has an \emph{a priori} privileged role), consensus on a single state (``single source of truth''), tamper-proof recording (validated transactions cannot feasibly be deleted or updated \emph{ex post}), once added to the blockchain), privacy preservation (publicly shared data, but secured and privatized by cryptographic techniques such as cryptographic hashing, private-public key cryptography, secret sharing for digital signatures \cite{JMV01,BDN18}, immutable self-certifying pointers \cite{CDMP05}, zero-knowledge proofs and more),  
and high availability (high degree of replication on nodes controlled by independent, non-colluding node operators) in an untrusted environment  \cite{BMC+15}.

Blockchain systems have tremendous application potential in various fields, such as Internet of Things (IoT) \cite{AVP+19,KS18}, 
cloud computing \cite{GGZ+20,YYS+19} (with a trusted data center provider),
smart cities \cite{XTH+19}, finance \cite{egelund2017automated,TBY17,Eyal17} (with authenticated legal entities), self-sovereign identity management\footnote{Such as ESSIF, part of the European Blockchain Services Initiative, Sovrin or a number of Hyperledger identity management frameworks.}, supply chain \cite{KHD17} and more. 

Blockchain technology develops very rapidly, but faces both fundamental and practical obstacles to wider applicability. The most critical issue is the trilemma of decentralization, security, and scalability.
To achieve decentralization, solutions need to support independent participants with varying assumptions on how participants may join or leave a network, from permissioned to permissionless blockchain systems \cite{NH19}.

As for security \cite{SSN+19,CEL+18}, a blockchain protocol has to be proved secure \cite{GKL15,PSS17}
to ensure it resists certain classes of attacks that may compromise its correct functioning vis a vis client. 

For improved scalability, existing approaches can be classified into \emph{off-chain} and \emph{on-chain} solutions \cite{HHS20}. The former employ a hierarchical architecture, where the core blockchain system (on-chain) only validates few aggregated resource transfers that are the net effect of many fine-grained payments, which in turn are managed separately in multiple unsynchronized subsystems.  
This includes technologies such as micropayment \cite{PD16,KL20}, payment channel networks \cite{MMK+17,KG17}, virtual payment channels \cite{DEF+19}, and sidechains \cite{GKZ19,KZ19,PB17}.  Conceptually, these solutions correspond to a (full-reserve) banking system, where
the central bank corresponds to the core blockchain system, and the channels to pop-up banks that have initially some accounts transferred from
the central bank, perform internal transactions without synchronization with other banks, and eventually transfer the final balances back to the central bank such that only the \emph{netted} result is stored in the blockchain.  
Off-chain solutions avoid the computational cost of a traditional blockchain system, which requires each honest node to receive, store and send all transactions and to come to an agreement amongst all nodes on a total order of all valid transactions.

Despite the applicability of off-chain solutions in some scenarios, on-chain solutions where \emph{all} transactions are recorded, validated, and retained without netting, are desired in many other application scenarios.  
A \emph{sharding} blockchain system is an on-chain solution that seeks to improve the scalability of a traditional blockchain system while achieving the same level of security and decentralization. 
In a sharding blockchain system, the nodes in the network are dynamically partitioned into \emph{shards} (subsets), where each shard is responsible for managing its own blockchain.  The basic idea is that, instead of storing a chain of transactions and replicating it across \emph{all} nodes, an acyclic graph of transactions is maintained, where each shard is only responsible for a specific part of the graph. 
As new nodes join the network, the cumulative transaction throughput can grow by increasing the number of shards \cite{YJZ+19}. Sharding technology was pioneered for database systems \cite{CDE+13}, where it describes methods for dynamically partitioning a database into parts, called shards, each managed by a different node in a distributed system.  The concept of partitioning data and their management, including the term sharding, was introduced to blockchain systems by ELASTICO \cite{LNZ+16}.

The design and implementation of sharding blockchain systems currently suffer from a number of problems, however.
First, the structure of sharding blockchain systems is complicated; they usually contain multiple key components,\footnote{We use the terms ``functional component'' and ``building block'' interchangeably.} such as the method for selecting shard members, the consensus algorithm inside a shard, and the processing method for transactions. 
Second, different sharding blockchain systems may adopt different models, such as network, adversary, and transaction models, without making these explicit enough to assess their design and (security) properties. 
These model assumptions engender a considerable set of design choices, which need to be characterized and compared in the context of their model assumptions.
Third, most sharding blockchain systems are presented as end-to-end systems, describing a specific point in a tremendous design space for
blockchain systems, without providing an architecture for exploring systematic modular design to rapidly and securely explore the design space for sharding blockchain systems. In particular, the functionality of components and their interfaces are not clear enough, which leads to difficulties in exploring various alternative designs for each component.

In this paper, we address these points: We provide a conceptual and technical framework for decomposing existing sharding blockchain systems into key functional components and describe a conceptual and technical modular architecture for composing them into full sharding blockchain systems. 
Besides, we propose a taxonomy for sharding blockchain systems from the dimensions of system models and components. 
Furthermore, we provide a systematic, in-depth analysis of each component, describing its input dependencies, functionality, and key properties. For each component, we classify existing approaches and solutions, identify open problems, and provide directions for future research.

\subsection{Our Contributions}
\label{subsec:contribution}
In summary, we make the following contributions in this paper.

\noindent \textbf{Decomposing sharding blockchains into functional components.} 
We decompose sharding blockchains into multiple functional components: node selection, epoch randomness, node assignment, intra-shard consensus, cross-shard transaction processing, shard reconfiguration, and motivation mechanism. 
For each component, we give its input, output, function, and property to be satisfied. 
Furthermore, we show how to compose these components into a complete sharding blockchain system.
The component decomposition provides a path to systematically developing yet unexplored sharding blockchain system designs.

\noindent \textbf{A detailed taxonomy for sharding blockchain systems.}
We provide a taxonomy for sharding blockchain systems in two dimensions: system models and components.
System models include network model, adversary model, and transaction model. We divide each model into categories that correspond to different types of sharding blockchain systems, such as eventual and instant sharding blockchain systems, or permissioned and permissionless sharding blockchain systems.
For each component of a sharding blockchain system, we classify solutions according to their principles and algorithms.

\noindent \textbf{In-depth analysis of components}. 
For each component of a sharding blockchain system, we first give its basic concepts, including its purpose, functionality, and essential procedures. We summarize and categorize existing approaches according to their underlying characteristics; the specific operational details of each method are expounded from multiple perspectives. In addition, we identify and analyze possible problems for every type of solution, including attacks an adversary might launch on security, throughput, and latency (transaction confirmation delays). Finally, we point out possible future research directions for each component.

\subsection{Paper Organization} 
\label{sub:paper_organization}
Section~\ref{sec:preliminaries} gives the background, definitions and notations that are useful in this paper. In Section~\ref{sec:decomposing} we decompose sharding blockchains into several components and discuss methodologies to derive sharding blockchain systems from their composition. In Section~\ref{sec:node_selection}, Section~\ref{sec:epoch_randomness}, and Section~\ref{sec:node_assignment}, basic concepts, existing approaches, open problems and future directions of node selection, epoch randomness, and node assignment are given, respectively. These three parts involve the method to confirm shard members. Then Section~\ref{sec:intra-shard_consensus} describes the classic state machine replication algorithms that are used as intra-shard consensus. In Section~\ref{sec:cross-shard}, cross-shard transaction processing methods are analyzed in detail, which is important to all sharding blockchain systems. Section~\ref{sec:shard_reconfiguration} and Section~\ref{sec:motivation_mechanism} give the existing approaches and potential problems about shard reconfiguration and motivation mechanisms. Section~\ref{sec:related_work} describes the related work. Finally, Section~\ref{sec:conclusion} summarizes this paper. 

\section{Preliminaries} 
\label{sec:preliminaries}
In this section, we first give the background of sharding blockchains. Then notations and definitions that are useful in the paper are described in detail.

\subsection{Background}
\label{subsec:background}
Sharding blockchains adopt a unique blockchain consensus mechanism. 
Next, we first describe the relevant background of blockchain consensus and then introduce sharding blockchains.

\subsubsection{Blockchain Consensus}
\label{subsec:overview_consensus}
Blockchain technology is introduced by Bitcoin \cite{N08} in 2008, which realizes the agreement of the ledger among distributed nodes through a specific consensus mechanism. The reason why the blockchain is called ``chain'' is because each block is linked to the previous one in a specific cryptographic way. The contents stored in a block mainly include the transactions generated in the network during each period of time.

Blockchain technology is currently a hot area of research and has great application potential since it has the following characteristics. The first characteristic is decentralization. Decentralization means that there is no trusted third party in the network, which is different from the traditional centralized transaction mode. The second characteristic is trustlessness, which means that nodes do not need to trust each other, and can finally reach consensus on the ledger through a specific consensus mechanism. The third characteristic is transparency. In a permissionless network (ref. Definition~\ref{def:permissionless}), all nodes can join and leave the protocol at any time, and nodes could obtain the historical ledger data of the blockchain at any time. The fourth is tamper-resistance. Under the assumptions of appropriate network and adversary models, historical data in the blockchain cannot be illegally tampered with, except for some special redactable blockchains \cite{DMT19,AMV+17}. Once a block is confirmed (strong consistency blockchains, ref. Definition~\ref{def:strong_consistency}), or a block reaches a certain depth (weak consistency blockchains, ref. Definition~\ref{def:weak_consistency}), the contents of the block can no longer be modified. The fifth is anonymity. Through some approaches (such as transaction graph analysis \cite{RH11,KL18}), deanonymization analysis on historical transaction data could be performed on some blockchains. However, by adopting some cryptographic technologies, such as blind signatures \cite{Chaum83,HBG16}, ring signatures \cite{ZK02,SALY17}, and zero-knowledge proofs \cite{BGG18,MGGR13}, privacy-protection blockchains are designed to prevent users' privacy from leakage.

\begin{figure*}[h!]
	\centering
	\includegraphics[width=0.7\linewidth]{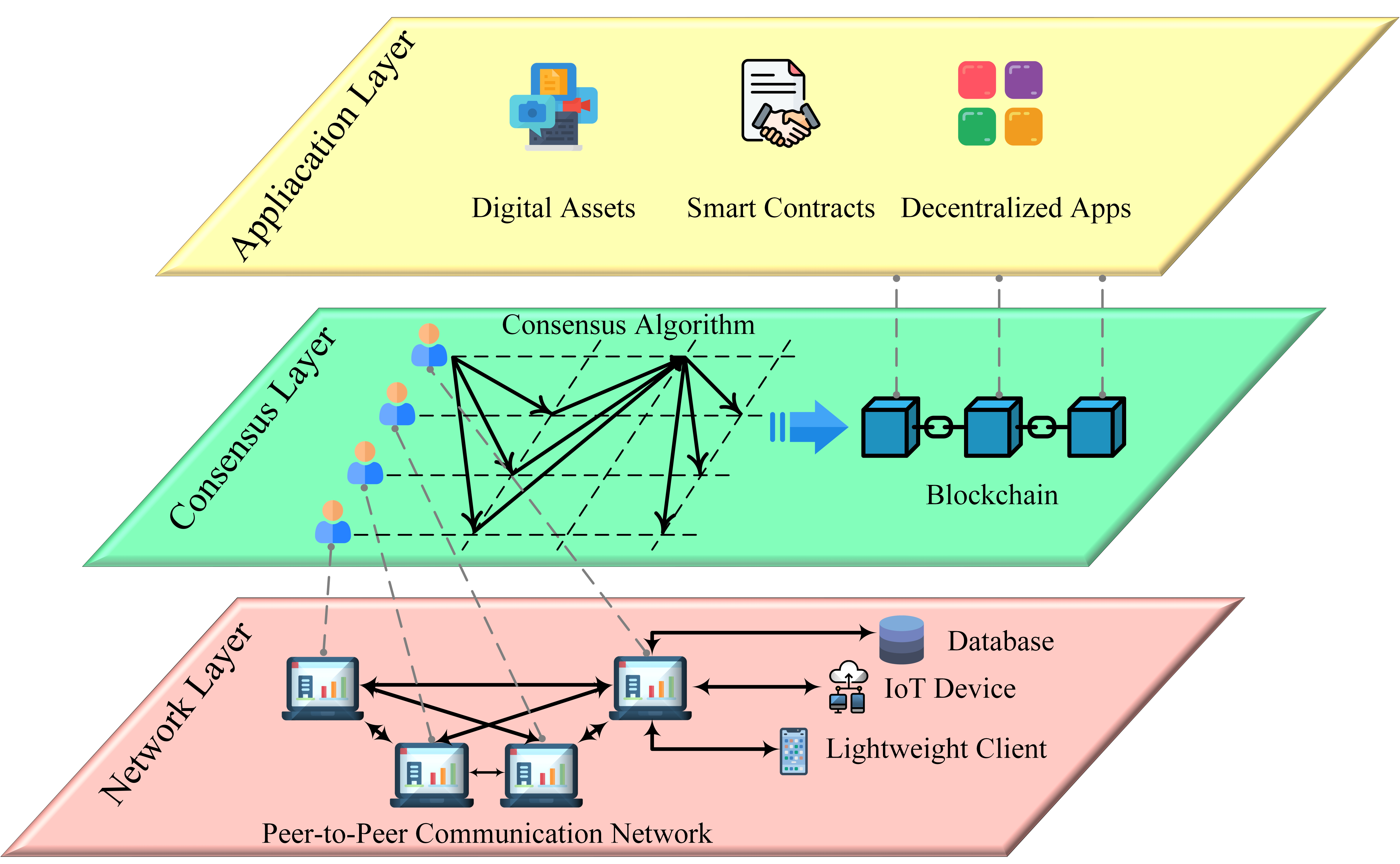}
	\caption{Blockchain Layers.}
	\label{fig:layers}
\end{figure*}

As shown in Fig.~\ref{fig:layers}, the architecture of blockchain could be divided into several layers, including a network layer, a consensus layer, and an application layer. In the network layer, participating nodes join a P2P communication network \cite{S01} to synchronize information with each other. In a P2P network, the nodes are distributed, and there is no central communication node in the network. The transfer and update of information are completed by the P2P communication between each node. Besides, there could be other kinds of connecting nodes in a blockchain network, such as databases, IoT devices, and lightweight clients. In the consensus layer, the participating nodes in the network with certain computing and communication capabilities act as consensus nodes, that is, block producers (we use the term ``block producer'' instead of ``miner'' to make the meaning more general), and generate blocks through certain consensus algorithms. Note that there are many types of consensus algorithms, and Fig.~\ref{fig:layers} shows a Byzantine Fault Tolerance (BFT) \cite{CL99} algorithm. Many applications, such as digital assets \cite{BAR+17}, smart contracts \cite{HLL+20}, and decentralized applications \cite{Ra16}, could be built based on a blockchain, together constituting the application layer.

A consensus mechanism is very critical to a blockchain system, and to a large extent determines the security and performance of the system. A blockchain consensus is to ensure the consistency and liveness of the system by formulating rules for nodes to participate in the blockchain protocol. Consistency means that all honest nodes ultimately have the same view of the blockchain. Liveness refers to that the transactions uploaded by users to the blockchain are sure to be processed after a certain period of time.

At present, the blockchain consensus mechanisms could be divided into the following categories, according to \cite{BSA+19}. First, it is the classic distributed consensus algorithm, represented by the Byzantine fault tolerance algorithms, which implements state machine replication (ref. Definition~\ref{def:state_machine_replication}) in a limited number of participating nodes. The second is proof-of-work (PoW) based consensus, including Bitcoin \cite{N08}, Bitcoin-NG \cite{EGSR16}, GHOST \cite{SZ15}, FruitChains \cite{PS17fruit}, SPECTURE \cite{SLZ16}, etc. Besides, there is proof-of-stake (PoS) based consensus, such as Casper FFG \cite{BG17}, Snow White \cite{DPS19}, Ouroboros \cite{KRD+17}, etc. Finally, the hybrid consensus, such as PeerCensus \cite{DSW16}, ByzCoin \cite{KJG+16}, Solida \cite{AMN+17}, etc., combines classic state machine replication algorithms and blockchain technology. According to the number of committees, the hybrid consensus mechanisms could be divided into single-committee and multi-committee hybrid consensus mechanisms. The multi-committee hybrid consensus is a kind of sharding consensus mechanism, which is introduced in the following.

\subsubsection{Sharding Blockchains}
\label{subsec:sharding_blockchains}
Sharding technology is first proposed and used in the field of databases \cite{CDE+13}. By dividing all participating nodes in the network into multiple shards, each shard is only responsible for maintaining its own corresponding data. In this way, the scalability of network processing capabilities could be achieved. As the number of nodes in the network increases, the enhancement of processing capabilities is realized by adding more shards.
The sharding blockchain is first proposed by ELASTICO \cite{LNZ+16}, which combines sharding technology and blockchain technology, with the purpose to increase the transaction throughput, i.e., the number of transactions processed per second. 
Since then, there have been many studies on sharding blockchains, such as Omniledger \cite{KJG+18}, Chainspace \cite{KJG+18}, RapidChain \cite{ZMR18}, and Monoxide \cite{WW19}.

In general, sharding blockchains have the following three characteristics. 
The first one is communication sharding. Participating nodes are divided into different shards where nodes in each shard only need internal communication most of the time. The clients and nodes within each shard could obtain the current state of the blockchain by communicating with the intra-shard nodes that are responsible for maintaining the blockchain, e.g., a committee.
The second one is computation sharding, which means that each shard is only responsible for processing its corresponding transactions. The distribution of transactions to shards is diversified, e.g., by selecting the corresponding shard according to the transaction ID. Generally speaking, according to the last several bits of the transaction ID, it is determined which shard the transaction belongs to. So transactions are handled by different shards in parallel. When the number of nodes in the network increases, more shards could be added to realize scalability. 
The third one is storage sharding. Storage sharding means that nodes of different shards only needs to store the data of its corresponding shard. The data includes transaction history and unspent transaction output (UTXO, ref. Definition~\ref{def:utxo}) data. Transaction history data exists in the form of a blockchain, while the UTXO data could be derived from transaction history data or could be stored separately to improve its processing efficiency. 
Storage sharding allows nodes to store a fraction of the entire blockchain system data, reducing the storage burden of nodes.

\begin{figure*}[h!]
	\centering
	\includegraphics[width=0.85\linewidth]{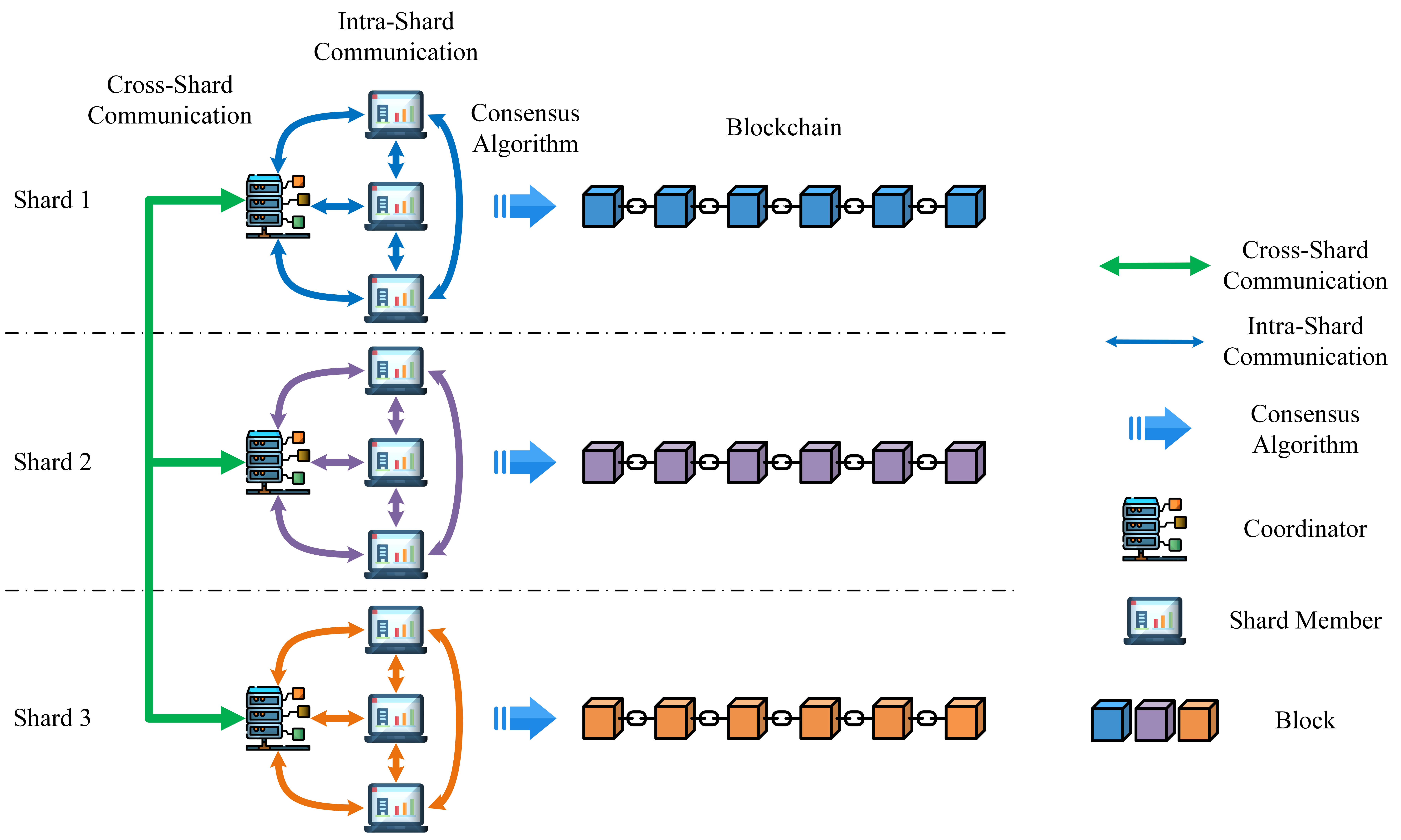}
	\caption{Communications in Sharding Blockchain Systems.}
	\label{fig:communication}
\end{figure*}

As shown in Fig.~\ref{fig:communication}, communications play an important role in sharding blockchain systems. Nodes in the same shard only need to execute intra-shard communication most of the time and send some key information to a coordinator of the shard. The coordinator is usually responsible for cross-shard communication as well as intra-shard consensus, and each shard has at least one coordinator. Generally speaking, the coordinator needs to have stronger communication capabilities than other intra-shard nodes.

Note that some methods, such as OHIE \cite{YNH+19} and Prism \cite{BKT+19}, are designed to improve the throughput of the blockchains, while in a strict sense, sharding technology is not used in their structures. Hence, we will not introduce this type of solutions, while focusing on the blockchains that use sharding technology.
A sharding blockchain usually includes several major components such as node selection, epoch randomness, node assignment, intra-shard consensus, cross-shard transaction processing, and shard reconfiguration.
Each component could be implemented by different methods, and then by composing all the components, a complete sharding blockchain could be obtained. We will give a detailed analysis of each component and their composition approach in this paper.



\subsection{Notations}
\label{subsec:notations}
We summarize the notations that are used in our paper in Table~\ref{tab:notations}. To make it compatible with other protocols, we use $n$ to denote the total number of participating nodes, while use $u$ to denote the number of members in a committee. So inside a committee, the security condition that should be satisfied could be represented by $u=2f+1$ or $u=3f+1$, where $f$ is the number of admissible failures. 
In addition, $\shard_i$ denotes the $i$-th shard, while the notation $\comm^i$ denotes the committee in $\shard_i$. These two concepts are different since there might be no committee in a shard in some kinds of sharding blockchains.
The notation $\logs$ means the output log, or ledger of a shard, while $\chains$ denotes a blockchain. The specific meaning of $\logs$ and $\chains$ are similar, while there are also differences. A blockchain $\chains$ might contain other contents and information, such as block headers and additional information in OP\_RETURN \cite{TD17}. $\logs$ usually means extracting key information from a blockchain, e.g., transaction data.

\renewcommand{\arraystretch}{1.4}

\begin{table}[h!]
	\begin{center}	
		\caption{Notations}
		\label{tab:notations}  
		\small
		\begin{threeparttable}
			\begin{tabular}{cp{13cm}}
				\midrule[.1em]
				\textbf{Symbol} & \textbf{Definition}   \\
				\midrule[.1em]
				$\tx$ & a transaction  \\	
				$\rm\Delta$ & the upper bound of the network's delay   \\	
				$\delta$ & the actual delay of the network \\	
				$u$ & the exact number of members in a shard committee\tnote{$\star$} \\	
				$f$ & the maximum number of malicious node in a shard committee \\ 
				$m$ & the total number of shards \\	
				$n$ & the total number of nodes that participate in the protocol \\	
				$\shard_i$ & the $i$-th shard \\ 
				$\comm^i_e$ & the $i$-th committee of epoch $e$\tnote{*} \\	
				$\comm^R_e$ & the reference committee of epoch $e$ \\
				$\mathsf{LOG}_i$ & the output log of the $i$-th shard's nodes \\
				$\chains$ & a blockchain \\ 
				$\rho$ & the fraction of the computational power that is held by an adversary \\
				$\tau$ & the corruption time parameter of an adversary \\
				$p$  & the probability that one node finds a PoW solution successfully in one single round \\	\midrule[.1em]
			\end{tabular}
			\begin{tablenotes}
				\item[$\star$] For the convenience of analysis, we assume that the number of members in a committee is fixed.
				\item[*] The concept of shard and committee is different. In committee-based sharding blockchains, there is a committee responsible for processing transactions in each shard, and we call it an ordinary committee. So the number of ordinary committees is also $m$. Besides, some sharding blockchains, e.g., RapidChain \cite{ZMR18}, utilize a reference committee to confirm committee members.
			\end{tablenotes}
		\end{threeparttable}
	\end{center}	
\end{table}

\subsection{Definitions}
\label{subsec:definitions}
\begin{figure}[h!]
	\centering
	\includegraphics[width=0.9\textwidth]{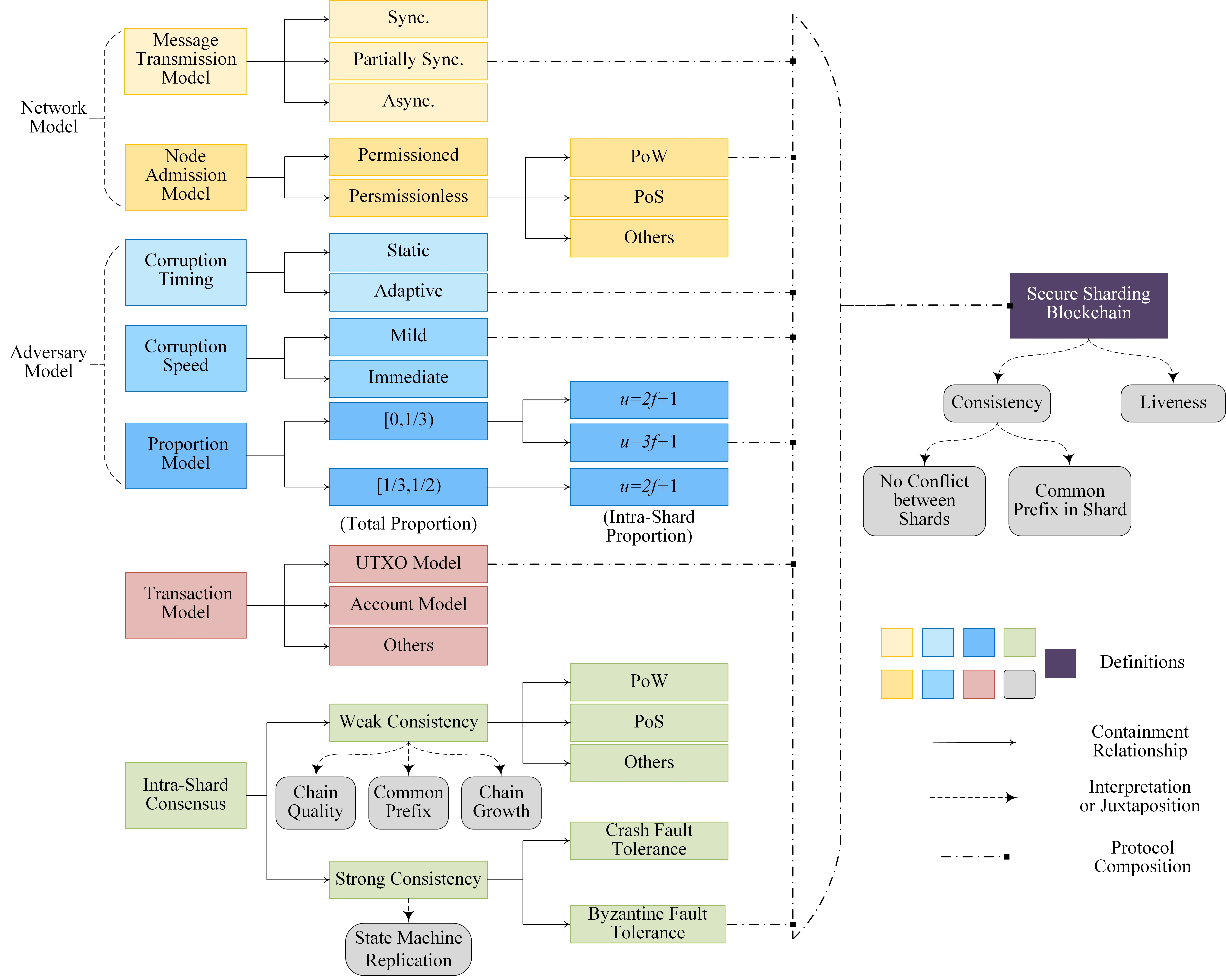}
	\caption{Definitions and their relationships.}
	\label{fig:definitions}
\end{figure}

To make our description clearer, we give the definitions that are useful in our paper and helpful in understanding sharding blockchains. 
As shown in Fig.~\ref{fig:definitions}, we divide the definitions into several major categories and give their relationships. 
Each category of definitions could be seen as a tree, where we can select one of the leaf nodes in each tree to form a sharding blockchain.
In the following, we introduce these definitions related to network model, adversary model, transaction model, intra-shard consensus, and sharding blockchains.

\subsubsection{Network Model}
\label{subsubsec:network_model}
To describe the network model more precisely, we divide network models into message transmission models\footnote{Note that in other work \cite{ZAZ+19,WSN+19}, network model mainly refers to message transmission model in our paper.} and node admission models as follows. 

\paragraph{Message transmission model}
We assume that nodes participate in the network with authentication. The messages sent by the nodes are signed, and an adversary cannot forge the signature of any honest node.
There are different models for the message delay rules. 
Next, we give the definitions of three different message transmission models.


\begin{definition}[Synchronous Network \cite{DLS88}]
	\label{def:syn_network}
	In a synchronous network, messages between honest nodes are propagated in rounds. In each round, the messages sent by honest nodes can reach all other honest users before the next round. Each round has a fixed length of time.
\end{definition}

Synchronous networks are relatively strong network models.

There are two kinds of definitions for a partially synchronous network. 
\begin{definition}[Partially Synchronous Network-Definition A\cite{DLS88}]
	\label{def:partial_syn_network}
	In a partially synchronous network, there is a certain upper bound $\Delta$ of message transmission delay in the network. The parameter $\Delta$ cannot be used as a parameter in the design of a protocol. 	
\end{definition}

\begin{definition}[Partially Synchronous Network-Definition B \cite{DDS87}]
	In a partially synchronous network, there is a known bound $\Delta$ and an unknown Global Stabilization Time (GST), such that after GST, all transmissions between two honest nodes arrive within time $\Delta$. 
\end{definition}

Under this definition, a protocol usually ensures safety and guarantees progress within a bounded duration after the GST.	
The partially synchronous network is a commonly used network model in the analysis of blockchain protocols.

\begin{definition}[Asynchronous Network \cite{DDS87}]
	\label{def:asyn_network}
	In an asynchronous network, there exists an adversary who can arbitrarily delay or reorder messages of honest nodes, as long as the messages of honest users can reach each other. There is no upper limit of message transmission delay.
\end{definition}

About the asynchronous network, the FLP impossibility theorem \cite{FLP85} argues that in an asynchronous system where the network is reliable but where crash failures are allowed, there is no deterministic consensus mechanism that can solve the consensus problem.

Most blockchain protocols adopt one of the above three models. However, in a sharding blockchain, two situations need to be considered. The first is the model of the entire network, and the second is the internal model of each shard. These two models could be the same or different and need to be analyzed according to actual conditions and requirements.

\paragraph{Node admission model}
Note that we assume that the nodes in the network are homogeneous, i.e., each node has close computation and communication capabilities. 
In blockchain protocols, the rules for nodes to participate in the protocols are different. We name these rules as node admission models and divide them into permissioned and permissionless networks as follows.

\begin{definition}[Permissioned Network]
	A permissioned network means that the nodes participating in the protocol must first complete identity authentication. 
\end{definition}

Identity authentication is usually done through a trusted third party, e.g., a certificate authority (CA). 
In a permissioned network, the number and identities of all nodes are known. This model is mostly adopted by some internal protocols of enterprises or units \cite{SMC+17}.

\begin{definition}[Permissionless Network]\label{def:permissionless}
	In a permissionless network:
	\begin{itemize}
		\item Any node can join or leave at any time;
		\item No identity authentication is required;
		\item The number of participating nodes varies at any time and is unpredictable.
	\end{itemize}
\end{definition}

So in a permissionless network, information about the number and identity of all participant nodes is unknown. 
The read and write rights of the data are generally open to all nodes, guaranteeing the decentralization property \cite{YJZ+19}.

The words ``permissioned'' and ``permissionless'' in the above two models can usually be combined with different terms, such as ``permissioned blockchain'', which means a blockchain protocol in a permissioned network, or ``permissionless consensus'', which means a consensus algorithm in a permissionless network.

\subsubsection{Adversary Model}
The adversary model describes the various capabilities of an adversary in a protocol. We divide the adversary model into the corruption model, the total proportion model, and the intra-shard proportion model.

\paragraph{Corruption model}
In this model, an adversary could completely control a target node and obtain its secret information, the input and output messages.

We describe an adversary's corruption ability from two aspects, i.e., corruption timing and corruption speed. First, according to the timing at which an adversary can launch a corruption attack, the corruption model could be divided into static and adaptive corruption. Second, according to the time taken by an adversary to complete the corruption attack, there are mild corruption and immediate corruption.

\begin{definition}[Static Corruption]
	A static corruption means that an adversary can only select its corruption targets before the protocol starts. Once the protocol starts running, it cannot choose other honest nodes to corrupt. 
\end{definition}

\begin{definition}[Adaptive Corruption \cite{CDD+99}]
	Adaptive corruption means that an adversary is able to dynamically corrupt target nodes according to the information collected during the operation of the protocol.
\end{definition}

The above two models are usually considered in some cryptographic protocols. In the context of sharding blockchains, it is more important to consider the corruption speed since this will affect the reconfiguration process of a sharding blockchain. Specifically, if the shard members remain unchanged, then an adversary may corrupt and control one of the shards after a period of time. Therefore, a sharding blockchain needs to update committee members at regular intervals, where the interval is called an epoch.
In order to make the description clearer, we introduce the definition of epoch here.


\begin{definition}[Epoch]
	An epoch in a sharding blockchain refers to the time period during which all shard members remain unchanged and continue to operate. Different epochs correspond to different shard member configurations.
\end{definition}

A reconfiguration refers to switching from one epoch to another, i.e., the process of updating the shard members.
The time length of an epoch is closely related to the time required for the adversary to complete the corruption. To better describe this, we give the following two definitions.

\begin{definition}[Mild Corruption \cite{PassS17hybrid}]
	Mild corruption means that it takes a certain amount of time which is usually denoted as $\tau$ for an adversary to corrupt a node. An adversary first issues the corruption command at time $t$ to the target node. After $\tau$ time, the target node is corrupted and becomes a malicious node. Before time $t+\tau$, the target node remains honest. 
\end{definition}

We call $\tau$ the corruption time parameter in this paper.

\begin{definition}[Immediate Corruption]
	Immediate corruption means that an adversary's corruption attack is effective immediately, i.e., $\tau=0$.
\end{definition}

At present, most sharding blockchains adopt the mild corruption model. As far as we know, the immediate corruption model is only used in few blockchain protocols, e.g., Algorand \cite{GHM+17}.

\paragraph{Total proportion model}
The total proportion model refers to a certain limit on the proportion of computational power or stake that an adversary can control in the whole network. For an entire blockchain protocol, a percentage or fraction is generally used. The common used total proportion model might be denoted by 25\%, 1/3, 49\%, etc. The adversary proportion is less than or equal to these specific percentages or fractions.

For the sake of consistency, we use fractions to indicate the total proportion of the adversary in this paper. According to its relationship with the intra-shard proportion, we divide the total proportion into $[0,1/3)$ and $[1/3,1/2)$, as shown in Fig~\ref{fig:definitions}.
$[0,1/3)$ means that the proportion is greater than or equal to $0$ and less than $1/3$. Similarly, $[1/3,1/2)$ refers to the proportion greater than or equal to $1/3$ and less than $1/2$.

\paragraph{Intra-shard proportion model}
Generally speaking, the representation of an adversary model in a shard is different, since the number of shard members is usually limited and fixed. 
Specifically, the relationship between the number of nodes controlled by an adversary and the total number of shard members is expressed in the form of an equation.
$f$ is used to represent the number of malicious nodes, and $u$ denotes the total number of nodes in a shard.\footnote{In other papers, $n$ is usually used to denote the number of shard members. In this paper, we use $n$ to represent the total number of nodes in the protocol. To avoid conflicts, we use $u$ to denote the number of members in a shard.} The intra-shard adversary proportion model could be described as follows.

\begin{definition}[$ u=2f+1 $]
	\label{def:2f}
	The proportion of malicious nodes that an adversary controls accounts for no more than 1/2 of the whole shard.
\end{definition}

\begin{definition}[$ u=3f+1 $]
	\label{def:3f}
	The proportion of malicious nodes that an adversary controls accounts for no more than 1/3 of the whole shard.
\end{definition}

These two models are usually utilized when there are committees running some BFT algorithms in the shards.
$ u=2f+1 $ is often in need in some synchronous BFT protocols, while partially synchronous BFT protocols usually require the model to be $ u = 3f + 1 $.

\subsubsection{Transaction Model}
The main function of most blockchain systems is to process transactions. A transaction usually contains information such as timestamp, input, output, signature, etc., and is often used to realize the transfer of property. Different blockchain systems may adopt different transaction models. The existing transaction models are mainly divided into the UTXO model, account model, and others.
The UTXO model and account model are the most commonly used models, and we term such a model a ``generic'' one. Others refer to some special transaction models. For instance, in Hyperledger Fabric \cite{SMC+17}, one can create transactions by not associating a balance with an account.

\paragraph{UTXO model}
The UTXO model is the most commonly used blockchain transaction model.

\begin{definition}[UTXO Model \cite{HZL+20}]\label{def:utxo}
	In the UTXO model, assets (money/coins/stakes) are stored in unspent transaction outputs (UTXOs). Each UTXO contains the public key address of the output and its value. Each transaction spends existing UTXOs and creates new UTXOs, essentially transferring assets from the input public key address to the output public key address. Besides, a valid transaction requires that the sum of values in the output UTXOs must be equal to that in the input UTXOs.
\end{definition}

\paragraph{Account model}
The account model is another common blockchain transaction model defined as follows.
\begin{definition}[Account Model \cite{HZL+20}]
	In the account model, each user has a fixed account. The account information includes the account address and account balance, and the account balance must be non-negative. A transaction is to transfer assets from one account to another account. A valid transaction requires that the balance in the input account is greater than or equal to the transaction amount.
\end{definition}

\subsubsection{Intra-Shard Consensus}
Intra-shard consensus is crucial to sharding blockchains. Next, we will introduce definitions related to intra-shard consensus in terms of weak and strong consistency.

\paragraph{Weak consistency}
Weak consistency is also known as eventual consistency, which is defined as follows.

\begin{definition}[Weak Consistency]\label{def:weak_consistency}
	Weak consistency means that different nodes might end up having different views of a blockchain, which leads to forks. A certain number of blocks at the end of the blockchain need to be truncated to obtain stable transactions.
\end{definition}

Specifically, based on the election method of a block producer, there might be two or more legal block producers in the same round. In this case, a short-term fork might appear in a blockchain. However, after a certain period of time, a final blockchain is determined according to some kind of decision rule, such as the longest chain rule of Bitcoin \cite{N08} or the heaviest chain rule of GHOST \cite{SZ15}. 
Other blockchain systems that have weak consistency property are PPCoin \cite{KN12}, Ouroboros \cite{KRD+17}, etc. 

The following three properties proposed by the Bitcoin backbone protocol \cite{GKL15} are suitable for blockchains with weak consistency property.

\begin{definition}[Common Prefix \cite{GKL15}]
	For any two blockchains $ \chains_1,\chains_2 $ output by any two honest nodes $P_1,P_2$ in any two rounds $ r_1,r_2 $, it holds that $ \chains_1^{\lceil k} \preceq \chains_2^{\lceil k} $ or $\chains_2^{\lceil k} \preceq \chains_1^{\lceil k}$ where $ k \in \mathbb{N} $ is the security parameter. That is to say, when $ r_1 \leq r_2 $ is satisfied, it holds that $ \chains_1^{\lceil k} \preceq \chains_2^{\lceil k}$; when $ r_2 \leq r_1 $ is satisfied, it holds that $\chains_2^{\lceil k} \preceq \chains_1^{\lceil k}$. $ \chains_1^{\lceil k}$ means removing the ending $k$ blocks of $ \chains_1 $, $ \chains_1 \preceq \chains_2$ denotes that $\chains_1$ is a prefix of $ \chains_2 $. $ P_1 , P_2 $ might be the same node. 
\end{definition}

The common prefix property could be understood in the following way. The blockchains held by honest nodes will eventually be consistent with each other, and the stable part of a blockchain will not be rewritten. After removing the last $k$ blocks, the remaining blockchain is regarded as the stable part. The value of $k$ is usually related to system security.

\begin{definition}[Chain Quality \cite{GKL15}]
	For any blockchain $\chains$ output by any honest node $ P $, after removing the latest $ k_0 $ blocks, the proportion of malicious blocks in any $ k $ consecutive blocks does not exceed $ \mu $. The security parameters are $ \mu \in \mathbb{R},k,k_0 \in \mathbb{N}$.  
\end{definition}

The chain quality property means that there must be a sufficient proportion of consecutive blocks generated by honest users. The last $ k_0 $ blocks are usually the ``unstable'' blocks at the end of a blockchain. 

\begin{definition}[Chain Growth \cite{GKL15}]
	For any round $ r$ $(r>r_0) $ and any honest node $ P $, if $P$ outputs $ \chains_1 $ and $ \chains_2 $ in round $ r $ and $ r+s $, respectively, then it holds that $ |\chains_2|-|\chains_1| \geq \tau \cdot s$. The security parameters are $\tau \in \mathbb{R},s,r_0 \in \mathbb{N}$. 
\end{definition}

The chain growth property means that a blockchain will continuously generate new blocks, and the block generation speed has a lower bound.

\paragraph{Strong consistency}
Strong consistency means that there is no need to wait for a block to reach a certain depth to confirm transactions. Blocks and transactions are confirmed immediately. 

\begin{definition}[Strong Consistency]\label{def:strong_consistency}
	Strong consistency means that the generation of each block is deterministic and instant. Besides, strong consistency has the following characteristics:
	\begin{itemize}
		\item There is no fork in a blockchain. By running a distributed consensus algorithm, state machine replication (ref. Definition~\ref{def:state_machine_replication}) is achieved;
		
		\item Transactions could be confirmed more quickly. As long as a transaction is written into a block, the transaction could be regarded as valid;
		
		\item Transactions are tamper-proof. As long as a transaction block is written to a blockchain, the transaction and block will not be tampered with and the block will remain on the chain at all times.
	\end{itemize}
\end{definition}

The blockchain systems that have strong consistency property are PeerCensus \cite{DSW16}, ByzCoin \cite{KJG+16} and its adaptation MOTOR \cite{Kogias19}, Hybrid Consensus \cite{PS17}, Solida \cite{AMN+17}, Omniledger \cite{KJG+18}, etc. 
In these blockchains, there is a committee or multiple committees that run distributed consensus algorithms, e.g., PBFT \cite{CL99}, to confirm transactions and generate new blocks. These distributed consensus algorithms achieve state machine replication, which is defined as follows.  


\begin{definition}[State Machine Replication \cite{S90}]\label{def:state_machine_replication}
	State machine replication is a general method for a set of servers, which include a single primary and other backups, to reach an agreement on a linearly-ordered log, where the following two security properties must be satisfied.
	\begin{itemize}
		\item Consistency, i.e., the views of all honest servers must be identical to each other.
		\item Liveness, i.e., whenever one piece of valid data is submitted to the servers, it will be written to the log within some bounded time.
	\end{itemize}
\end{definition}

State machine replication is also known as atomic broadcast, and it has been studied for decades in the area of distributed systems. State machine replication is often used to synchronize large databases. For example, Google and Facebook use it for the synchronization of core parts of their databases \cite{TS16}.

State machine replication needs to tolerate a specific proportion of faulty nodes. 
A faulty node might be a crashed node or a Byzantine node. A crashed node refers to a node that does not respond. The node may have a system failure or be offline. A crashed node is a relatively simple faulty node compared with a Byzantine node defined as follows.


\begin{definition}[Byzantine Node]
	\label{def:byzantine_node}
	A node is called a Byzantine node if it could behave arbitrarily as follows \cite{Fischer83}.
	\begin{itemize}
		\item It does not respond to messages sent to it;
		
		\item It sends different messages to different nodes when such messages were supposed to be identical.
	\end{itemize}
	Byzantine nodes are considered to be controlled by an adversary. A Byzantine node must observe some restrictions, which are also restrictions on the adversary, usually given in the network model and the adversary model.
\end{definition}

The Byzantine node model is a commonly used fault model in distributed systems, and an algorithm that can tolerate such nodes is called a Byzantine Fault Tolerance (BFT) algorithm, which is defined as follows.

\begin{definition}[Byzantine Fault Tolerance \cite{Fischer83}]
	\label{def:BFT}
	A set of nodes achieve state machine replication and satisfy consistency and liveness in the presence of Byzantine nodes. 
\end{definition}

There is usually a certain constraint on the proportion of Byzantine nodes.
For instance, in a partially synchronous network model, the system is usually able to tolerate at most $1/3$ Byzantine nodes. In a synchronous network, the largest tolerated fraction of Byzantine nodes is below $1/2$.



\subsubsection{Sharding Blockchains}
After determining the network model, adversary model, transaction model, intra-shard consensus, and other components, a sharding blockchain is constructed.
We give a formal definition of a secure sharding blockchain. Since this definition is based on that of a ``public ledger'' by Garay \textit{et al.} \cite{GKL15}, we first introduce the definition of a ledger.

\begin{definition}[Ledger]
	A ledger refers to a credible ``bulletin board'' that meets the following properties: 
	\begin{itemize}
		\item Persistence. Persistence means that for any honest node, if the node outputs a transaction $\tx$ at a certain position in his ledger, such as the $ j $-th transaction of the $ i $-th block, then $\tx$ must appear in the identical position in the ledgers of all honest nodes.
		\item Liveness. Liveness means that if a valid transaction $\tx$ is uploaded at time $t$, then after a certain time, $\tx$ must appear in the ledgers output by all honest nodes.
	\end{itemize}
	The participating nodes are able to write some data on a ledger if the data meets the specific rules of the ledger. 
\end{definition}

Note that in the original Bitcoin backbone protocol \cite{GKL15}, the definition of a public ledger is given. In order to make the concept of ``ledger'' better applicable to various network models, we make a slight modification and directly define a ledger. Then different node admission models correspond to different types of ledgers, i.e., a private ledger in a permissioned network, and a public ledger in a permissionless network.


\paragraph{A secure sharding blockchain}
A sharding blockchain could be regarded as a special type of ledger. In order to make the description clearer, we give a relatively formal definition of a secure sharding blockchain in the following.
\begin{definition}[A Secure Sharding Blockchain]
	\label{def:secure_sharding_blockchain}
	Let $(\adv,\env)$ be an adversary and environment pair w.r.t. a sharding consensus protocol $\rm\Pi$.  
	$T_{\text{initial}}$ denotes the time for a sharding blockchain protocol to start up, including the production of genesis blocks and initial committees. $T_{\text{liveness}}$ denotes the transaction confirmation delay parameter, i.e., the time required to commit a transaction. 
	We say $\rm\Pi$ is secure w.r.t. $(\adv,\env)$ with parameters $T_{\text{initial}}, T_{\text{liveness}}$ if the following properties hold with an overwhelming probability:
	\begin{itemize}[leftmargin=*]
		\item Consistency. Consistency includes the following two properties:
		\begin{itemize}[leftmargin=*]
			\item Common prefix inside a shard: For any two honest nodes $i,j \in \shard_c$ where $c \in [1,m]$, node $i$ outputs $\logs_i$ to $\env$ at time $t$, and node $j$ outputs $\logs_j$ to $\env$ at time $t'$, it holds that either $\logs_i \preceq \logs_j$ or $\logs_j \preceq \logs_i$. 
			\item No conflict between shards: For any two honest nodes $i \in \shard_c, j \in \shard_{c'}$ where $c,c' \in [1,m]$ and $c \neq c'$, node $i$ outputs $\logs_i$ to $\env$ at time $t$, and node $j$ outputs $\logs_{j}$ to $\env$ at time $t'$. For any transaction $\tx_1 \in \logs_i$ and $\tx_2 \in \logs_{j}$ where $\tx_1 \neq \tx_2$, it holds that $\tx_1$ and $\tx_2$ don't conflict with each other, i.e., there is no input that belongs to $\tx_1$ and $\tx_2$ simultaneously.
		\end{itemize}
		\item $T_{\text{liveness}}$-Liveness: For any honest node from any shard, if it receives a transaction $\tx$ at time $t_0 \geq T_{\text{initial}}$ from $\env$, then at time $t_0 + T_{\text{liveness}}$, $\tx$ must be accepted or rejected.
	\end{itemize}
\end{definition}

Note that in the description of ``no conflict between shards'', we require that the condition $\tx_1 \neq \tx_2$ holds since different shards might record the same cross-shard transaction in their blockchain, respectively. The essence of ``no conflict between shards'' is that no double-spending happens for any UTXO.
This paper uses the term "sharding blockchain" and "secure sharding blockchain" interchangeably for the same meaning.

\paragraph{Transaction confirmation}
Generally, the main purpose of a blockchain is to complete transaction processing and confirmation. In the following, the definition of transaction confirmation delay and responsiveness are given, both of which are very important evaluation indicators.

\begin{definition}[Transaction Confirmation Delay]
	For a transaction $\tx$, if it is submitted by a client at some time $t$, and it appears in an honest node's ledger at time $t'$, then $t'-t$ is the transaction confirmation delay. $t'-t$ could also be regarded as the liveness parameter of a ledger. 
\end{definition}

Transaction confirmation delay refers to the time needed for a valid transaction to be confirmed by a blockchain. Namely, it ranges from the time that the transaction is submitted by a client, to the time that the transaction appears in an honest node's ledger. 

Besides, Pass and Shi \cite{PS17} propose the concept of responsiveness. 

\begin{definition}[Responsiveness \cite{PS17}]
	Responsiveness means the confirmation time of a transaction is only related to the actual network delay $\delta$, but not to a priory upper bound $\Delta$.
\end{definition}

The concept of responsiveness is used in much related research \cite{ANR+20,MCK20} now as an evaluation indicator for transaction confirmation.

The systematic introduction of these definitions aims at supporting the concept of composability of components in Section~\ref{sec:decomposing}. Specifically,
as Fig.~\ref{fig:definitions} indicates, we could select one of the leaf nodes from each ``tree'' on the left to form a complete sharding blockchain system. For instance, in Fig.~\ref{fig:definitions}, we have selected the ``Partially Sync'' in the ``Message Transmission Model'', the ``PoW'' and ``Permissionless'' in the ``Node Admission Model'', the ``Adaptive'' in the ``Corruption Timing'', the ``Mild'' in the ``Corruption Speed'', the ``UTXO Model'' in the ``Transaction Model'', and the ``Byzantine Fault Tolerance'' in the ``Intra-Shard Consensus'' trees. In this way, we obtain a complete sharding blockchain system that is very similar in its assumptions to Omniledger \cite{KJG+18}.

Interestingly, through this choice-combination approach, we could obtain a ``sharding blockchain system generator'' which could be used to produce a variety of different sharding blockchain systems. Considering the practical combinations, the corruption timing and speed is usually set to ``Adaptive'' and ``Mild'', respectively. For the node admission model, intra-shard consensus, and transaction model, if we exclude the ``Others'' leaf options, then there are $3*3*1*1*3*2*4=216$ types of different secure sharding blockchain systems that might be composed. 
The constraints between some models have been taken into consideration, such as the total proportion and the intra-shard proportion.

\section{Decomposing Sharding Blockchains into Functional Components}
\label{sec:decomposing}
In this section, we decompose sharding blockchains into functional components. Section~\ref{subsec:decomposing} describes the inputs, outputs, basic functions, and properties of each component. Section~\ref{subsec:composing} provides concrete methods to compose different components into a complete sharding blockchain system. Section~\ref{subsec:summary} summarizes existing sharding blockchain schemes.

\subsection{Decomposition of Sharding Blockchains}
\label{subsec:decomposing}
\begin{figure}[h!]
	\centering
	\includegraphics[width=0.7\textwidth]{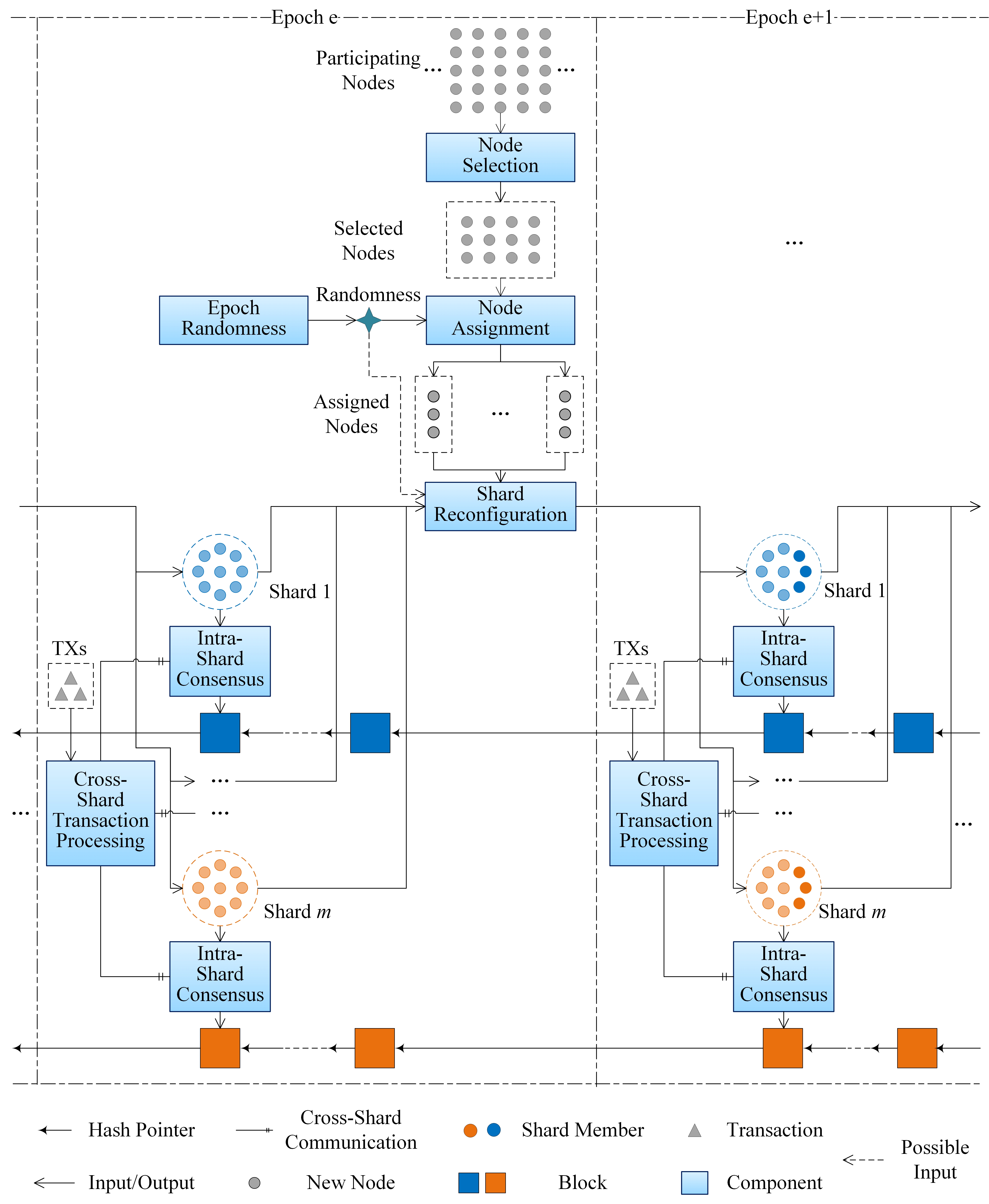}
	\caption{The schematic diagram of a sharding blockchain.}
	\label{fig:diagram}
\end{figure}

Inspired by the concept of a framework in \cite{FSJ99}, we propose a framework for sharding blockchains and decompose it into functional components, including node selection, node assignment, epoch randomness, intra-shard consensus, cross-shard transaction processing, shard reconfiguration, and motivation mechanism. These components could be composed into a complete sharding blockchain system. The schematic diagram of a sharding blockchain is shown in Fig.~\ref{fig:diagram}.

Next, we give an informal description of each component. In each component, we describe the interfaces (inputs and outputs), basic functions, and properties to be satisfied.

\subsubsection{Node Selection} 
As shown in Component~\ref{alg:ns}, $\ns$ is a subset selection component.
$\ns$ takes in $pnodes$ as an input, which represents a set of participating nodes in the network, and outputs $snodes$, which denotes a set of selected nodes. 
$\{P_i\}_{|km|}$ denotes that there are $km$ nodes in the set.
$snodes$ is a subset of $pnodes$, and $snodes$ might be used by the $\na$ component. $k$ denotes the number of old members that each shard needs to replace during each reconfiguration (equal to the number of newly added members of each shard). So $km$ represents the number of selected nodes required for each reconfiguration. 
In order to select $km$ nodes, the total number of participating nodes $n$ must be greater than $km$.


\begin{component}[h!]
	\justifying
	\noindent	\textbf{input:} a set of participating nodes $pnodes = \{P_1, \cdots, P_n\}$. 
	
	\noindent	\textbf{output:} a subset of $pnodes$ containing selected nodes $snodes=\{P_i\}_{|km|}$.
	
	\noindent	\textbf{function:} select a certain number of qualified nodes from all participating nodes. 
	
	\noindent	\textbf{property:} fairness; robustness.
	
	\caption{$\ns$ (Node Selection)}
	\label{alg:ns}
\end{component}

The basic function of $\ns$ is to select qualified shard members from all participating nodes. In a permissionless network, each node could be a member of $pnodes$. $\ns$ might make use of some mechanisms, such as PoW and PoS, to defend against the Sybil attacks \cite{Douceur02}, where an adversary tries to increase his proportion in $snodes$ by creating fake identities. 
In a permissioned network, the function of $\ns$ is implemented by a trusted third party, e.g., CA. The CA completes the confirmation of $snodes$ by providing the identity registration service.

The properties that need to be satisfied by $\ns$ are fairness and robustness. The definition of fairness is given in Definition~\ref{def:fair_selection}, which is mainly used to limit the proportion of the adversary's nodes in $snodes$. Robustness means that even with the participation of the adversary, $snodes$ will still be confirmed by honest nodes.
\subsubsection{Epoch Randomness}
As shown in Component~\ref{alg:er}, $\er$ is usually an interactive and distributed component where each participant creates a private input, namely $x_1,x_2,\cdots,x_q$. We use $q$ to denote the number of participants. The epoch randomness is denoted by $\xi_e$ where $e$ represents the epoch number.

\begin{component}[htp]
	\justifying
	\noindent	\textbf{input:} $q$ private inputs $x_1,x_2,\cdots,x_q$.
	
	\noindent	\textbf{output:} an epoch randomness $\xi_e$.
	
	\noindent	\textbf{function:} participants run a randomness generation protocol to generate a secure randomness.
	
	\noindent	\textbf{property:} public-verifiability; unpredictability; bias-resistance; availability.
	
	\caption{$\er$ (Epoch Randomness)}
	\label{alg:er}
\end{component}

The basic function of $\er$ is to enable each honest node to obtain an identical randomness through the interactions with participants. The randomness must be secure, i.e., satisfies the following properties: public-verifiability, unpredictability, bias-resistance, and availability. Public-verifiability means that every node can verify the correctness of $\xi_e$. Unpredictability indicates that no one can obtain the randomness in advance. Bias-resistance means that the adversary?s participation will not affect the result of the randomness and availability is to ensure that the randomness is sure to be generated.
The concrete explanation of these properties for a randomness is given in Section~\ref{subsec:er_bp}. 
An epoch randomness is usually utilized as a seed to assign nodes randomly into shards and used as a fresh puzzle in PoW mining. 

\subsubsection{Node Assignment}
As shown in Component~\ref{alg:na}, $\na$ represents the node assignment component, which takes in $snode$ and $\xi_e$ as inputs, and outputs $anodes$. $anodes$ denotes an assigned node list which might contain $m$ groups of nodes. $\{P_i\}_{|k|}$ denotes that there are $k$ elements in the set.
$\na$ is to map the $km$ nodes in $snodes$ to a set $anodes$ that includes $m$ different subsets $a_1,\cdots,a_m$, and each subset $a_j$ contains $k$ nodes. 

\begin{component}[htp]
	\justifying
	\noindent	\textbf{input:} selected nodes $snodes$, epoch randomness $\xi_e$. 
	
	\noindent	\textbf{output:} assigned nodes $anodes=\{a_1,\cdots,a_m\}$ where $a_j = \{P_i\}_{|k|}$ for every $j=1$ to $m$.
	
	\noindent	\textbf{function:} assign selected nodes into $m$ different subsets randomly based on the epoch randomness $\xi_e$. 
	
	\noindent	\textbf{property:} random distribution; robustness.
	\caption{$\na$ (Node Assignment)}
	\label{alg:na}
\end{component}

The basis function of $\na$ is to assign the selected new nodes randomly to multiple shards. The random distribution of new nodes is required to prevent an adversary from centralizing the nodes controlled by himself into a certain shard. Similarly, robustness is to ensure the final execution of the assignment operation. Generally speaking, the epoch randomness $\xi_e$ is treated as a random seed for a pseudorandomness generator \cite{Luby96}, to produce multiple pseudorandomnesses for each new node as a reference to be assigned. 

Note that the list $anode$ is not the same as the list of nodes participating in the entire protocol in the next epoch, since different sharding blockchains might have different replacement rules to substitute the old nodes in each shard with new nodes. The replacement rule is determined by the $\sr$ component and will be introduced in Section~\ref{subsubsec:sr}.


\subsubsection{Intra-Shard Consensus}
Component~\ref{alg:isc} shows the interfaces, functions and properties of the intra-shard consensus $\isc$. $\isc$ takes continuous proposals as inputs, and each proposal could be represented by $p$ with a different subscript. In normal blockchains, $\isc$ is used to process transactions, which means the proposals are transactions. In sharding blockchains, a proposal $p$ could be a transaction, a transaction input, or other values to be committed. $\isc$ outputs $\langle p \rangle$, where the notation $\langle \cdot \rangle$ denotes that the value is committed.

\begin{component}[htp]
	\justifying
	\noindent	\textbf{input:} a proposal $p$.
	
	\noindent	\textbf{output:} a committed $\langle p \rangle$. 
	
	\noindent   \textbf{function:} shard members run some consensus algorithm to commit proposals, i.e., reach agreement on proposals.
	
	\noindent	\textbf{property:} consistency; liveness.
	
	\caption{$\isc$ (Intra-Shard Consensus)}
	\label{alg:isc}
\end{component}

The basic function of $\isc$ is to process and commit proposals. $\isc$ might be divided into strong consistency (ref. Definition~\ref{def:strong_consistency}) and weak consistency (ref. Definition~\ref{def:weak_consistency}), depending on the specific algorithm adopted. Strong consistency corresponds to classic distributed consensus algorithms, such as BFT-type algorithms, while weak consistency corresponds to PoW, PoS, and other methods.

Regardless of the specific implementation method adopted by $\isc$, $\isc$ should meet the consistency and liveness properties. Consistency ensures that all honest nodes have an identical view of the commitment values and liveness guarantees that $\isc$ could process any proposal within a period of time.
In the asynchronous network transmission model, since there is no GST, when the network condition is bad, the protocol can only choose either consistency or liveness. As an intra-shard consensus algorithm for sharding blockchains, the assurance of consistency is more important.

\subsubsection{Cross-Shard Transaction Processing}
As shown in Component~\ref{alg:cstp}, $\cstp$ takes in a transaction package $TXs$ as input. Note that in practical situations, transactions may be uploaded individually by users, and different transactions are submitted to the corresponding shard. The output of $\cstp$ is a transaction log denoted by $\logs_c$ which is specific to the current shard. 

\begin{component}[htp]
	\justifying
	\noindent	\textbf{input:} a transaction package $TXs$. 
	
	\noindent	\textbf{output:} $m$ committed transaction logs $\logs_1,\cdots,\logs_m$. 
	
	\noindent	\textbf{function:} $\cstp$ invokes $\isc$ inside each shard to process and commit cross-shard transactions through interactions and communication with other related shards. In each shard, $\cstp$ outputs the corresponding transaction log or blocks. 
	
	\noindent	\textbf{property:} common prefix inside a shard; no conflict between shards; liveness.
	
	\caption{$\cstp$ (Cross-Shard Transaction Processing)}
	\label{alg:cstp}
\end{component}

$\cstp$ includes and makes use of $\isc$. The basic function of $\cstp$ is to process cross-shard transactions. Note that cross-shard transactions occupy most of the transactions in a sharding blockchain, and intra-shard transactions could also be processed by $\cstp$ as special cross-shard transactions. In most sharding blockchains, the processing of cross-shard transactions could be divided into two phases. In the first phase, input shards generate proofs to prove if the transaction inputs are available or not and send the proofs to related shards. In the second phase, all related shards verify if the transaction is valid through the proofs received. 
Please see Section~\ref{sec:cross-shard} for a detailed analysis.

The properties that $\cstp$ needs to satisfy are the same as those of a secure sharding blockchain defined in Definition~\ref{def:secure_sharding_blockchain}, that is, consistency and liveness. Due to the special scenario of the sharding blockchain, the consistency property includes common prefix inside a shard and no conflict between shards.

\subsubsection{Shard Reconfiguration}
As shown in Component~\ref{alg:sr}, $\sr$ takes in the assigned nodes $anodes$ and the list of epoch $e$ $\lists_e$ as inputs, and outputs the shard member list $\lists_{e+1}$ of epoch $e+1$. $\comm=\{P_i\}_{|u|}$ is a set including $u$ nodes.
Note that we use $\comm$ to denote committee members when there is a committee in a shard, while $\comm$ could also be used to denote shard members when there is no committee in a shard.

\begin{component}[htp]
	\justifying
	\noindent	\textbf{input:} assigned nodes $anodes$, a shard member list $\lists_e$ of epoch $e$.
	
	\noindent	\textbf{output:} a shard member list of epoch $e+1$: $\lists_{e+1}=\{\comm^1_{e+1},\cdots,\comm^m_{e+1}\}$ where $\comm_{e+1}^j=\{P_i\}_{|u|}$ for every $j=1$ to $m$. 
	
	\noindent	\textbf{function:} confirm the shard member list of epoch $e+1$ based on $\lists_e$ and $anodes$, i.e., determine which old nodes of each shard are replaced by new nodes; specify the details of bootstrapping when new nodes join the shard. 
	
	\noindent   \textbf{property:} honest shard; liveness.
	\caption{$\sr$ (Shard Reconfiguration)}
	\label{alg:sr}
\end{component}
\label{subsubsec:sr}

Due to the adversary corruption attack, shards or committees need to be updated after a certain period of time, or an adversary might control a shard. The basic functions of $\sr$ are to determine which nodes participate in the protocol in epoch $e+1$, namely the members of each shard. Generally, in order to ensure that the blockchain can still process transactions normally during reconfiguration, only part of the old members are replaced with new ones during reconfiguration. The replacement process may be random (epoch randomness $\xi_e$ is useful in this case, as shown in Fig.~\ref{fig:diagram}), or rely on other special rules. Please refer to Section~\ref{subsec:sharding_blockchains} for details. In addition, $\sr$ needs to design the bootstrapping details when a new node joins the shard, such as downloading historical transaction data and UTXO/account data.

One of the properties that $\sr$ needs to satisfy is to ensure that each shard is honest. An honest shard means that the honest member proportion in each shard exceeds the preset safety threshold, which is determined by the $\isc$ component inside the shard. For instance, $u \geq 3f+1$ should be satisfied if $\isc$ adopts PBFT \cite{CL99} as its basic algorithm.

\subsubsection{Motivation Mechanism}
As shown in Component~\ref{alg:mm}, $\mm$ does not have clear inputs and outputs, since $\mm$ is usually determined by the entire blockchain protocol, rather than as a local algorithm that could be called by nodes.

\begin{component}[htp]
	\justifying
	\noindent	\textbf{function:} reward nodes who participate the protocol positively and honestly;
	punish nodes who behave negatively and maliciously.
	
	\noindent	\textbf{property:} fairness.
	\caption{$\mm$ (Motivation Mechanism)}
	\label{alg:mm}
\end{component}

Generally, a motivation mechanism includes an incentive mechanism to reward the active and honest nodes, as well as a punishment mechanism to collect fines from nodes who behave maliciously or go offline. The penalty mechanism might first require each participating node to pay a certain deposit, and all nodes could report the malicious behaviors of other nodes.

$\mm$ needs to meet the fairness of reward distribution, i.e., assuming that the nodes participating in the protocol are rational, the level of rewards for nodes should correspond to their workload. For example, committee leaders usually have higher computation and communication costs and deserve a higher reward.


\subsection{Composing Separate Components into Sharding Blockchain Systems}
\label{subsec:composing}
The previous section gives the components of sharding blockchains, including their interfaces, basic functions, and properties. Now, we could utilize these components to compose a complete sharding blockchain system. 

\subsubsection{General Methods to Compose a Sharding Blockchain System}
\label{subsubsec:general_methods}
A complete sharding blockchain system includes all the building blocks described earlier in the previous section. In the following, we discuss how to compose all these building blocks into a complete sharding blockchain system.

\begin{figure*}[h!]
	\centering
	\includegraphics[width=0.95\textwidth]{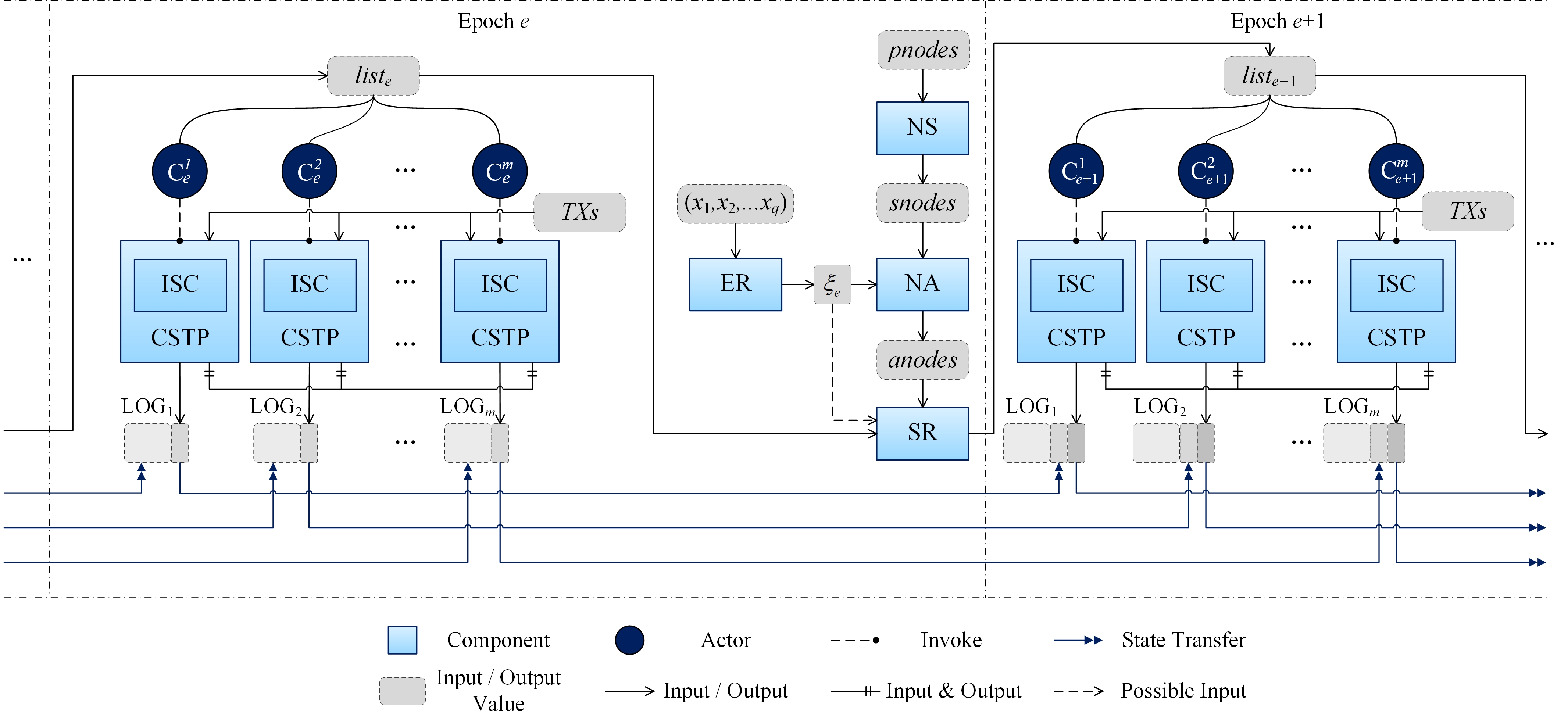}
	\caption{Component composition diagram of a sharding blockchain.}
	\label{fig:components}
	\caption*{\scriptsize For $\cstp$ and $\isc$, it is clear that the actors are committees. For other components, i.e., $\ns,\ \er, \ \na$, and $\sr$, different blockchain protocols might have different actors. For example, in RapidChain, there is a reference committee as an actor. In Omniledger, it is all the participants in the protocol as actors. So we do not mark the actors of these components in the figure.}
\end{figure*}



A complete sharding blockchain protocol $\Pi$ is a composition of the $\ns,\na,\er,\sr,\isc,\cstp,$ and $\mm$ components, and the composition approach is shown in Fig.~\ref{fig:components}. Besides, Fig.~\ref{fig:components} describes the epoch transition operations of a sharding blockchain, i.e., from epoch $e$ to epoch $e+1$. Since $\mm$ is employed by the entire protocol rather than a local algorithm of nodes, $\mm$ is not indicated in Fig.~\ref{fig:components}.
We divided the whole operations of a sharding blockchain into the following two parts.

The first part is the confirmation of the shard member list for epoch $e+1$, including the $\ns, \na, \er$ and $\sr$ components. First, $\ns$ adopts a certain method, such as PoW, PoS and CA, to selects a certain number of qualified nodes $snodes$ among all the participating nodes $pnodes$. Second, at the end of epoch $e$, $\er$ is run to generate a secure epoch randomness $\xi_e$. Third, $\na$ uses $\xi_e$ to randomly allocate all new nodes of $snode$ into $m$ different groups, corresponding to $m$ shards. Note that at this time, $anodes$ only includes new nodes, which means $anodes$ is not equivalent to the list of shard members for the new epoch. At last, $\sr$ takes in $list_e$ and $anodes$ as inputs, determines which old members are to be replaced by new nodes in each shard, and finally generates a shard member list for epoch $e+1$. Note that $\ns$ might take place during the execution of the entire epoch $e$, while $\na$, $\er$, and $\sr$ are usually executed during epoch changes. Besides, the above operations apply for every epoch.

The second part is related to transaction processing, including $\isc$ and $\cstp$. Each shard runs the $\isc$ component. If the $\isc$ has the strong consistency property, then there is a committee inside each shard to run $\isc$. $\cstp$ processes transactions by invoking $\isc$. Note that the inputs from $\cstp$ to $\isc$ are not always transactions, but also other kinds of proposals, e.g., transaction inputs. In addition, inter-shard communication is required among shards to complete the commitments of cross-shard transactions. Finally, each shard outputs its own transaction log, i.e., $\logs_1, \cdots, \logs_m$.

\subsubsection{Distinct Combinations of System Models and Components}
\label{subsubsec:combinations}
In the process of constructing a sharding blockchain system, by choosing distinct kinds of system models and components, we could obtain different types of sharding blockchain systems. 

\paragraph{Distinct system models}
As shown in Fig.~\ref{fig:definitions}, system models include network models, adversary models, and transaction models.

For the message transmission model, a sharding blockchain usually adopts a partially synchronous or a synchronous model. Note that the message transmission model for the entire network and inside each shard in a sharding blockchain could be different.

Regarding the node admission model, by choosing different models, we could obtain permissionless and permissioned sharding blockchains.
In a permissioned network, nodes that participate in the protocol first need to register their identity in a CA. The process of identity registration by the CA could be regarded as a particular way for the permissioned sharding blockchains, e.g., \cite{ACC+18,AAA19,AAA19b}, to implement the $\ns$ component.

For the adversary corruption model, a sharding blockchain usually assumes an adaptive and mild adversary model, since an immediate corruption adversary is much too powerful and unrealistic. For the proportion model, the total and intra-shard proportion models are related. The intra-shard proportion model should first be determined according to $\isc$. Then, the total proportion should be lower than the intra-shard proportion, since in the process of $\na$ and $\ns$, an adversary might increase its proportion through various attacks.

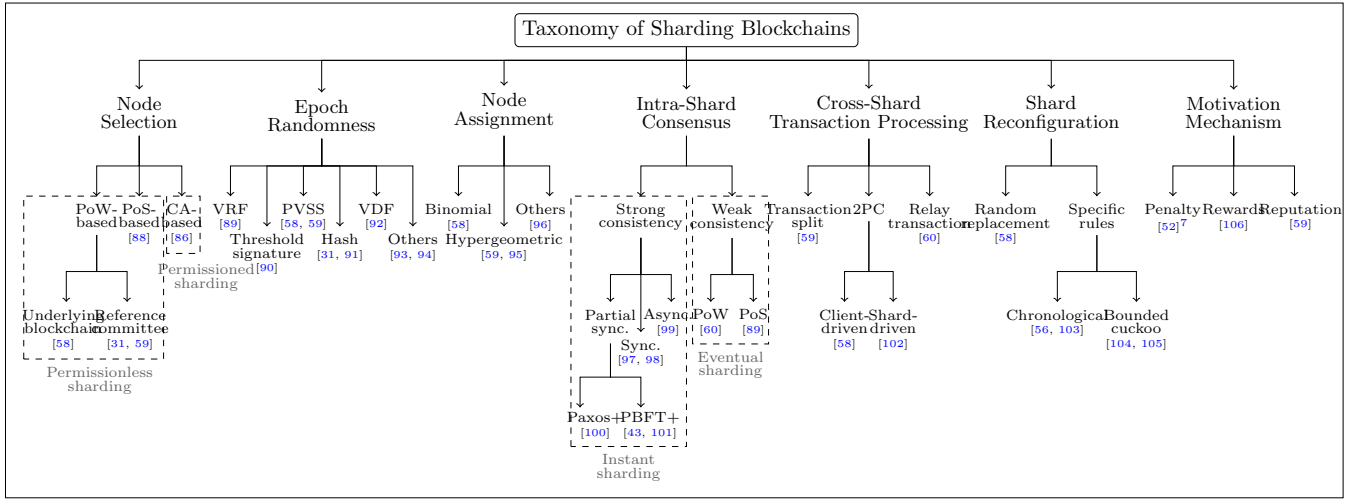
\begin{figure*}
	\centering
	\scalebox{0.8}{
		\begin{tikzpicture}
			\draw (-11.2,-1) rectangle (10.8,7.2);
			\node[draw=black,rounded corners = 2pt] (header) at (0,6.75) {Taxonomy of Sharding Blockchains};
			\node[above,align=center,execute at begin node=\setlength{\baselineskip}{2ex}] (nodeSelection) at (-8-1,5) {\small Node\\\small Selection};
			\node[above,align=center,execute at begin node=\setlength{\baselineskip}{2ex}] (epochRandomness) at (-5-1,5-0.05)  {\small Epoch\\\small Randomness};
			\node[above,align=center,execute at begin node=\setlength{\baselineskip}{2ex}] (nodeAssignment) at (-2-1,5) {\small Node\\\small Assignment};
			\node[above,align=center,execute at begin node=\setlength{\baselineskip}{2ex}] (intraShard) at (-2+3-1,5) {\small Intra-Shard\\\small Consensus};
			\node[above,align=center,execute at begin node=\setlength{\baselineskip}{2ex}] (crossShard) at (1+3-1,5-0.05) {\small Cross-Shard\\\small Transaction Processing};
			\node[above,align=center,execute at begin node=\setlength{\baselineskip}{2ex}] (shardRecon) at (4+3-1,5-0.05) {\small Shard\\\small Reconfiguration};
			\node[above,align=center,execute at begin node=\setlength{\baselineskip}{2ex}] (motivation) at (7+3-1,5) {\small Motivation\\\small Mechanism};
			\draw[->] (header)--(0,6.25)--(-9,6.25)--(nodeSelection);
			\draw[->] (header)--(0,6.25)--(-6,6.25)--(epochRandomness);
			\draw[->] (header)--(0,6.25)--(-3,6.25)--(nodeAssignment);
			\draw[->] (header)--(0,6.25)--(0,6.25)--(intraShard);
			\draw[->] (header)--(0,6.25)--(3,6.25)--(crossShard);
			\draw[->] (header)--(0,6.25)--(6,6.25)--(shardRecon);
			\draw[->] (header)--(0,6.25)--(9,6.25)--(motivation);

			\draw[->] (nodeSelection)--(-9,4.5)--(-10+0.3,4.5)--(-10+0.3,4);
			\node[below,align=center,execute at begin node=\setlength{\baselineskip}{1.5ex}] (pow) at (-10+0.3,4) {\scriptsize PoW-\\\scriptsize based};
			\draw[->] (nodeSelection)--(-9,4.5)--(-9,4.5)--(-9,4);
			\node[below,align=center,execute at begin node=\setlength{\baselineskip}{1.5ex}] (pos) at (-9,4) {\scriptsize PoS-\\\scriptsize based \\ \tiny \cite{LJK19}};
			\draw[->] (nodeSelection)--(-9,4.5)--(-8-0.3,4.5)--(-8-0.3,4);
			\node[below,align=center,execute at begin node=\setlength{\baselineskip}{1.5ex}] (ca) at (-8-0.3,4) {\scriptsize CA-\\\scriptsize based \\ \tiny \cite{AAA19}};
			\draw[->] (pow)--(-10+0.3,3.25-0.5)--(-10.5+0.3,3.25-0.5)--(-10.5+0.3,3-0.5-0.25);
			\node[below,align=center,execute at begin node=\setlength{\baselineskip}{1.5ex}] (underlying) at (-10.55+0.3,3-0.5-0.25) {\scriptsize Underlying\\\scriptsize blockchain \\ \tiny \cite{KJG+18}};
			\draw[->] (pow)--(-10+0.3,3.25-0.5)--(-9.5+0.3,3.25-0.5)--(-9.5+0.3,3-0.5-0.25);
			\node[below,align=center,execute at begin node=\setlength{\baselineskip}{1.5ex}] (underlying) at (-9.45+0.3,3-0.5-0.25) {\scriptsize Reference\\\scriptsize committee \\ \tiny \cite{LNZ+16,ZMR18}};
			
			\draw[->] (epochRandomness)--(-6,4.5)--(-7.5,4.5)--(-7.5,4);
			\draw[->] (epochRandomness)--(-6,4.5)--(-7.5+0.6,4.5)--(-7.5+0.6,4-0.5);
			\draw[->] (epochRandomness)--(-6,4.5)--(-7.5+1.2,4.5)--(-7.5+1.2,4);
			\draw[->] (epochRandomness)--(-6,4.5)--(-7.5+1.8,4.5)--(-7.5+1.8,4-0.5);
			\draw[->] (epochRandomness)--(-6,4.5)--(-7.5+2.4,4.5)--(-7.5+2.4,4);
			\draw[->] (epochRandomness)--(-6,4.5)--(-7.5+3,4.5)--(-7.5+3,4-0.5);
			\node[below,align=center,execute at begin node=\setlength{\baselineskip}{1.5ex}] (vrf) at (-7.5,4) {\scriptsize VRF \\ \tiny \cite{FGK+18}};
			\node[below,align=center,execute at begin node=\setlength{\baselineskip}{1.5ex}] (threshold) at (-7.5+0.6,4-0.5) {\scriptsize Threshold\\\scriptsize signature\\ \tiny \cite{HMW18}};
			\node[below,align=center,execute at begin node=\setlength{\baselineskip}{1.5ex}] (pvss) at (-7.5+1.2,4) {\scriptsize PVSS \\ \tiny \cite{KJG+18,ZMR18}};
			\node[below,align=center,execute at begin node=\setlength{\baselineskip}{1.5ex}] (hash) at (-7.5+1.8,4-0.5) {\scriptsize Hash \\ \tiny \cite{LNZ+16,Zilliqa17}};
			\node[below,align=center,execute at begin node=\setlength{\baselineskip}{1.5ex}] (vdf) at (-7.5+2.4,4) {\scriptsize VDF \\ \tiny  \cite{Buterin17}};
			\node[below,align=center,execute at begin node=\setlength{\baselineskip}{1.5ex}] (vdf) at (-7.5+3,4-0.5) {\scriptsize Others \\ \tiny  \cite{NNL+19,ZKC+19}};
			
			\draw[->] (nodeAssignment)--(-3,4.5)--(-3.75,4.5)--(-3.75,4);
			\draw[->] (nodeAssignment)--(-3,4.5)--(-3,4-0.5);
			\draw[->] (nodeAssignment)--(-3,4.5)--(-2.25,4.5)--(-2.25,4);
			\node[below,align=center,execute at begin node=\setlength{\baselineskip}{1.5ex}] (bin) at (-3.75,4) {\scriptsize Binomial \\ \tiny  \cite{KJG+18}};
			\node[below,align=center,execute at begin node=\setlength{\baselineskip}{1.5ex}] (hyper) at (-3,4-0.5) {\scriptsize Hypergeometric \\ \tiny  \cite{ZMR18,DDL+19}};
			\node[below,align=center,execute at begin node=\setlength{\baselineskip}{1.5ex}] (others) at (-2.4,4) {\scriptsize Others \\ \tiny  \cite{HHS19}};
			
			\draw[->] (intraShard)--(0,4.5)--(-1+0.25,4.5)--(-1+0.25,4);
			\draw[->] (intraShard)--(0,4.5)--(1-0.25,4.5)--(1-0.25,4);
			\node[below,align=center,execute at begin node=\setlength{\baselineskip}{1.5ex}] (strong) at (-1+0.25,4) {\scriptsize Strong\\\scriptsize consistency};
			\node[below,align=center,execute at begin node=\setlength{\baselineskip}{1.5ex}] (weak) at (1-0.25,4) {\scriptsize Weak\\\scriptsize consistency};
			\draw[->] (strong)--(-1+0.25,3.2-0.5)--(-1.5+0.25,3.2-0.5)--(-1.5+0.25,3-0.5-0.25);
			\draw[->] (strong)--(-1+0.25,3.2-0.5)--(-0.5+0.25,3.2-0.5)--(-0.5+0.25,3-0.5-0.25);
			\draw[->] (strong)--(-1+0.25,3.2-0.5)--(-1+0.25,3-0.5-0.5-0.25);
			\node[below,align=center,execute at begin node=\setlength{\baselineskip}{1.5ex}] (partial) at (-1.5+0.25,3-0.5-0.25) {\scriptsize Partial\\\scriptsize sync.};
			\node[below,align=center,execute at begin node=\setlength{\baselineskip}{1.5ex}] (sync) at (-1+0.25,3-0.5-0.5-0.25) {\scriptsize Sync. \\ \tiny \cite{Bazzi00,AMN+20}};
			\node[below,align=center,execute at begin node=\setlength{\baselineskip}{1.5ex}] (async) at (-0.5+0.2,3-0.5-0.25) {\scriptsize Async. \\ \tiny \cite{MXC+16}};
			\draw[->] (partial)--(-1.5+0.25,3.25-1.25-0.25-0.5-0.25)--(-2+0.25,3.25-1.25-0.25-0.5-0.25)--(-2+0.25,3-1.25-0.25-0.5-0.25-0.25);
			\draw[->] (partial)--(-1.5+0.25,3.25-1.25-0.25-0.5-0.25)--(-1+0.25,3.25-1.25-0.25-0.5-0.25)--(-1+0.25,3-1.25-0.25-0.5-0.25-0.25);
			\node[below,align=center,execute at begin node=\setlength{\baselineskip}{1.5ex}] at (-2+0.5,3-1.2-0.25-0.5-0.25-0.25) {\scriptsize Paxos+ \\ \tiny  \cite{L98}};
			\node[below,align=center,execute at begin node=\setlength{\baselineskip}{1.5ex}] at (-1+0.4,3-1.2-0.25-0.5-0.25-0.25) {\scriptsize PBFT+ \\ \tiny  \cite{CL99,YMR+19}};
			\draw[->] (weak)--(0.75,3.2-0.5)--(0.4,3.2-0.5)--(0.4,3-0.5-0.25);
			\draw[->] (weak)--(0.75,3.2-0.5)--(1.1,3.2-0.5)--(1.1,3-0.5-0.25);
			\node[below,align=center,execute at begin node=\setlength{\baselineskip}{1.5ex}] at (0.4,3-0.5-0.25) {\scriptsize PoW \\ \tiny \cite{WW19}};
			\node[below,align=center,execute at begin node=\setlength{\baselineskip}{1.5ex}] at (1.1,3-0.5-0.25) {\scriptsize PoS \\ \tiny \cite{FGK+18}};
			
			%
			\draw[->] (crossShard)--(3,4.5)--(2,4.5)--(2,4);
			\draw[->] (crossShard)--(3,4.5)--(3,4);
			\draw[->] (crossShard)--(3,4.5)--(4,4.5)--(4,4);
			\node[below] (2pc) at (3,4) {\scriptsize 2PC};
			\node[below,align=center,execute at begin node=\setlength{\baselineskip}{1.5ex}] at (2,4) {\scriptsize Transaction\\\scriptsize split \\ \tiny \cite{ZMR18}};
			\node[below,align=center,execute at begin node=\setlength{\baselineskip}{1.5ex}] at (4,4) {\scriptsize Relay\\\scriptsize transaction \\ \tiny \cite{WW19}};
			\draw[->] (2pc)--(2+1,3.25-0.5)--(1.6+1,3.25-0.5)--(1.6+1,3-0.5-0.25);
			\draw[->] (2pc)--(2+1,3.25-0.5)--(2.4+1,3.25-0.5)--(2.4+1,3-0.5-0.25);
			\node[below,align=center,execute at begin node=\setlength{\baselineskip}{1.5ex}] at (1.6+1,3-0.5-0.25) {\scriptsize Client-\\\scriptsize driven \\ \tiny \cite{KJG+18}};
			\node[below,align=center,execute at begin node=\setlength{\baselineskip}{1.5ex}] at (2.4+1,3-0.5-0.25) {\scriptsize Shard-\\\scriptsize driven \\ \tiny \cite{ASB+18}};
			
			\draw[->] (shardRecon)--(6,4.5)--(5+0.25,4.5)--(5+0.25,4);
			\draw[->] (shardRecon)--(6,4.5)--(7-0.25,4.5)--(7-0.25,4);
			\node[below,align=center,execute at begin node=\setlength{\baselineskip}{1.5ex}] at (5+0.25,4) {\scriptsize Random\\\scriptsize replacement \\ \tiny \cite{KJG+18}};
			\node[below,align=center,execute at begin node=\setlength{\baselineskip}{1.5ex}] (specific) at (7-0.25,4) {\scriptsize Specific\\\scriptsize rules};
			\draw[->] (specific)--(7-0.25,3.25-0.5)--(6.1,3.25-0.5)--(6.1,3-0.5-0.25);
			\draw[->] (specific)--(7-0.25,3.25-0.5)--(7.4,3.25-0.5)--(7.4,3-0.5-0.25);
			\node[below,align=center,execute at begin node=\setlength{\baselineskip}{1.5ex}] at (6.1,3-0.5-0.25) {\scriptsize Chronological \\ \tiny \cite{KJG+16,CPS18}};
			\node[below,align=center,execute at begin node=\setlength{\baselineskip}{1.5ex}] (specific) at (7.4,3-0.5-0.25) {\scriptsize Bounded\\\scriptsize cuckoo  \\ \tiny \cite{HWC+19,ZLC+20}};
			
			\draw[->] (motivation)--(9,4.5)--(8,4.5)--(8,4);
			\draw[->] (motivation)--(9,4.5)--(9,4);
			\draw[->] (motivation)--(9,4.5)--(10,4.5)--(10,4);
			\node[below,align=center,execute at begin node=\setlength{\baselineskip}{1.5ex}] at (8,4) {\scriptsize Penalty \\  \tiny \cite{BG17}\footnotemark};
			\node[below,align=center,execute at begin node=\setlength{\baselineskip}{1.5ex}]  at (9,4) {\scriptsize Rewards \\  \tiny \cite{WW19lever}};
			\node[below,align=center,execute at begin node=\setlength{\baselineskip}{1.5ex}] at (10.1,4) {\scriptsize Reputation \\  \tiny  \cite{ZMR18}};
			
			\draw[dashed] (-1.9,1.1-1.25) rectangle (0,4);
			\node[below,align=center,execute at begin node=\setlength{\baselineskip}{1.5ex}] at (-1.9/2,1.1-1.25) {\scriptsize \color{black!60}Instant\\\scriptsize \color{black!60}sharding};
			\draw[dashed] (0.1,2.3-0.75) rectangle (1.35,4);
			\node[below,align=center,execute at begin node=\setlength{\baselineskip}{1.5ex}] at (1.45/2,2.3-0.75) {\scriptsize \color{black!60}Eventual\\\scriptsize \color{black!60}sharding};
			
			\draw[dashed] (-10.9,1.1+0.2) rectangle (-8.6,4);
			\node[below,align=center,execute at begin node=\setlength{\baselineskip}{1.5ex}] at (-9.65,1.1+0.2) {\scriptsize \color{black!60}Permissionless\\\scriptsize \color{black!60}sharding};
			\draw[dashed] (-8.55,2.3+0.75) rectangle (-8,4);
			\node[below,align=center,execute at begin node=\setlength{\baselineskip}{1.5ex}] at (-7.9,2.3+0.70) {\scriptsize \color{black!60}Permissioned\\\scriptsize \color{black!60}sharding};
			
		\end{tikzpicture}
	}
	\caption{Taxonomy of each component.}
	\label{fig:taxonomy}
\end{figure*}

For the transaction model, the UTXO model and the account model could be commonly used. The UTXO model supports multiple input and multiple output transactions. The account model usually only supports single-input single-output transactions.

\paragraph{Distinct components}
For each component, we could choose specific and different implementation methods. In Section~\ref{sec:node_selection}-\ref{sec:motivation_mechanism}, we will classify the possible implementation methods of each building block and introduce the basic concepts, existing approaches, and possible problems. Here, we give the taxonomy of each building block in Fig.~\ref{fig:taxonomy}.

In Fig.~\ref{fig:taxonomy}, it is worth noting that in the $\isc$ part, according to the specific algorithm used, the sharding blockchains could be divided into instant sharding blockchains and eventual sharding blockchains.

\begin{definition}[An Instant Sharding Blockchain]
	\label{def:instant}
	In a sharding blockchain, if there is a committee running a consensus algorithm with strong consistency inside each shard to process transactions, then it is called an instant sharding blockchain. 	
\end{definition}

In instant sharding blockchains, the transaction confirmation is instant, due to the strong consistency property of the intra-shard consensus algorithm. ELASTICO \cite{LNZ+16}, Chainspace \cite{ASB+18}, Omniledger \cite{KJG+18}, RapidChain \cite{ZMR18}, RSCoin \cite{DM16}, etc. are all instant sharding blockchains.

\begin{definition}[An Eventual Sharding Blockchain]
	\label{def:eventual}
	In a sharding blockchain, if there is no committee inside a shard, and the intra-shard consensus algorithm satisfies weak consistency, then it is called an eventual sharding blockchain. 	
\end{definition}

Eventual sharding blockchains are relative to instant ones, where each shard still relies on PoW, PoS, or other methods to confirm transactions. Transactions or blocks are not confirmed instantly, and several blocks at the end of a blockchain must be removed to obtain stable states. The number of blocks is determined by the system security parameter. The cross-shard transaction processing in eventual sharding blockchains is different from the one in instant sharding blockchains due to its weak consistency property in each shard. Eventual sharding blockchains include Monoxide \cite{WW19}, Parallel Chains \cite{FGK+18}, etc.

\footnotetext{Note that Casper FFG is not a sharding blockchain. We refer to it here since as far as we know, there is currently no sharding blockchain with a penalty mechanism, and the penalty mechanism of Casper FFG could be seen as a reference.}

\subsubsection{Instantiation of Composing Components into a Sharding Blockchain System}
\label{subsubsec:instantiation}
Next, we take Omniledger \cite{KJG+18} as an example to illustrate how to use our proposed functional components to compose a complete sharding blockchain system.

The system models should first be analyzed. Since Omniledger uses a partially synchronous BFT algorithm within the shard, the entire message transmission model is a partially synchronous network. Besides, Omniledger allows nodes to join the protocol dynamically, so the node admission model is a permissionless network. Regarding the adversary model, Omniledger assumes the adaptive and mild adversary adopted by most blockchains. The intra-shard proportion model is $u=3f+1$, which is determined by $\isc$. As a result, the total proportion is limited to $[0,1/3)$. Note that since Omniledger utilizes an underlying PoW blockchain to realize $\ns$, the honest chain quality decreases due to the selfish mining attack. Consequently, the actual total computational power proportion of the adversary is constrained to $[0,1/4]$. This will be analyzed in detail in Section~\ref{subsubsec:pow_based_ns}.

Next is the implementation method for each component. The first is the $\ns$ component. In epoch $e$, Omniledger requires all nodes that want to participate in epoch $e+1$ to find a PoW solution, and broadcast the found results and their own identities. In this way, nodes complete the registration on the identity blockchain, i.e., the underlying PoW blockchain. The block producers are treated as selected nodes $snodes$. Second, Omniledger uses Randhound (described in detail in Section~\ref{sec:epoch_randomness}) as the randomness generation component $\er$. A leader of Randhound is elected by Verifiable Random Functions (VRF), and then all participating nodes execute PVSS for multiple rounds of interactive communication to generate a secure epoch randomness $\xi_e$. Third, the $\na$ component takes in $H(0||\xi_e)$ as a seed to compute a pseudorandom permutation $\pi_{0,e}$, and assigns the selected nodes into $m$ different groups to obtain $anodes$ based on $\pi_{0,e}$. Fourth, the shard reconfiguration component $\sr$ stipulates that $\log n/m$ old members in each shard are replaced. Similarly, $\sr$ uses $H(c||\xi_e)$ as a seed to generate $m$ pseudorandom permutations $\pi_{1,e},\cdots,\pi_{m,e}$ for each shard, and then determines which old members are randomly replaced by the assigned new nodes in $anodes$.
The following parts are about transaction processing. Omniledger employs an improved version of PBFT which is called Omnicon, as the implementation of $\isc$. Regarding $\cstp$, Omniledger utilizes a client-driven 2PC method to process cross-shard transactions, where the client acts as a coordinator to complete the collection and forwarding of proofs for transaction inputs \cite{KJG+18}.

In this way, we obtain a complete sharding blockchain protocol. We argue that our components and their outlined composition are suitable for most sharding blockchain systems. The independent design and composability of each component allow our framework to be used to conceptualize and develop new, yet unexplored sharding blockchain systems.

\subsection{Summary}
\label{subsec:summary}
We provide a summary of sharding blockchain systems in Table~\ref{tab:summary}.

\newcommand{\tabincell}[2]{\begin{tabular}{@{}#1@{}}#2\end{tabular}}  
\renewcommand{\arraystretch}{2.6}
\begin{table*}[h!]	
	
	\centering
	\LARGE
	\caption{Summary of sharding blockchain systems.}
	\label{tab:summary}
	\resizebox{\textwidth}{!}{
		
		\begin{threeparttable}
			\begin{tabular}{x{2.0cm}x{3.7cm}x{3.4cm}x{3.5cm}x{3.5cm}x{3.5cm}x{3.5cm}x{3.5cm}x{2.9cm}x{3.5cm}x{3.5cm}x{3.3cm}x{3.3cm}}
				\midrule[.1em] 
				\multicolumn{2}{c}{\multirow{2}{*}{\textbf{System}}} & \textbf{ELASTICO} \cite{LNZ+16} & \textbf{Omniledger} \cite{KJG+18} & \textbf{RapidChain} \cite{ZMR18} & \textbf{Chainspace} \cite{ASB+18} & \textbf{SGX-Sharding}\tnote{$\sharp$} \cite{DDL+19}& \textbf{ZILLIQA} \cite{Zilliqa17} & \textbf{PoSBP}\tnote{$\sharp$} \cite{LJK19} & \textbf{Ethereum} \cite{Buterin17} & \textbf{Monoxide} \cite{WW19} & \textbf{Parallel Chains} \cite{FGK+18} & \textbf{RSCoin} \cite{DM16} \\	\midrule[.1em] 
				
				\multicolumn{1}{c}{\multirow{4}{*}{\rotatebox{90}{\textbf{Classification}}}} &	\centering \textbf{Node Admission Model} & Permissionless & Permissionless & Permissionless & Permissionless & Permissioned & Permissionless & Permissionless & Permissionless & Permissionless & Permissionless & Permissioned \\	\addlinespace \cmidrule(l){2-13}
				~ & \centering \textbf{Instant or Eventual} & Instant & Instant & Instant & Instant & Instant  & Instant & Instant & Instant & Eventual & Eventual & Instant  \\  \midrule[.1em]

				\multicolumn{1}{c}{\multirow{5}{*}{\rotatebox{90}{\textbf{System Model}}}} &	\centering \textbf{Network Model} &  Partially Sync. &  Partially Sync. & Partially Sync./Sync.\tnote{*} & Partially Sync. & Partially Sync.  & Partially Sync. & Partially Sync. & Partially Sync. & Partially Sync. & Partially Sync. & - \\ 	\cmidrule(l){2-13}
				~ & \centering \textbf{Adversary Model} & $\leq \frac{1}{4}$ & $\leq \frac{1}{4}$ & $\leq \frac{1}{3}$ & - & $\leq \frac{1}{3}$  & $\leq \frac{1}{4}$ & $\leq \frac{1}{4}$ & $\leq \frac{1}{4}$  & $\leq \frac{1}{2}$ & $\leq \frac{1}{2}$ & - \\ \cmidrule(l){2-13}
				~ & \centering \textbf{Transaction Model} & UTXO & UTXO & UTXO & Account & Generic\tnote{$\natural$}  & Account & UTXO & Account & Account & UTXO & Account \\ \midrule[.1em] 
				
				\multicolumn{1}{c}{\multirow{5}{*}{\rotatebox{90}{\textbf{\makecell[c]{Node Selection and \\Assignment}}}}} &\centering \textbf{Sybil attacks Resistance}  & PoW & PoW & PoW & - & SGX  & PoW & PoS & PoS & PoW & PoS & -\\ 	\cmidrule(l){2-13}
				~ &\centering \textbf{Basic Method}  & Reference Committee & Underlying Blockchain & Reference Committee & - & SGX   & Reference Committee & Hash Func. & VDF & - & VRF & - \\ 	\cmidrule(l){2-13}
				~ &\centering \textbf{Distribution Model} & - & Binomial & Hypergeometric & - & Hypergeometric  & - & - & - & - & - & -\\ 	\midrule[.1em] 
				
				\multicolumn{2}{c}{\centering \textbf{Epoch Randomness}}  & Hash Func. & RandHound (PVSS)  & VSS &  -  & SGX  & Hash Func. & - & RANDAO+VDF & - & VRF & -\\	\midrule[.1em] 

				\multicolumn{1}{c}{\multirow{3}{*}{\rotatebox{90}{\textbf{\makecell[c]{Intra-Shard\\ Consensus}}}}} &\centering \textbf{Adversary Model}  &  $u=3f+1$ &  $u=3f+1$ &  $u=2f+1$ &  $u=3f+1$ &  $u=2f+1$ & $u=3f+1$ & $u=3f+1$ & $u=3f+1$ & $\leq \frac{1}{2}$ & $\leq \frac{1}{2}$ & $u=3f+1$ \\ 	\cmidrule(l){2-13}
				~ &\centering \textbf{Consensus Algorithm}  & PBFT  & Omnicon (BFT) & Sync BFT & PBFT & AHL (BFT) & PBFT & BFT-DPoS & BFT & PoW & PoS & BFT\\ 	\midrule[.1em] 
				
				\multicolumn{1}{c}{\multirow{3}{*}{\rotatebox{90}{\textbf{\makecell[c]{Cross-Shard \\Scheme}}}}} &\centering \textbf{Basic Algorithm} & - & 2PC & Split & 2PC & 2PC  & - & - & Relay Transaction & Relay Transaction & - & 2PC \\	\addlinespace \cmidrule(l){2-13}
				~ &\centering \textbf{Coordinator} & - & Client-Driven & Shard-Driven & - & Shard-Driven  & - & - & - & - & - & Client-Driven\\ 	\midrule[.1em] 
				
				\multicolumn{1}{c}{\multirow{3}{*}{\rotatebox{90}{\textbf{\makecell[c]{Shard \\Reconfiguration}}}}} &\centering \textbf{Basic Rule} & - & Random Replacement & Bounded Cuckoo Rule  & - & Random Replacement & - & - & - & - & - & -\\	\cmidrule(l){2-13}
				~ &\centering \textbf{Update Fraction} & - & $\log u$  & $\frac{1}{2}$ & - & $\log u$  & - & - & - & - & - & -\\	\midrule[.1em] 
				
				
				\multicolumn{1}{c}{\multirow{3}{*}{\rotatebox{90}{\textbf{Performance}}}} &\centering \textbf{Responsiveness} & \xmark & \xmark & \cmark & \cmark & \xmark & \xmark & \xmark & \xmark & \xmark & \xmark & \xmark\\ \addlinespace	\cmidrule(l){2-13}
				~ &\centering	\textbf{Scalability} & \xmark\tnote{$\diamond$} & \cmark & \cmark & \cmark & \cmark  & \xmark & \xmark & \cmark & \xmark\tnote{$\diamond$} & \xmark & \xmark \\ \midrule[.1em] 
			\end{tabular}
			\begin{tablenotes}   
				\item  The notation ``\cmark'' means that the system has the property; ``\xmark'' means the system does not have the property; ``-'' denotes that the property does not apply to the system.              
				\item[$\sharp$] We use ``SGX-Sharding'' to represent the system proposed in \cite{DDL+19} which uses the trusted hardware Intel SGX. Similarly, PoSBP represents the system described in \cite{LJK19}.   
				\item[*] The network model for RapidChain is partially synchronous for the whole network and synchronous inside a committee.
				\item[$\diamond$] ELASTICO is not scalable since all transactions will be processed by a final committee. Monoxide is not scalable since all miners have to verify all transactions in the network, which is analyzed in reference \cite{AKW19}.
				\item[$\clubsuit$] ``Malicious leader resistance'' refers to the ability to prevent a malicious adversary from providing false input availability certificates (ref. Definition~\ref{def:ac}) during cross-shard transaction processing. 
				\item[$\natural$] As explained in Section~\ref{sec:preliminaries}, ``Generic'' means a UTXO model or an account model.  
			\end{tablenotes}
		\end{threeparttable}
	}
\end{table*}

The notation ``\cmark'' means that the system has the property; ``\xmark'' means the system does not have the property; ``-'' denotes that the property does not apply to the system.
We use ``SGX-Sharding'' to denote the system proposed in \cite{DDL+19} which utilizes the trusted hardware Intel SGX. Similarly, PoSBP represents the system described in \cite{LJK19}.

The network model for RapidChain is partially synchronous for the whole network and synchronous inside a committee.
In the ``scalability'' line, ELASTICO is not scalable since all transactions will be processed by a final committee. Monoxide is not scalable since all miners have to verify all transactions in the network, which is analyzed in Reference \cite{AKW19}.

\section{Node Selection}
\label{sec:node_selection}
In this section, we first introduce basic concepts to realize node selection for sharding blockchains in Section~\ref{subsec:ns_bp}. Then existing approaches to select new nodes are classified into PoW-based ones and PoS-based ones in Section~\ref{subsec:ns_es}. In addition, PoW-based methods consist of using an underlying blockchain and using a reference committee. Finally, we analyze potential problems in the process of node selection in Section~\ref{subsec:ns_pp}.

\subsection{Basic Concepts}
\label{subsec:ns_bp}
Node selection is necessary for sharding blockchains to select qualified shard members. 
The problems to be solved in the node selection process are as follows. First, all nodes should have a consistent view of the selected result, i.e., $snode$. Second, the specific requirements that honest node number proportion in $snode$ should meet for different application scenarios need to be analyzed. Third, to against various attacks, strict security proofs should be given for the node selection process..

In permissioned networks, the node selection process is completed with the participation of CA through providing the identity registration service for nodes. 
In permissionless networks, the node selection process is more complicated, so we discuss this situation in detail.

During the selection process in a permissionless network, PoW or PoS is utilized to prevent Sybil attacks \cite{Douceur02,RMD+20}, that is, an adversary increases the probability of becoming a shard member by creating fake identities. If a PoW-based node selection method is adopted, a certain mining difficulty needs to be set carefully, such that enough nodes could find PoW solutions in each period to replace the corresponding old ones. In a PoS-based node selection method, a certain number of nodes need to be selected randomly as new shard members according to the stake held by each node.

The node selection process usually causes a decline in the proportion of honest nodes for some reason. In order to measure the degree of decline, we introduce an honest fraction decline degree parameter $\omega_d$, which is described in Definition~\ref{def:honest_fraction}.

\begin{definition}[Honest Fraction Decline Degree]
	\label{def:honest_fraction}
	Assume that the honest fraction (computational power or stake) is $\beta$. After a node selection process, let $\newnodes$ denote the selected node list. Assume the honest node fraction in $\newnodes$ to be $(1-\omega_d)\beta$, then $\omega_d$ is said to be the honest fraction decline degree parameter. 
\end{definition} 

In order to ensure that the proportion of honest nodes in the node selection process will not decrease too much, we describe the concept of fair selection, which is defined in Definition~\ref{def:fair_selection}.
\begin{definition}[Fair Selection \cite{LLZ+20}]
	\label{def:fair_selection}
	Let $Q_f$ denote the fraction of honest nodes in a selected node list $\mathsf{newnodes}$ and let $\beta$ denote the honest fraction (computational power or stake). We say that a node selection process for shard members is $(k_f,\omega_d)$-fair if for all $\beta > 0$, there exists some negligible function $\mu_f(k)$ such that for every $k \geq k_f, 0 \leq \omega_d < 1$, the following condition holds
	\begin{equation*}
		\Pr[Q_f \geq (1-\omega_d)\beta] \geq 1-\mu_f(k)
	\end{equation*}
\end{definition}

The definition of ``fair selection'' in \cite{LLZ+20} is inspired by \cite{PS17fruit}, while it has some obvious differences from the definition of ``fairness'' in FruitChains. Fairness in FruitChains only applies to its own specially designed mining process, where a fruit and a block are mined simultaneously through a single 2-in-1 mining function. So the analysis in FruitChains is specific. Definition~\ref{def:fair_selection} gives a more general description of the node selection process, which could be used to evaluate the fairness of a selection result.

\subsection{Existing Approaches}
\label{subsec:ns_es}
According to the underlying technologies of node selection methods, we divide related approaches into three categories, i.e., PoW-based, PoS-based, and CA-based node selection methods. 
As mentioned in Section~\ref{subsec:composing}, the first two methods are suitable for permissionless networks, and the latter method is used in permissioned networks.

\subsubsection{PoW-Based Node Selection}
\label{subsubsec:pow_based_ns}
A PoW-based node selection approach utilizes PoW mining to select qualified nodes where any node who wants to take part in the protocol must find a PoW solution. There are currently two PoW-based node selection methods, namely, using an underlying blockchain and using a reference committee.

\paragraph{Using an underlying blockchain}

\begin{figure}[h!]
	\centering
	\includegraphics[width=0.6\textwidth]{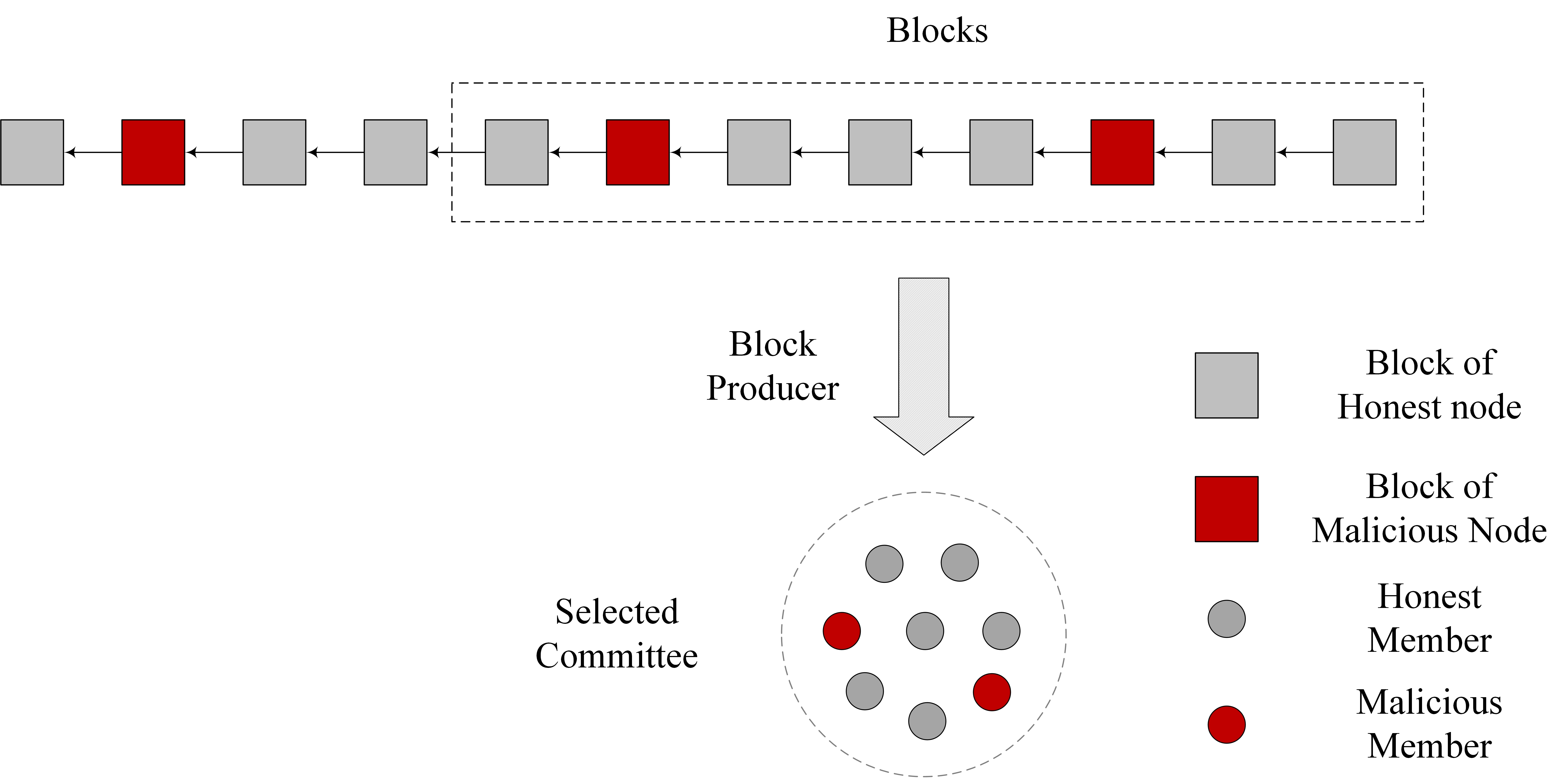}
	\caption{PoW-based node selection: using an underlying blockchain.}
	\label{fig:node_selection1}
\end{figure}

As shown in Fig.~\ref{fig:node_selection1}, an underlying blockchain is a chain similar to that of Bitcoin \cite{N08}. All blocks are connected by hash pointers. Nodes mine on the basis of the last block, and verify if the hash value meets the requirements according to the following condition:
\begin{equation*}
	h=\hash(str,nonce,pk_i) < D
\end{equation*}

Here, $str$ is the hash of the last block, and $nonce \in \{0,1\}^\lambda$ denotes the potential solution of the PoW. $\lambda$ is a security parameter where $\lambda \in \mathbb{N}$. $pk_i$ is the public key of the node $P_i$. $D$ is a difficulty parameter where $D = p \cdot 2^{\lambda}$ and for all $(str,nonce,pk_i)$, we have $\Pr[\hash(str,nonce,pk_i) < D] = p$. $p$ is the probability that one node finds a PoW solution successfully in one single round.

After finding a nonce that meets the requirement, a node broadcasts the block, i.e., $(str,nonce,pk_i)$. Then the nodes receiving the block verify its legitimacy. If the requirements are met, then the nodes will continue to mine, using $\hash(str,nonce,$ $pk_i)$ as a new $str$. The block producers, that is, nodes that find valid PoW solutions successfully, are considered as new shard members. When there are enough shard members confirmed, shards could launch a reconfiguration to update members.

Note that the node selection process is necessary for all hybrid consensus blockchains, which combines classical distributed consensus algorithms and the blockchain consensus, such as PeerCensus \cite{DSW16}, ByzCoin \cite{KJG+16}, Solida \cite{AMN+17}, Hybrid Consensus \cite{PS17}, Thunderella \cite{PS18}, and Algorand \cite{GHM+17}. In Solida \cite{AMN+17} and Byzcoin \cite{KJG+16}, a committee conducts a reconfiguration whenever a new block producer is confirmed through the above steps.

Omniledger \cite{KJG+18} uses an underlying blockchain to select new members. Specifically, in epoch $e$, the node that wants to participate in epoch $e + 1$ tries to find a PoW solution and mines on an identity blockchain. The identity blockchain in Omniledger plays the role of an underlying blockchain. When enough number of nodes for the next epoch is registered on the identity blockchain, a reconfiguration begins and the protocol enters a new epoch.

\begin{figure}[h!]
	\centering
	\includegraphics[width=0.6\textwidth]{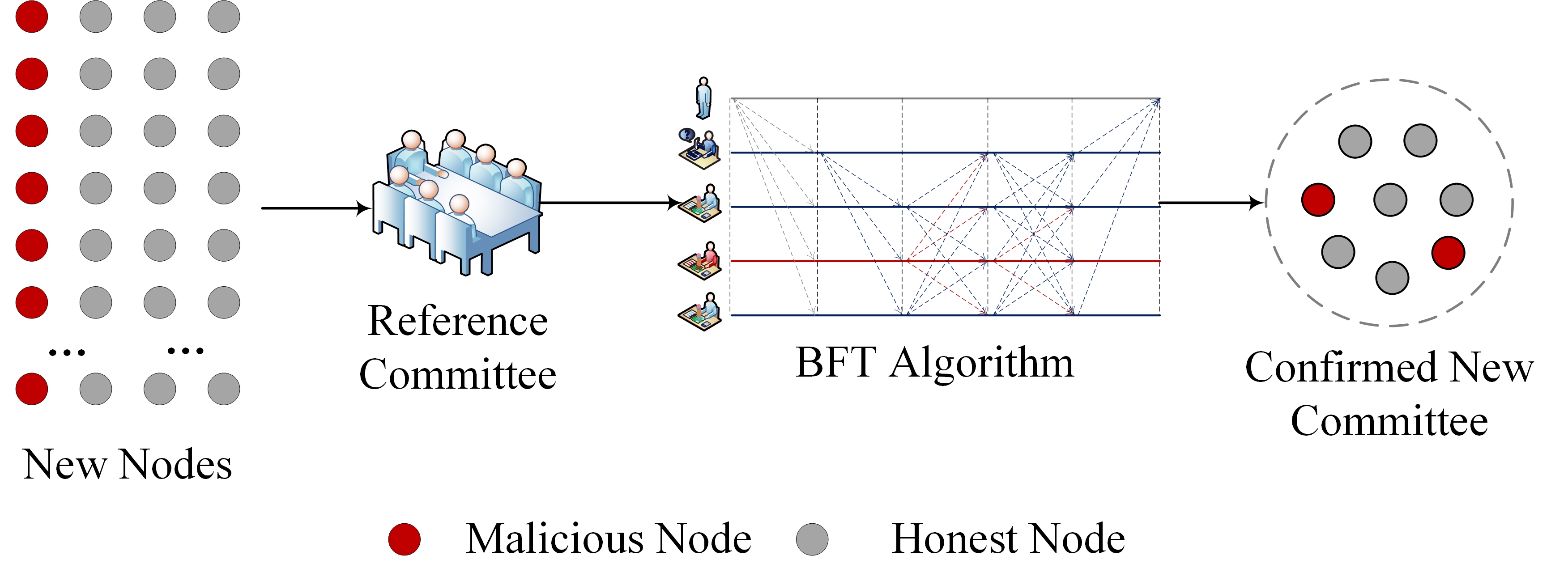}
	\caption{PoW-based node selection: using a reference committee.}
	\label{fig:node_selection_refcomm}
\end{figure}

\paragraph{Using a reference committee}
Another approach uses a reference committee and a fixed PoW puzzle to select new nodes, which is shown in Fig.~\ref{fig:node_selection_refcomm}. 

This approach includes two steps. First, a mining step. In a current epoch, nodes use a puzzle specially set to mine. The mining equation is similar to that of using an underlying blockchain. 
\begin{equation*}
	h=\hash(puzzle,nonce,pk_i) < D
\end{equation*}

A node submits his solution to a reference committee $\comm^R$ after finding a PoW solution successfully, yet the found PoW solutions do not form a chain. That is, in an entire epoch, the $puzzle$ remains unchanged until a sufficient number of PoW solutions are found. Note that in this case, a miner might use a different public key to continue mine after finding a PoW solution. This will not influence the system security as long as the adversary's computational power is limited. 

Second, a new node list confirmation step. After enough number of PoW solutions are submitted, the reference committee runs an intra-committee consensus to confirm the new node list, denoted by $\newnodes$. The intra-committee consensus might be a BFT style consensus algorithm, where a leader is responsible for proposing a value, and other nodes vote for the proposal. After an agreement is reached upon the new node list, the reference committee broadcasts the committed $\newnodes$ to the entire network. Nodes that are on the committed list are regarded as valid members of the next epoch.

RapidChain utilizes a reference committee to select new nodes \cite{ZMR18}. The puzzle in RapidChain is a fresh randomness generated by the reference committee based on verifiable secret share (VSS). An adversary could not precompute a PoW solution ahead of honest nodes since the randomness is unpredictable. Besides, the PoW mining process is done offline without influencing the normal operations of the whole protocol. At the beginning of every epoch, the reference committee reaches an agreement on a reference block which includes the new node lists for the epoch.

\subsubsection{PoS-Based Node Selection}
PoS-based node selection approaches have the following characteristic: the more stakes a user has, the higher the probability of being selected. In general, the coins in the system need to be divided into small units, and each unit has the right to participate in the selection of committee members. In this way, it is ensured that the probability of being selected as a committee member will increase if a node has more stakes. At the same time, when judging whether each unit is selected, the verifiable random function (VRF) could be employed. The unique serial number of each unit could be regarded as the input of VRF to generate the random output and its corresponding proof. Whether a unit is selected as a committee member could be verified publicly by judging if the random output satisfies a certain threshold condition.

Similar to PoW-based node selection approaches, there are two fundamental ways to use PoS to select committee members, i.e., using an underlying blockchain or not. 

\paragraph{With an underlying blockchain}
Using an underlying blockchain means that nodes still rely on PoS to generate blocks first, and then block producers during a certain time range are considered to be committee members. 
Some PoS-based underlying blockchains, such as Ouroboros \cite{KRD+17}, can be used to confirm committees in a sharding blockchain. The basic process is that the block producers within a period of time are confirmed as new nodes, and the new nodes are randomly allocated to multiple different shards based on a randomness. 

\paragraph{Without an underlying blockchain}
If there is no underlying blockchain, using PoS to select shard members requires selecting multiple nodes at once, such as a committee or multiple committees.

Lee \textit{et al.} \cite{LJK19} propose a sharding blockchain system that uses PoS to select committee members. The selection process is quite simple, which uses the following formula $\hash(v\text{'s address}||$ $ \hash(b)) \mod k$. $\hash$ is a hash function, $v$ represents a validator, $b$ denotes the last block in the previous epoch, and $k$ is the number of shards. In this way, validators in \cite{LJK19}, i.e., nodes, are assigned into $k$ shards randomly.  
Ethereum sharding \cite{Buterin17} adopts the PoS-based node selection method to select shard members.

Parallel Chains \cite{FGK+18} is an eventual sharding blockchain system that uses PoS to directly generate blocks, while there is no committee in each shard. The blockchain in each shard is built on top of that of Ouroboros Praos \cite{DGK+18}.

\subsubsection{CA-Based Node Selection}
\label{subsubsec:ca_based_ns}
In the permissioned network, the function of the $\ns$ component is realized by a CA. Every node that wants to participate should first complete identity authentication via the CA. When there are enough nodes (in the initial phase of the protocol), or at the end of each epoch (during epoch reconfiguration), the CA will publish a list of nodes participating in the protocol, i.e., $snode$, which contains the public key information of nodes. Only the nodes on the list published by the CA can take part in further operations of the protocol, i.e., join different shards and process transactions. 
The permissioned sharding blockchains that use CA-based node selection approaches include \cite{ACC+18,AAA19,AAA19b}.

\subsection{Problems and Future Directions}
\label{subsec:ns_pp}
We summarize problems that might occur during the process of node selection adopting both PoW-based and PoS-based node selection approaches in the following.

\subsubsection{PoW-Based Node Selection}
We analyze the potential problems in each kind of PoW-based node selection approach. First, possible problems of using an underlying blockchain approach are introduced, including the impact of attacks (e.g., selfish mining, stubborn mining, etc.). Then potential threats of using a reference committee are analyzed, including an adversary's potential mining advantages and node censorship attacks by malicious leaders.

\paragraph{Using an underlying blockchain}
When an underlying blockchain is used to select nodes, an adversary could launch the attacks such as selfish mining \cite{EyalS18,Eyal15}, stubborn mining \cite{NKM+16,LHX+20}, block withholding \cite{BRS17}, and fork after withholding \cite{KKS+17}, to increase his proportion of blocks and get more advantages to be selected. In this case, the chain quality of honest nodes decreases severely. 

\begin{figure}[h!]
	\centering
	\includegraphics[width=0.6\textwidth]{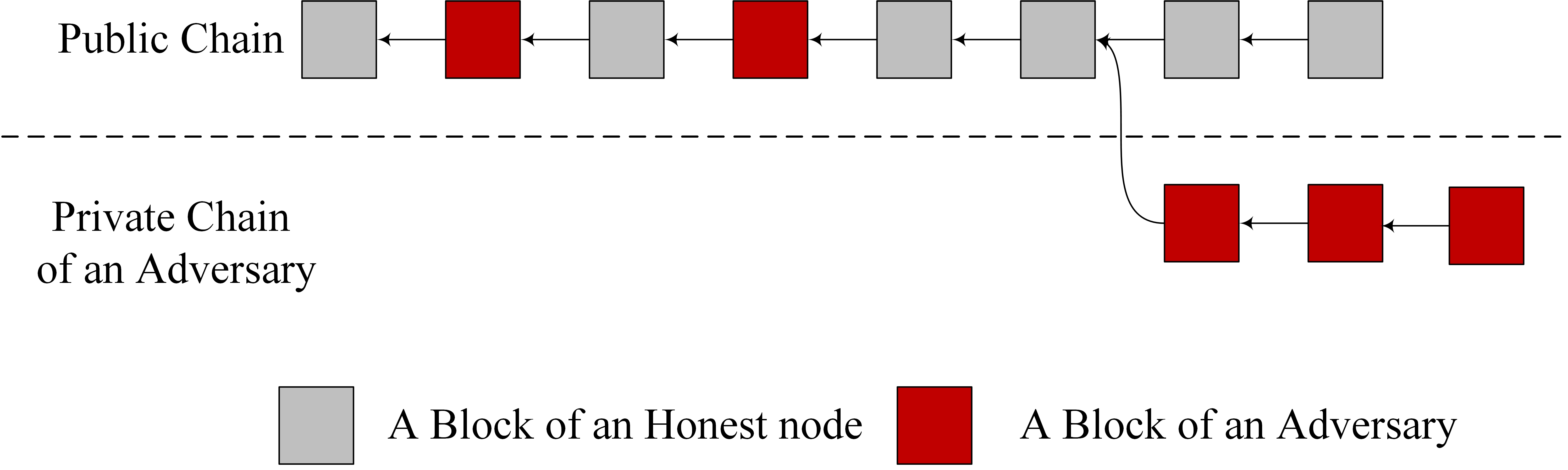}
	\caption{Schematic diagram of selfish mining.}
	\label{fig:selfish-attack}
\end{figure}

The selfish mining attack is shown in Figure~\ref{fig:selfish-attack}. In a selfish mining attack, an adversary and honest nodes mine at the same time. After finding a PoW solution (i.e., a block), an honest node broadcasts the block to the entire network immediately. On the contrary, an adversary does not broadcast a block after finding a PoW solution, yet adopts different strategies to reveal his blocks in different situations. The blockchain that is known to all honest nodes is called the public chain, while we name an adversary's privately controlled blockchain as a private chain. When the adversary's private chain is longer than the public chain, the adversary does not immediately announce the block after finding a new PoW solution, i.e., withhold the block. When the length of the public chain catches up with that of the private chain, the adversary publishes a certain number of private blocks, making the new chain longer than the public chain. In this way, honest nodes will choose to continue mining at the end of the newly disclosed chain. Selfish mining increases the proportion of blocks controlled by an adversary. Intuitively, an adversary could waste honest computational power. When the adversary's private chain is longer, he does not publish the block. At this time, even if a valid PoW solution is found by honest nodes, it will be replaced by an adversary's longer chain.

The stubborn mining attack \cite{NKM+16} provides an attacker with more advantages in PoW mining. Stubborn mining is actually an extension of selfish mining, which wastes honest computational power by creating more opportunities for competition. Three basic strategies and their different combinations are proposed and analyzed in depth \cite{NKM+16}. 
In addition, the block withholding attack \cite{BRS17} and fork after withholding attack \cite{KKS+17} are proposed. We regard the attacks above as the same type of attack as selfish mining. By formulating different mining strategies, an adversary could occupy a higher proportion of all blocks generated than it deserves. In other words, the chain quality of honest nodes decreases.

Besides, the eclipse attack \cite{HKZ+15,MHG18} and network partition attack \cite{AZV17} could also be used to enhance the effects of selfish mining type attacks \cite{NKM+16}. The key idea of the eclipse attack is to control all incoming and outgoing connections of a node \cite{HKZ+15}. The network is divided into three partitions, i.e., an attacker, a victim, and an honest one's part. Although some measures could detect intrusion \cite{MTW+18}, eclipse attacks are still appealing for attackers.

These kinds of attacks decreases the honest chain quality while increasing the fraction of blocks belonging to an adversary. A formal analysis is given in the notable Bitcoin backbone protocol \cite{GKL15} which points out that the ratio of blocks contributed by an adversary is bounded by
\begin{equation*}
	\frac{t}{n-t}
\end{equation*}
where $t$ and $n$ is the computational power of an adversary and the entire network, respectively. For example, when $t=1/3$ (we use fraction to denote simplify the description), $t/(n-t)$ equals to $1/2$. When $t=1/4$, $t/(n-t) = 1/3$. Hence, a $3/4$ honest computational power is required to achieve a $2/3$ fraction of honest nodes when using an underlying blockchain to select new nodes.

\paragraph{Using a reference committee}
In a PoW-based member selection process, two important factors influence the selection results, i.e., the time advantages of an adversary to mine, and the confirmation of a new node list. We give a detailed analysis in the following.

The time advantages of an adversary to mine mainly refer to the network latency. The adversary is usually responsible for network transmission who might have a certain advantage in mining. In a partially synchronous network, the adversary could delay messages sent by honest nodes for at most $\Delta$ time, which is the upper bound of the network delay. On the one hand, the adversary could acquire a mining puzzle in advance so that he could start mining ahead of the honest nodes. On the other hand, an adversary could delay honest miners' PoW solutions submitted to a reference committee. The adversary's network advantage is a key factor that must be considered when strictly analyzing the mining process and results. In the analysis of the mining step, the speed to find a PoW solution needs to be calculated. The mining step could be treated as independent binary random variables. The value for each variable is $1$ with probability $p$. So the PoW solutions found by honest nodes or the entire network could be estimated by the Chernoff bound \cite{MotwaniR95}. 	




In any case, the adversary could have a certain time advantage over honest nodes to obtain a mining puzzle. Since even if an unpredictable randomness (see Section~\ref{subsec:er_bp}) is used, an adversary has the advantage of network latency. If there is no randomness used as a puzzle, the adversary might learn a puzzle in advance through various other means such as withholding his PoW solutions. An adversary's time advantage must be considered in the analysis of the mining process. Because the adversary will produce more PoW solutions than expected. To ensure that the proportion of honest nodes in the finally found PoW solutions exceeds a certain safety threshold, it is usually required that the total number of solutions exceeds a certain lower bound.
A concrete analysis of the mining step is referred to \cite{LLZ+20}.

The other factor to consider is the confirmation of a new node list. After enough number of PoW solutions are found, a reference committee run an intra-shard consensus to confirm $\newnodes$. The most commonly employed intra-shard consensus is the BFT-type algorithm. In such a BFT algorithm, there is a leader who proposes a new node list. A malicious leader could ``censor'' and exclude some honest nodes, to involve more malicious nodes, and thus harm the system security.

The specific procedures of the node censorship attack are as follows. Let $\newnodes$ and $\newnodes'$ denote two member lists proposed by a leader and held by an honest node in a reference committee $\comm^R$, respectively. A malicious leader might replace a certain number of honest nodes in $\newnodes$ with the same number of malicious nodes who find the PoW solution after the replaced honest nodes \cite{LLZ+20}.

In this case, if honest members in $\comm^R$ vote without checking the validity of $\newnodes$, then safety will be ruined since the malicious nodes proportion on $\newnodes$ will exceed the specified limit. As a result, $\adv$ will control the committee in the next epoch. On the contrary, if an honest member only votes when $\newnodes$ equals to $\newnodes'$ held by himself, then the liveness property will be broken with a high probability, since the new member list held by different honest members may have some differences due to the network latency. 

The above node-censorship attack could be handled by a threshold-vote strategy proposed in \cite{LLZ+20}. A proper threshold $k_T$ is chosen for the differences between $\newnodes$ received from the leader and $\newnodes'$ held by an honest committee member. A BFT algorithm needs to be modified in moderation to be compatible with the threshold-vote strategy.
The specific threshold-vote rule is as follows. 
An honest committee member votes for a list $\newnodes$ if and only if the number of different nodes between $\newnodes$ and $\newnodes'$ is less than or equals to $k_T$.
The threshold-vote strategy might introduce some new problems. For specific analysis, please refer to \cite{LLZ+20}.

The future research directions for PoW-based node selection approaches are as follows. First, design a more fair node selection approach to make the honest fraction decline degree lower. Second, analyze various attacks that may be encountered in the process of using PoW such as selfish mining attacks. Third, use a strict analysis process to analyze the security of the entire mining process, and the requirement of each parameter. Fourth, fully consider the characteristics of the sharding blockchain, and design a method for selecting sharding members more suitable for the sharding blockchain.

\subsubsection{PoS-Based Node Selection}
In a PoS-based node selection process, some vulnerabilities or attacks might happen. We separately describe the possible problems in the two types of PoS-based node selection approaches.

\paragraph{With an underlying blockchain}
In the following, we describe the attacks against the PoS-based node selection approaches using an underlying blockchain.

Nothing at stake \cite{N14} refers to that an attacker tries to mine on different forks of the chain to obtain higher benefits. In a PoS-based blockchain, to generate a fork is not as costly as that in a PoW-based blockchain, where a huge amount of computational power might be required. In a PoS-based blockchain, if there is no protective mechanism, when the blockchain has a fork, a node will try to mine on both forks to increase the probability of obtaining profit. 
Nothing at stake could be prevented by introducing punishment mechanisms in the PoS consensus, that is, punish the nodes that generate blocks on different forks.

A grinding attack \cite{C16} means that in some PoS consensus mechanisms, the selection of a block producer in the $r+1$ round is affected by a block in a certain round, e.g., round $r$, or multiple blocks in previous rounds. Namely, the selection results are not random and might be biased. In some PoS consensus mechanisms, the block producer in round $r+1$ is usually selected according to the block generated in round $r$. If the block producer in round $r$ is controlled by the adversary, to become the block producer in round $r+1$, the adversary might try to continuously generate different new blocks in round $r$, i.e., ``grind'' the generated block until it is conducive to make the adversary become the next block producer. 
Grinding attacks could be prevented by using an unbiasable randomness such as RandHound \cite{SJK+17}, to determine a block producer.

A long range attack \cite{J14} means that when an offline node or a new node joins the network, an adversary forges a blockchain from the genesis block to the latest block, trying to make the newly joined node accept this forged blockchain and mine on it. In a PoW-based blockchain, the longest chain rule or the heaviest chain rule \cite{SZ15} is usually used to determine which blockchain is accepted as the valid main chain by all participants. Assuming $A$ is the main chain, the adversary wants to create a fake chain $B$ to make newly joined nodes believe that $B$ is the main chain. In a PoW-based blockchain, new nodes can easily judge $A$ as the main chain by verifying the difficulty of mining in the two chains, since the mining difficulty of the blocks in $A$ must be significantly higher than that of $B$. If the adversary wants to forge a chain with similar mining difficulty, he needs a huge amount of computational power, and the attack cost will greatly exceed the benefits. However, in a PoS-based blockchain, it is much easier to forge a main chain $A$. The adversary could bribe the nodes to sell the important private keys used in the past without spending too much to forge a fake chain $B$, convincing the newly joined nodes that chain $B$ is the main chain. Long-range attacks could be prevented by the checkpoint mechanism \cite{BG17}.

Ga?i, Kiayias, and Russell \cite{GKR18} propose a stake bleeding attack against the PoS consensus mechanism. The stake bleeding attack is mainly implemented after a successful long-range attack. For a PoS consensus system that does not use a checkpoint mechanism, after an attacker launches a long-range attack, the newly joined nodes believe that the adversary's chain $B$ is the current main chain, and the transaction generated by the new nodes will be submitted to chain $B$ for processing. The adversary has full control of chain $B$, and could earn a lot of transaction fees and even launches a double-spending attack \cite{KAC12,KAR+15}.

\paragraph{Without an underlying blockchain}
The problems that might be encountered in the PoS-based node selection approach without an underlying blockchain mainly include the following two aspects.
First, the application of some cryptographic techniques, e.g., VRF, might bring more computation and communication overhead. Second, the existence of network delay might affect a node's view of the selected new nodes.

The future research directions for PoS-based node selection are as follows. 
First, the impacts of various attacks against PoS on the node selection process and results should be fully considered. Second, the computation and communication overhead of some cryptographic tools on practical applications need to be analyzed. For instance, analyze the time cost required to complete a PoS-based node selection, and evaluate the impact on system liveness during reconfiguration.


\section{Epoch Randomness}
\label{sec:epoch_randomness}
In this section, we first introduce basic concepts about epoch randomness in Section~\ref{subsec:er_bp}. Then existing approaches to generate randomness are divided into VRF, PVSS, etc. in Section~\ref{subsec:er_es}. Additionally, we compare the state-of-the-art distributed random beacon protocols in Section~\ref{subsec:er_c}. Finally, we analyze potential problems and future directions about epoch randomness in Section~\ref{subsec:er_pp}.

\subsection{Basic Concepts}
\label{subsec:er_bp}
Epoch randomness is important in sharding blockchains. A randomness could be used as a fresh puzzle for mining and as a seed to achieve random node allocation.
The problems that the $\er$ component needs to solve are as follows. 
First, it is necessary to determine which nodes are responsible for running $\er$ to generate the randomness. 
Second, in each epoch, the time point to invoke $\er$ needs to be confirmed. 
Third, the running time, system overhead, and failure rate of the epoch randomness generation protocol need to be fully considered. 
Fourth, the properties of the randomness generated by $\er$ need to be analyzed to make it applicable for the sharding blockchain.

Randomness generation could be regarded as an independent research field and has been studied for a long time.
In 1983, Blum \cite{B83} first proposed a coin-tossing protocol that aims at generating random values between two untrusted parties. In the same year, Rabin \cite{R83} formalized the concept of a random beacon, which generates fresh random numbers at regular intervals. For a group of nodes in need of continuous random numbers, such protocols can be executed to obtain a reliable source of randomness.


When a group of untrusted nodes is involved in a consensus protocol, an important issue is how to fairly generate public randomness without trusted third parties. However, there may be several malicious nodes trying to bias the outputs or forcing the protocol to restart to their advantage. Therefore, the distributed randomness protocols need some cryptographic building blocks to ensure fairness and security. 
Considering distributed approaches, random beacon protocols need to meet with the following properties: public-verifiability, unpredictability, bias-resistance, and availability (liveness), as outlined in \cite{WSN+19}, \cite{SJK+17}, \cite{SJS+20}.
\begin{itemize}
	\item Public-verifiability: Any third party not directly participating in the protocol should also be able to verify the generated values. As soon as a new random beacon value becomes available, all parties can verify the correctness of the new value using public information only.
	\item Unpredictability: Any node (either honest or malicious)
	should not be able to predict (precompute) future random beacon values.
	\item Bias-resistance: Neither a single node nor colluding nodes can bias (influence) the output value to their benefit.
	\item Availability/Liveness: Neither a single node nor colluding nodes can obstruct the progress.
\end{itemize}

In addition to the above four properties, some distributed random beacon protocols also have the property of guaranteed output delivery \cite{SJS+20}, \cite{CD17}, which means that any adversary cannot interfere or prevent honest nodes from obtaining the randomness output. 

In recent years, there has been a substantial amount of new research related to the generation of distributed randomness in academia and industry, which are introduced below.

\subsection{Existing Approaches}
\label{subsec:er_es}
Current random beacon protocols employ various cryptographic techniques to generate secure randomness. These techniques are mainly divided into several categories, namely VRF \cite{MRV99}, threshold signature \cite{B03}, publicly verifiable secret sharing (PVSS) \cite{S96},\cite{S99} and verifiable delay functions (VDF) \cite{BBB+18}, etc. In this section, we detail those randomness generation methods.

\subsubsection{VRF}
The concept of VRF comes from a pseudorandom oracle \cite{GGM86}, which simulates a random oracle from an $a$-bit string to a $b$-bit string using a seed. Formally, there exists a polynomial-time algorithm $F(\cdot,\cdot)$ such that $f_s = F(s,\cdot):\{0,1\}^a\rightarrow\{0,1\}^b$ always holds, where $s$ denoted the seed. Intuitively, such a pseudorandom oracle is not verifiable. Therefore, Micali \textit{et al.} \cite{MRV99} proposed a new type of pseudorandom oracle, named VRF. That is, given an input $x$, the seed-owner should be able to compute the value $v=f_s(x)$ and a proof proving the correctness of $v$ in polynomial time. The result $v = f_s(x)$ is unique and computationally indistinguishable from a truly random string $v'$ of equal length. With application to the blockchain, the main idea of the VRF is that all nodes (i) use their private keys as part of the input to generate random numbers, (ii) output random numbers along with zero-knowledge proofs, and (iii) verify the correctness of received randomness. Each node combines the output of a VRF with other variables (i.e., round numbers), then signs by its own private key. If the resulted randomness is smaller than a pre-defined threshold, the node can know privately that it is selected as a leader or a committee member.

In general, the purpose of a VRF is to generate verifiable and unpredictable random values locally. Combined with a consensus protocol like PoS, it can dynamically adjust the weight of all nodes. So this strategy is scalable and suitable for different applications. That is why there are many well-known public blockchain projects using the VRF as their randomness source, such as Algorand \cite{GHM+17}, Ouroboros Praos \cite{DGK+18}, and DVRFs \cite{GLO+20}

\subsubsection{Threshold Signature} The idea behind threshold cryptographic schemes is to split secret information (i.e., a secret key) and computation (i.e., signature generation or decryption) among multiple parties in order to remove the risk of a single point of failure. The difference between a threshold signature and a general digital signature is that the former is no longer completed by an individual, but by a threshold set of participants. In a $(t,n)$ threshold signature scheme, $n$ represents the total number of participants, and $t$ is the threshold. When any subset of $t$ (or more) participants jointly sign the same message, they can get a signature representing the whole group, but any $t-1$ or fewer participants cannot get a valid signature. Also, anyone can verify the correctness of the signature using the pre-fixed public key. The general process about threshold signatures in randomness generation is that all parties (i) provide a signature share on a common message, (ii) verify the received signatures shares, and (iii) integrate the valid shares to obtain a random output.

There are two methods to ensure the secure key distribution process, one is an initial trusted setup (i.e., a trusted dealer), and the other is an interactive protocol among all parties (i.e., distributed key generation protocol (DKG) \cite{GJK+07}). The former relies on trust assumptions which are easy to understand, so we discuss the latter briefly. The DKG protocol allows multiple participants to work together to initialize the cryptosystem securely and generate its public and private keys. While the public key is output in the clear, the private key is shared by participants through a secret sharing scheme which can be used in group-oriented cryptosystems. In summary, the threshold signature can avoid misuse of power and achieve ``fairness''. Cachin \textit{et al.} \cite{CKS05} and Dfinity \cite{HMW18} both employ threshold signatures in their constructions. 

\subsubsection{PVSS}
Secret sharing was first proposed by Shamir \cite{S79} in 1979, which enables a dealer to split a secret among a group of participants, each participant obtains a secret share. Shares can be combined to reconstruct the secret through polynomial interpolation. Note that the secret sharing scheme has an important precondition: both dealers and participants are honest. If some parties are malicious and send invalid shares, the honest parties may not reconstruct the secret. To deal with a corrupted dealer or invalid shares in the reconstruction phase, verifiable secret sharing (VSS) \cite{F87} is proposed. In 1996, Stadler \cite{S96} proposed PVSS, where anyone (including participants and third parties) can verify the correctness of shares through public information only. During the distribution phase, a dealer computes an encrypted share along with a non-interactive zero-knowledge proof (NIZK) \cite{BDM+91} for each participant to ensure the validity of encryption. During the reconstruction phase, the participants recover the original secret by publishing the properly decrypted shares and the NIZK proof showing its correct decryption. The general idea of PVSS-based schemes is that each node (i) privately generates a random secret value, (ii) broadcasts a commitment and shares of this secret to all nodes, and (iii) reveals this secret after verification. If a node fails to do so, other honest nodes can jointly recover the secret from the received shares.

The schemes including HydRand \cite{SJS+20}, Scrape \cite{CD17}, Rand$^{*}$ protocol family \cite{SJK+17} and ALBATROSS \cite{CD20} are all based on PVSS. RandHerd \cite{SJK+17} also uses collective signing (CoSi) \cite{STV+16} and ALBATROSS \cite{CD20} is the first secure random beacon protocol under the universal composability (UC) framework \cite{Canetti01}. 


\subsubsection{Hash Functions}
Several existing solutions generate hash values as randomness through leveraging resources of existing systems. For example, PoW \cite{N08} relies on block hashes as a source of public randomness, proof-of-delay \cite{BGB17} employs a delay function on top of the PoW block hash, and Caucus \cite{AMM18} is designed in the form of a hash chain, which is implemented within a smart contract on Ethereum.

\subsubsection{VDF}
VDF requires a specified number of sequential steps to compute whether or not it is executed on multiple processors, then, produces a unique output that can be efficiently and publicly verified. VDF is useful for constructing randomness beacons from sources such as PoW-based blockchains, in which powerful miners could potentially manipulate the beacon result by refusing to post blocks, resulting in producing an unfavorable beacon output. Therefore, VDFs with a suitable time delay would be sufficient to prevent attacks, miners will not be able to determine the beacon output from a given block before it becomes stale \cite{BBB+18}. Lenstra and Wesolowski proposed Unicorn \cite{LW15} with a sufficiently long delay parameter (longer than the time period during which values may be submitted), even the last party to publish its random value cannot predict the final beacon outcome \cite{BBB+18}. RandRunner \cite{SJH+20} implements a trapdoor VDF with strong uniqueness and does not require an agreement protocol for the VDF inputs, which achieves much lower communication overhead. Additionally, the Ethereum research team \cite{Buterin17} plans to use RANDAO \cite{randao17} and VDF in the Ethereum beacon chain to randomly select block producers. Chia Network \cite{CP19} plans to use VDFs to support their proof-of-space and time. 

\subsubsection{Others}
In addition to the above typical types, there exists some other solutions that use various encryption schemes to generate randomness, namely homomorphic encryption (HE) \cite{CDN01} and multi-authority ciphertext-policy attribute-based encryption (MA CP-ABE) \cite{RW15}, etc. The homomorphic property of cryptosystems allows nodes to operate the ciphertext directly without decryption. Both HE-Rand\footnote{We name the protocol proposed in the paper as ``HE-Rand''.} \cite{NNL+19} and HERB \cite{CSS19} implement the threshold version of ElGamal as homomorphic encryption to generate randomness. Moreover, Zhang \textit{et al.} \cite{ZKC+19} define a threshold MA CP-ABE protocol and use it as a commit-and-reveal scheme to construct ABERand as a public distributed randomness beacon.


\subsection{Comparison of Distributed Random Beacon Protocols}
\label{subsec:er_c}
In the following, we provide a specific comparison of the state-of-the-art distributed random beacon protocols. Our comparison mainly focuses on the cryptographic primitives, network models, randomness properties, and complexity evaluation. The results are presented in Table~\ref{comparison} wherein $n$ denotes the number of participants in the network, $c$ denotes the size of a subset in the specific protocol. Note that this comparison not only considers protocols specifically targeted at implementing random beacons, but also includes approaches that provide a random beacon functionality as a byproduct of their intended application \cite{WSN+19},\cite{SJS+20}.

\begin{table*}[!t]
	\renewcommand{\arraystretch}{2.3}
	
	\caption{Comparison of distributed random beacon protocols}
	\label{comparison}
	\centering
	\small
		\resizebox{\textwidth}{!}{
	\begin{threeparttable}
		\begin{tabular}{cccccccccc}
			\hline
			\rotatebox{90}{\textbf{Typical exsisting solutions}}& \rotatebox{90}{\textbf{Cryptographic
					primitive(s)}} & \rotatebox{90}{\textbf{Network model}} & \rotatebox{90}{\textbf{Trusted dealer or
					DKG required  }} & \rotatebox{90}{\textbf{Liveness /
					failure probability}} & \rotatebox{90}{\textbf{Unpredictability}} & \rotatebox{90}{\textbf{Bias-Resistance}} & \rotatebox{90}{\textbf{Communication complexity}} & \rotatebox{90}{\textbf{Computational complexity}} & \rotatebox{90}{\textbf{Verification complexity}} \\
			\hline
			
			\textbf{Cachin \textit{et al.} \cite{CKS05}} & Threshold Sig. & Async.& yes & \cmark & \cmark & \cmark & $O(n^2)$ & $O(n)$ & $O(1)$ \\
			\hline
			\textbf{Dfinity \cite{HMW18}} & Threshold Sig. + VRF & Sync.& yes\tnote{$\sharp$} & $10^{-12}$ & \cmark & \cmark & $O(cn)$ & $O(c)$ & $O(1)$ \\
			\hline
			\textbf{Algorand \cite{GHM+17}} & VRF & Partially Sync.& no & $10^{-12}$ & $\nearrow$ & \xmark & $O(cn)$\tnote{*} & $O(c)$\tnote{*} & $O(1)$\tnote{*} \\
			\hline
			\textbf{Ouroboros Praos \cite{DGK+18}} & VRF & Partially Sync.& no & \cmark & $\nearrow$ & \xmark & $O(n)$\tnote{*} & $O(1)$\tnote{*} & $O(1)$\tnote{*} \\
			\hline
			\textbf{Nguyen-Van \textit{et al.} \cite{NNL+19}} & HE + VRF & Sync.& no & \cmark & \cmark & \cmark & $O(n)$ & $O(1)$ & $O(n)$ \\
			\hline
			\textbf{RandRunner \cite{SJH+20}} & VDF & Async. & no & \cmark &\xmark\tnote{$\S$} &\cmark & $O(n)$\tnote{$\clubsuit$} & $O(T)$\tnote{\textdaggerdbl} &$O(\log T)$\tnote{\textdaggerdbl}\\
			\hline
			\textbf{Ouroboros \cite{KRD+17}} & PVSS & Sync.& no & \cmark & \cmark & \cmark & $O(n^3)$ & $O(n^3)$ & $O(n^3)$ \\
			\hline
			\textbf{RandShare \cite{SJK+17}}& PVSS & Async.& no & \xmark & \cmark & \cmark &  $O(n^3)$ & $O(n^3)$ & $O(n^3)$ \\
			\hline
			\textbf{RandHound \cite{SJK+17}} & PVSS & Sync.& no & 0.08\% & \cmark & \xmark & $O(c^2n)$\tnote{$\diamond$} & $O(c^2n)$ & $O(c^2n)$ \\
			\hline
			\textbf{RandHerd \cite{SJK+17}} & PVSS + CoSi & Sync.& yes\tnote{$\sharp$} & 0.08\% & \cmark & \cmark & $O(c^2\log n)$\tnote{$\diamond$} & $O(c^2\log n)$ & $O(1)$ \\
			\hline
			\textbf{Scrape \cite{CD17}} & PVSS & Sync.& no & \cmark & \cmark & \cmark & $O(n^3)$ & $O(n^2)$ & $O(n^2)$ \\
			\hline
			\textbf{HydRand \cite{SJS+20}} & PVSS & Sync.& no & \cmark & $\nearrow$\cmark & \cmark & $O(n^2)$ & $O(n)$ & $O(n)$ \\
			\hline
			\textbf{Proof-of-work \cite{N08}} & Hash Func. & Sync.& no & \cmark & $\nearrow$ & \xmark & $O(n)$ & very high\tnote{$\star$} & $O(1)$ \\
			\hline
			\textbf{Proof-of-delay \cite{BGB17}} & Hash Func. & Sync.& no & \cmark & \cmark & \cmark & $O(n)$ & very high\tnote{$\star$} & $O(\log\Delta)$\tnote{$\dagger$} \\
			\hline
			\textbf{Caucus \cite{AMM18}} & Hash Func. & Sync.& no & \cmark  & $\nearrow$ & \xmark & $O(n)$ & $O(1)$ & $O(1)$ \\
			\hline
			
		\end{tabular}
		
		\begin{tablenotes}   
			\footnotesize              
			\item[$\sharp$] In Dfinity and RandHerd, nodes are devided into smaller groups, and within each of these groups a DKG protocol is required.        
			\item[*] In Algorand and Ouroboros Praos, the approaches for generating randomness require additional communication and verification steps for the underlying consensus protocols or the implementation of a bulletin board. Here we do not take the additional steps into consideration.
			\item[$\diamond$] In RandHound and RandHerd, $c$ is a security parameter and depends on $n$. If $c$ is constant, RandHound thereby reduces the asymptotic cost to $O(n)$ and RandHerd further reduces the cost of producing successive beacon outputs to $O(\log n)$ per server.
			\item[$\star$] In proof-of-work and proof-of-delay, the computational complexity is not dependent on the number of nodes $n$.
			\item[$\dagger$] In proof-of-delay, the verification is executed within a smart contract via an interactive challenge/response protocol which has logarithmic verification complexity $O(\Delta)$ in the security parameter $\Delta$.
			\item[$\S$] In RandRunner, only unpredictability is affected by network asynchrony while all other properties remain unchanged. After the network conditions normalize, unpredictability is restored and the recovery time increases linearly with the duration of the asynchronous period \cite{SJH+20}.
			\item[$\clubsuit$] In RandRunner, if all nodes execute the protocol properly and the network is reliable, the communication complexity is $O(n)$. When concerning an adversarial leader, the communication complexity changes to $O(n^2)$ (reliable broadcast) or $O(n\log n)$ (gossip protocol).
			\item[\textdaggerdbl] $T$ is the correct nodes? upper bound for the computation time of a VDF.
		\end{tablenotes}
	\end{threeparttable}
}
\end{table*}

\subsubsection{Network Model} We divide the network models of the analyzed protocols into three categories, namely synchronous, partially synchronous, and asynchronous models (ref. Definition~\ref{def:syn_network}-\ref{def:asyn_network}). 

PoW-based blockchains, such as Bitcoin and Ethereum, assume a synchronous network model. Proof-of-delay \cite{BGB17} and Caucus \cite{AMM18} are also synchronous since they are built on top of such PoW-based blockchains. In \cite{SJK+17}, RandHound and RandHerd are designed within a synchronous setting while RandShare is asynchronous. HE-Rand \cite{NNL+19} is also synchronous as the protocol is described in rounds. RandRunner is designed for practical deployment scenarios with bounded network delay, while it still ensures liveness, public-verifiability, and bias-resistance even under periods of full asynchrony \cite{SJH+20}.

\subsubsection{Randomness Properties}
As for randomness properties, ``\cmark'' in Tabel~\ref{comparison} denotes that protocols achieve corresponding properties unconditionally. ``\xmark'' means the protocol does not satisfy such properties.

In regard to liveness property, Algorand and Dfinity consider failure probabilities of at most $10^{-12}$ \cite{GHM+17}, \cite{DGK+18}, RandHound and RandHerd are 0.08\% \cite{SJK+17} while RandShare does not guarantee liveness under full asynchrony since malicious nodes might never send messages. 

For unpredictability, ``$\nearrow$'' denotes probabilistic guarantees for unpredictability, which quickly (exponentially in the waiting time) get stronger the longer a client waits after it commits to using a future protocol output \cite{SJS+20}. In Algorand \cite{GHM+17}, Ouroboros Praos \cite{DGK+18}, PoW \cite{N08} and Caucus \cite{AMM18}, the new randomness depends on the miner's or the leader's secret value. As long as malicious nodes mine a sequence of blocks or are selected repeatedly as leaders, prediction becomes possible. This problem can be solved by letting honest nodes participate in block production or be selected as leaders. Therefore, the probability of a successful prediction decreases exponentially with the number of rounds. Moreover in HydRand \cite{SJS+20}, nodes are not allowed to be leaders again within $f$ rounds ($f$ is the maximum number of byzantine parties) which only achieves unpredictability after $f + 1$ rounds. As for RandRunner, unpredictability relies on a synchronous network model. When nodes cannot disseminate messages within a bounded network delay, unpredictability is weakened. After the network conditions normalize, unpredictability is restored and the recovery time increases linearly with the duration of the asynchronous period \cite{SJH+20}.

Finally, for bias-resistance, Algorand \cite{GHM+17} and Ouroboros Praos \cite{DGK+18} do not provide this property. As mentioned before, miners in PoW and Caucus protocols could arbitrarily manipulate the beacon result (i.e., block timestamps), that is, the result is biasable.

\subsubsection{Complexity Evaluation}
Complexity evaluation includes communication complexity (the overall bits
transmitted by all nodes per round), computational complexity (the number of operations by one node per round), and verification complexity (the number of operations by an external verifier per round).

Obviously, in proof-of-work \cite{N08}, proof-of-delay \cite{BGB17}, and Caucus \cite{AMM18} protocols, a miner only has to perform one broadcast which leads to a communication complexity of $O(n)$. For Ouroboros Praos, communication complexity is not provided in the original work \cite{DGK+18}, here we refer to \cite{SJS+20} which infers that Ouroboros Praos has a communication complexity in $O(n)$, because the protocol only provides guarantees for eventual consensus and is based upon many of the design principles of PoW blockchains. 
Other protocols like Ouroboros \cite{KRD+17}, Scrape \cite{CD17}, and RandShare \cite{SJK+17} are all based on the PVSS scheme, where each node commit to a secret value and broadcast a message of size $O(n)$ to all other nodes, leading to a communication complexity of $O(n^3)$. HydRand \cite{SJS+20} reduces communication complexity to $O(n^2)$ because only a single node (leader) has to perform the distribution of PVSS shares per round. RandRunner \cite{SJH+20} achieves a communication complexity of $O(n)$ if all nodes follow the protocol and the network is reliable. When concerning an adversarial leader, it assumes two possible strategies for message dissemination, namely reliable broadcast and gossip protocol. The former means every honest node sends any valid message it received to all other nodes, resulting in a communication complexity of $O(n^2)$, while the latter is suitable for a large number of nodes and the complexity is $O(n\log n)$.

As for computational and verification complexity, proof-of-work \cite{N08} and proof-of-delay \cite{BGB17} achieve high computational complexity for the reason that both of them rely on solving cryptographic puzzles. The VRF-based approaches such as Algorand \cite{GHM+17}, Ouroboros Praos \cite{DGK+18}, Caucus \cite{AMM18} (after the initial setup) as well as HE-Rand \cite{NNL+19} are efficient, because they only require the verification of a VRF or hash preimage. The verification of RandRunner \cite{SJH+20} only requires two hash functions and one verification algorithm with $2t$ as exponentiation ($T = 2^t$ where $T$ is the time parameter of the VDF).

To sum up, proof-of-work \cite{N08} and proof-of-delay \cite{BGB17} approaches are suitable for larger and dynamic sets of participants. RandRunner \cite{SJH+20} is very resilient to temporary network delays or network breaks. Ouroboros \cite{KRD+17}, RandShare \cite{SJK+17}, Scrape \cite{CD17} and HydRand \cite{SJS+20} are more suitable for smaller groups due to their high communication complexity with the increasing number of nodes. Nguyen-Van \textit{et al.} \cite{NNL+19} use homomorphic encryption (ElGamal on elliptic curves) to encrypt shares, while they do not consider a colluded user (``Requester''). Cachin \textit{et al.} \cite{CKS05} come with a formal security proof in the asynchronous network model, but it is based on elliptic curve pairings which are not yet well-established. Both RandHound \cite{SJK+17} and RandHerd \cite{SJK+17} divide all the participants into small groups. However, RandHound does not offer bias-resistance and RandHerd needs DKG protocol during setup and requires additional ``view-change'' \cite{CL99} if the current leader fails to take adequate steps. Dfinity \cite{HMW18} provides strong bias-resistance but relies on DKG protocol during initialization which increases the communication complexity. Algorand \cite{GHM+17} and Ouroboros Praos \cite{DGK+18} achieve better scalability while weakening bias-resistance. Caucus \cite{AMM18} can be easily deployed in a smart contract, but it is easily biased. 

\subsection{Problems and Future Directions}
\label{subsec:er_pp}
In the following, we analyze potential problems and future
directions related to randomness generation in sharding
blockchains from the following three perspectives, i.e., security requirements,
performance improvements and rigorous analysis.

\subsubsection{Security Requirements} 
As shown in Table~\ref{comparison}, some of these schemes do not guarantee either the generation of perfectly unpredictable random values or that a value will be generated regardless of adversarial behavior. Specifically, an adversary might cause a liveness failure, try to bias the randomness, predict future random beacon outputs before the honest nodes get the values, or deceive a third party into accepting an invalid randomness. Besides, VDF-based approaches also have problems that their security depends on very accurate estimates of the average concrete complexity of certain computational processes \cite{CD20} (i.e., the time delay parameter), which are hard to obtain in practice. 

Therefore, how to ensure the security guarantees (randomness properties mentioned in Section~\ref{subsec:er_bp}) still needs to be studied in the future.

\subsubsection{Performance Improvements} Generally speaking, the more nodes participating in the consensus, the more secure a system will be. But on the other hand, the communication overhead also increases with the number of nodes. In Table~\ref{comparison}, the methods that do
have perfect security guarantees suffer from higher computational and communication complexity, especially, some PVSS schemes (Ouroboros \cite{KRD+17}, Scrape \cite{CD17}) have a complexity of $O(n^3)$. Moreover, the approaches based on threshold cryptographic schemes need to execute the DKG protocol during the setup phase, which may increase the communication complexity. Besides, most approaches do not support frequent changes within the nodes set. When some new nodes join the network, it takes additional time and requires transferring more data than the original process.

Consequently, on the premise of ensuring the availability of the sharding blockchains, how to reasonably design the randomness generation process, organize participation in node communication, balance scalability, and security requirements, and achieve efficient implementation of the protocol are issues that need to be resolved in the future.

\subsubsection{Formal Security Analysis} 
The rigorous analysis of a randomness generation protocol mainly includes formal definitions, precise assumptions, and formal security proofs. Formal definitions mean the definitions of adversary model, network model, and randomness properties that a protocol satisfies, while precise assumptions refer to the assumptions of the underlying cryptographic schemes. Formal security proofs should be strictly logical, which prove that under certain definitions and assumptions, no adversary can successfully break the scheme with overwhelming probability.

As we analyzed above, most approaches are under synchronous models that are relatively strong and might be temporarily violated. For example, in HydRand, any leader which (temporarily) fails to deliver required messages is excluded from further participation, and the round duration parameter has to be carefully selected to avoid liveness failures \cite{SJS+20}. Thus, how to address the current limitations should be considered in future works.


Rand$^{*}$ family \cite{SJK+17}, HydRand \cite{SJS+20}, and Scrape \cite{CD17} are typical representatives of stand-alone protocols, which are specifically designed to generate randomness. Namely, these protocols are not used in isolation but as building blocks of more complex systems. Therefore, the composability of these protocols is an important issue. In particular, a UC secure protocol ensures that it can be used as a building block for more complex systems while retaining its randomness properties, which is essential for randomness beacons.

\section{Node Assignment}
\label{sec:node_assignment}
In this section, node assignment is analyzed. We first give the basic concepts of node assignment in Section~\ref{subsec:na_bp}. Then in Section~\ref{subsec:na_es}, existing approaches are classified into binomial distribution and hypergeometric distribution. In addition, potential problems and future directions are discussed in Section~\ref{subsec:na_pp}. 

\subsection{Basic Concepts}
\label{subsec:na_bp}

The new nodes need to be randomly assigned to multiple shards. Otherwise, an adversary might gather the colluding nodes into a certain shard, thereby controlling the entire shard. In order to achieve the security of the node assignment, randomness must be unpredictable, unbiased, and public verifiable. The definitions of the specific properties of randomness is given in Section~\ref{subsec:er_bp}.
The problems in the node assignment process are as follows. 
First, it is necessary to ensure that the entire allocation process is random, that is, the adversary cannot bias the allocation process. This requires that the random number is safe, and a pseudo-random number generator is used to generate a corresponding random number for each node. 
Second, the parameters need to be set reasonably to ensure that the number of honest nodes in each group meets the standard. This requires the use of a certain mathematical model to strictly analyze the final distribution result when the node is a non-infinite pool.


\subsection{Existing Approaches}
\label{subsec:na_es}
Let $n$ denote the total number of nodes participating in a protocol. Let $m$ represent the number of shards, and let $u$ denote the number of nodes in a single shard. 

\subsubsection{Binomial Distribution}
The node assignment process is regarded as a random sampling problem. Under the following assumption, the binomial distribution can be used.

The nodes before the distribution process are assumed to form an infinite pool. In other words, each time a node is selected from the infinite pool, the probability that the node being honest or malicious remains constant.

The probability here refers to an adversary's computational power, which is denoted by $\rho$. Assume that the target honest fraction in a committee is $Q_0$, which might be $2/3$ or $1/2$. Let $X$ denote the number of times that picking a malicious node. The probability that an adversary's proportion in a selected committee is exactly a certain value, e.g., $1-Q_0$, is calculated through the binomial distribution as in Equation~\ref{equ:binomial_single}.
\begin{equation}\label{equ:binomial_single}
	\Pr[X=u(1-Q_0)]=\binom{u}{u(1-Q_0)}\rho^{u(1-Q_0)}(1-\rho)^{uQ_0}
\end{equation}

When a selected committee is malicious, we say a distribution is failed, i.e., the fraction of malicious nodes exceeds a predefined target value $1-Q_0$. So the cumulative binomial distribution could be adopted to calculate the failure probability, where $X$ is supposed to be greater than the value $u(1-Q_0)$. Hence, the failure probability could be calculated as shown in Equation~\ref{equ:binomial_cumulative}.
\begin{equation}\label{equ:binomial_cumulative}
	\Pr[X\geq \lceil u(1-Q_0) \rceil]=\sum_{x=\lceil u(1-Q_0) \rceil}^{u}\binom{u}{x}\rho^{x}(1-\rho)^{u-x}
\end{equation}

Omniledger \cite{KJG+18} employs the cumulative binomial distribution described above. 

\subsubsection{Hypergeometric Distribution}
The infinite pool assumption in binomial distribution means the member selection process does not influence the probability of being honest or malicious in the selected node. The selection is done with node replacement, i.e., the selected node has to be replaced back to the pool. On the contrary, hypergeometric distribution does not assume an infinite pool, which means the distribution is done without replacement. 

The failure probability is calculated by the cumulative hypergeometric distribution, as shown in Equation~\ref{equ:hypergeometric}.
\begin{equation}\label{equ:hypergeometric}
	\Pr[X\geq \lceil (1-Q_0)u \rceil] = \sum_{x=\lceil u(1-Q_0) \rceil}^{u} \frac{\binom{\rho n}{x}\binom{n(1-\rho)}{u-x}}{\binom{n}{u}}
\end{equation}

RapidChain \cite{ZMR18} and SGX sharding \cite{DDL+19} analyze the epoch security utilizing the cumulative hypergeometric distribution. Hafid \textit{et al.} \cite{HHS19} carry out a probabilistic security analysis of sharding blockchains using tail inequalities such as Hoeffding inequality \cite{Hoeffding94} to approximate the upper bound of the failure probability for each epoch.

\subsubsection{Other Distribution}
In fact, both binomial distribution and hypergeometric distribution are based on random allocation of participating nodes, while some existing schemes use special node distribution rules. 

In PolyShard \cite{LYY+21}, nodes store and compute on a coded shard of the same size that is generated by linearly mixing uncoded shards. PolyShard uses the Lagrange Coded Computing \cite{YRK+19} technology to code shards. 

\subsection{Problems and Future Directions}
\label{subsec:na_pp}
The processes of node selection and node assignment are connected together, so we analyze the problems in the whole procedures.

\begin{figure}[h!]
	\centering
	\includegraphics[width=0.6\textwidth]{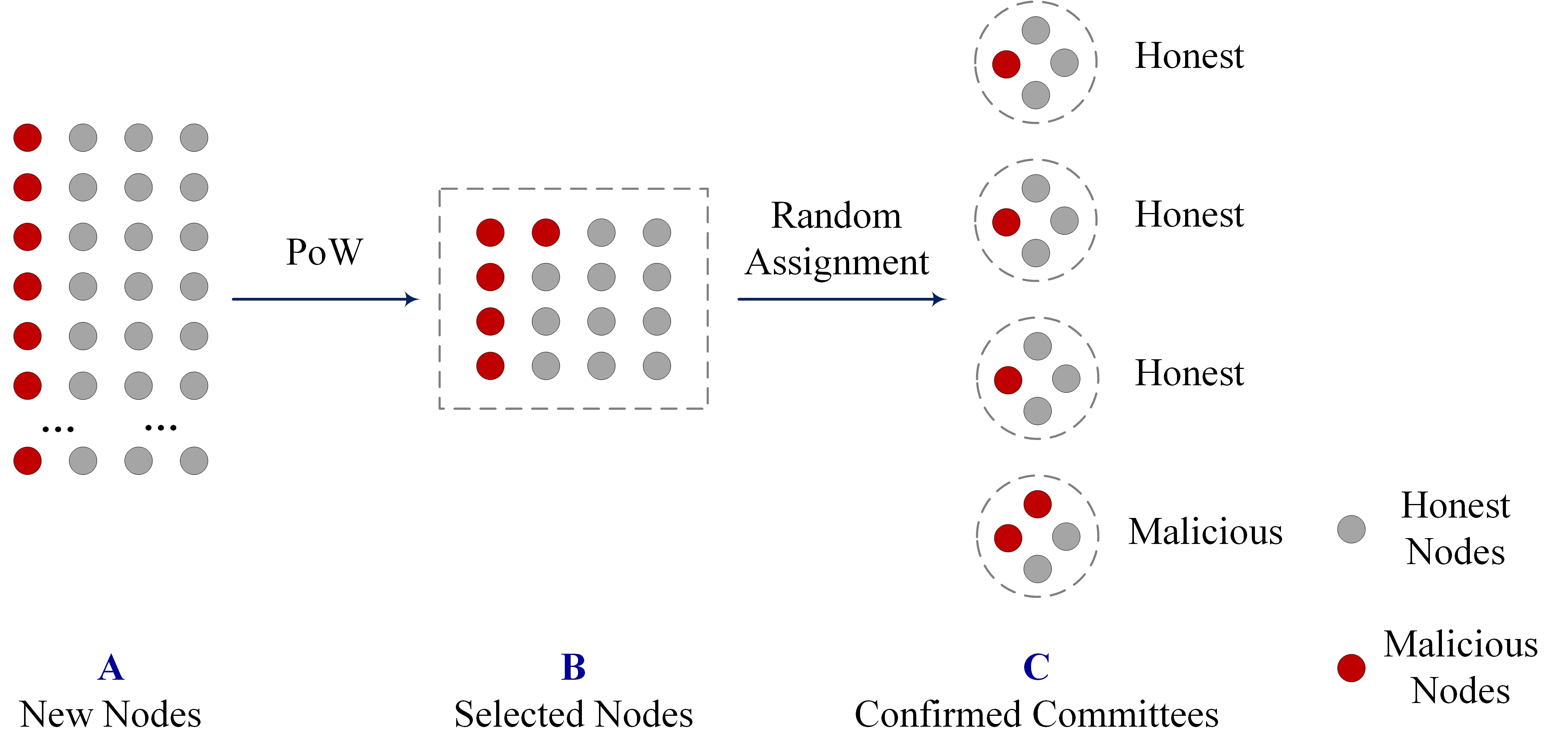}
	\caption{The process of node selection and node assignment.}
	\label{fig:node_assignment}
\end{figure}

As shown in Figure~\ref{fig:node_assignment}, $\biga$ is used to denote new nodes, i.e., all nodes that want to participate in the protocol. $\bigb$ represents selected new nodes and $\bigc$ refers to confirmed committees. From $\biga$ to $\bigb$, there must be a mechanism such as PoW to defend against the Sybil attacks \cite{Douceur02}. Furthermore, from $\bigb$ to $\bigc$, a secure randomness is in need to assign selected nodes into multiple committees. 

\subsubsection{The analysis from $\biga$ to $\bigb$ is ignored}
In Omniledger \cite{KJG+18} and RapidChain \cite{ZMR18}, the protocol contains the steps from $\biga$ to $\bigb$ and then from $\bigb$ to $\bigc$. However, they do not consider the changes in the adversarial proportion from $\biga$ to $\bigb$. In fact, the adversarial proportion in $\bigb$ is larger than that of $\biga$ due to several reasons. First, an adversary might have an advantage in message transfer so that he could start mining in advance of honest nodes. Second, if a leader in a reference committee is malicious, then he could launch a node censorship attack as described in Section~\ref{subsec:ns_pp} to increase the proportion of nodes under his control. Assume that the computational power proportion of an adversary is $\rho$, then the proportion of malicious nodes in $\bigb$ is greater than $\rho$ with a high probability. The analysis in Omniledger and RapidChain still regards $\rho$ as an initial input, which would lead to false results after random allocation. 
As a result, the confirmed committees in $\bigc$ could be malicious. 

\subsubsection{The infinite pool assumption is not accurate}
The node assignment process is to randomly allocate selected nodes $\bigb$, while the number of nodes in $\bigb$ is limited. Whenever a node is selected, the proportion of malicious nodes in $\bigb$ changes. Therefore, assuming an infinite pool is inaccurate, the probability obtained in this way will have a deviation. 

\subsubsection{The failure rate with cumulative hypergeometric distribution is imprecise}
When computing the failure probability of an epoch in RapidChain, the failure rates of all committees is calculated through Equation~\ref{equ:hypergeometric}, which is imprecise. In fact, Equation~\ref{equ:hypergeometric} could only be used to compute the failure rate of the first committee. However, after the current committee is confirmed, the subsequent selection of committees will be affected by the current one, that is to say, the parameters of the cumulative hypergeometric distribution have been changed at this time. Therefore, the calculation of the epoch failure probability is also inaccurate. The impact of the first committee on subsequent allocation parameters is not considered.

\section{Intra-Shard Consensus}
\label{sec:intra-shard_consensus}
As described in the modular design in Section~\ref{sec:decomposing}, the intra-shard consensus is the key component for every sharding blockchain. 
In this section, we research intra-shard consensus protocols. In Section~\ref{subsec:isc_bp}, basic concepts of intra-shard consensus protocols are given. Based on this, we divide sharding blockchains into instant and eventual sharding blockchains according to their intra-shard consensus and give their definitions, respectively. Section~\ref{subsec:isc_es} introduces the state machine replication algorithms that may be used in sharding blockchains from the aspects of different network models. Finally, Section~\ref{subsec:isc_pp} summarizes potential problems that might occur in intra-shard consensus protocols. 

\subsection{Basic Concepts}
\label{subsec:isc_bp}
The main purpose of the intra-shard consensus is to efficiently process the transactions within a shard. In a sharding blockchain system, the intra-shard consensus needs to cooperate with other shards to commit cross-shard transactions. This requires the intra-shard consensus algorithm to provide availability certificates of the relevant transaction inputs, that is, to generate proofs of inputs. The proofs appear in the form of signatures. In addition, in some sharding blockchains that adopt reference committees, the intra-shard consensus is also used to confirm the list of new committee members. The intra-shard consensus algorithm greatly affects the efficiency of transaction processing.

Several issues need to be considered for intra-shard consensus. First, the scalability of the intra-shard consensus algorithm should be taken into account. The scalability is related to the communication and computational complexity inside the shard. 
The transaction processing capabilities might sharply decrease as the number of nodes in the shard increases.
Second, in order to meet the special needs of the sharding blockchain scenario, the intra-shard consensus algorithm needs to handle different types of proposals. 
Third, the relationship between the intra-shard and the entire network transmission model needs to be taken into account.

Intra-shard consensus algorithms could be divided into two categories, i.e., strong consistency consensus algorithms and weak consistency consensus algorithms. In a weak consistent (ref. Definition~\ref{def:weak_consistency}) blockchain, a block producer is a single node, determined by PoW or PoS within each shard. We call such blockchains as eventual sharding blockchains (ref. Definition~\ref{def:eventual}). In a strong consistent blockchain (ref. Definition~\ref{def:strong_consistency}), each shard runs a committee, and the committee acts as the block producer, running a distributed consensus algorithm, e.g., PBFT, to confirm transactions and generate blocks. We call such blockchains as instant sharding blockchains (ref. Definition~\ref{def:instant}).


\subsection{Existing Approaches}
\label{subsec:isc_es}
In the following, we discuss the algorithms that could be used as intra-shard consensus in sharding blockchains, including strong consistent and weak consistent algorithms.

\subsubsection{Strong Consistency}
The intra-shard consensus algorithms for instant sharding blockchains are usually some classical distributed consensus algorithms or some adaptions of them which realize strong consistency. 

In classical distributed consensus algorithms, a group of nodes in a permissioned network realizes state machine replication (ref. Definition~\ref{def:state_machine_replication}), achieving consistency and liveness.

In general, classical distributed consensus algorithms have different assumptions on the situation of nodes in the network, such as whether crash or Byzantine nodes (ref. Definition~\ref{def:byzantine_node}) exist. It is usually considered that the behaviors of Byzantine nodes include those of crash nodes, so protocols that tolerate the Byzantine nodes are more applicable and robust. In the context of blockchain, various BFT (ref. Definition~\ref{def:BFT}) protocols have been proposed. We mainly introduce the research on BFT below since BFT protocols could be well combined with blockchain to form a so-called hybrid consensus.

According to the network model assumptions, classical distributed consensus algorithms could be divided into the following three categories: classical distributed consensus algorithms in synchronous networks, asynchronous networks, and partially synchronous networks. 


\paragraph{Synchronous networks}
In the following, we first introduce some representative distributed consensus algorithms under synchronous networks, then we describe their applications to sharding blockchains.

\emph{(1) Distributed consensus algorithms:}
As described in Definition~\ref{def:syn_network}, in a synchronous network, the messages sent by honest nodes in a certain round must reach each other before the next round. As a result, the message transmission delay $\Delta$ is used as a parameter in related protocols, which simplifies the protocol design to some extent.

The Byzantine quorum system \cite{Bazzi00} first proposes the concept of a ``quorum'', which could be seen as the minimum number of votes required for honest nodes to agree on a proposed value in a voting round. The quorum is to prevent a malicious leader from equivocating, that is, sending different proposals to different honest members in the same round. The concept of a quorum is applicable in both synchronous and partially synchronous networks. Specifically, in a partially synchronous network where the adversary model is $u=3f+1$, a quorum is set to $2f+1$, which means in a voting round, an honest member considers this voting round to be successful only after collecting at least $2f+1$ votes (including its own). The reason is as follows. Assuming that the malicious leader sends different proposals $p$ and $p'$ to different members, it is proved by contradiction that $p$ and $p'$ cannot be committed at the same time. Assume that both $p$ and $p'$ get $2f+1$ votes from $u$ members. Then there are at least $2f+1+2f+1-(3f+1)=f+1$ members voting for both $p$ and $p'$, which is contradictory to the assumption that there are only $f$ malicious members, since honest members will only vote for one proposal value in a round.

Sync HotStuff \cite{AMN+20} assumes a synchronous network and utilizes the pipeline technology to improve proposals. Different from HotStuff, Sync HotStuff adopts a two-phase leader-based method to process proposals. The transaction confirmation delay is declared as $2\Delta$ in a steady state where $\Delta$ denotes the upper bound of message transmission delay. 


Other distributed consensus algorithms under the synchronous network model include XFT \cite{LVC+16}, practical synchronous Byzantine consensus \cite{RNA+17}, Ouroboros-BFT \cite{KR18}, Flexible BFT \cite{MNR19}, Hybrid BFT \cite{MCK20}, PiLi \cite{CPS18pili}, etc. 
These schemes could be applied to sharding blockchains by adding specific interfaces.

\emph{(2) Combined With Sharding Blockchains:} 
Note that in a sharding blockchain, the message transmission model of the entire network might be different from that inside a shard. That is to say, the network model within the shards could be a synchronous network, while the entire network is synchronous or partially synchronous.

RapidChain \cite{ZMR18} uses a variant of the practical synchronous Byzantine consensus proposed in \cite{RNA+17}. The adversary model is still $u=2f+1$, namely, it could tolerant nearly $1/2$ Byzantine nodes. A leader is selected according to the randomness generated by the reference committee. The intra-shard consensus mainly contains four rounds: \textsf{propose}, \textsf{echo}, \textsf{pending}, and \textsf{accept}. A leader broadcasts the message with its hash value in the \textsf{propose} round. Then every node receiving the message broadcasts its hash value among the network in the \textsf{echo} round, to ensure every honest node obtains all messages sent by the leader. If a malicious leader sends more than one message to different nodes in the propose round, then honest nodes will detect it during the \textsf{echo} round, mark the conflicting messages with a pending tag, and broadcast the pending messages. In this case, the malicious leader will be replaced. In the normal case, honest nodes that have received $f+1$ valid echo messages consider the proposal as valid and broadcast the hash value with an accept tag. RapidChain allows a leader to propose a new block while re-proposing the headers of the pending blocks to facilitate the processing of blocks. Here, pending blocks refer to the blocks that is not accepted by honest nodes in some round. 

RapidChain uses a synchronous network model to achieve a fault tolerance of nearly $1/2$ within the committee, while it sacrifices certain transaction processing performance. In the intra-shard consensus protocol, every node needs to wait for a fixed $\Delta$ time in each round of communication. This is also one of the differences between synchronous and partially synchronous network model protocols. The parameter of the upper limit of network delay $\Delta$ could be directly used in the synchronous network model protocols, but not in the partially synchronous and asynchronous network model protocols. Since every node needs to wait for a fixed time in each round, the transaction confirmation time of the protocol is not related to the actual delay of the network, so that the property of responsiveness is not achieved. Meanwhile, the synchronous network model puts high requirements on the network status and is not particularly applicable in reality.

\paragraph{Partially synchronous networks}
The partially synchronous network is a model adopted by most blockchain systems, and it is also a model closer to the network in reality. Consequently, there are many studies in this field. In the following, we first describe several typical schemes, then we introduce the combination of distributed consensus algorithms and sharding blockchains.

\emph{(1) Distributed consensus algorithms:} The distributed consensus algorithms in partially synchronous networks are represented by Paxos, PBFT and its improvements.

\emph{(i) Paxos}

Paxos \cite{L98} is designed for database maintenance in distributed systems.
In Paxos, a primary node sends a prepare message to more than $1/2$ backup nodes of the entire network;
a backup node verifies the legitimacy of the message, and returns a commit message to the primary node after verification;	the primary node forms a commit certificate after collecting enough commit messages;
the primary node sends an accept message containing the commit certificate to the backup nodes;	
backup nodes verify the legitimacy of the accept message;
the node returns an accepted message (corresponding to the accept message) to the primary node after the verification is passed.

Paxos could tolerate $ f $ crash nodes in the $ u=2f+1 $ model (ref. Definition~\ref{def:2f}).
Since most blockchain systems adopt the Byzantine node model, Paxos is combined with blockchains in only a few studies \cite{BW17,CAD18}. 

\emph{(ii) PBFT and its improvements}

In most sharding blockchains or committee-based blockchains, the consensus algorithm within a committee is the PBFT algorithm or its adapted version, so the PBFT algorithm is of vital importance. 
Hence, in the following, we introduce in detail the basic process of the PBFT algorithm and some recent research on its improvement.

\begin{figure}[h!]
	\centering
	\includegraphics[width=0.6\textwidth]{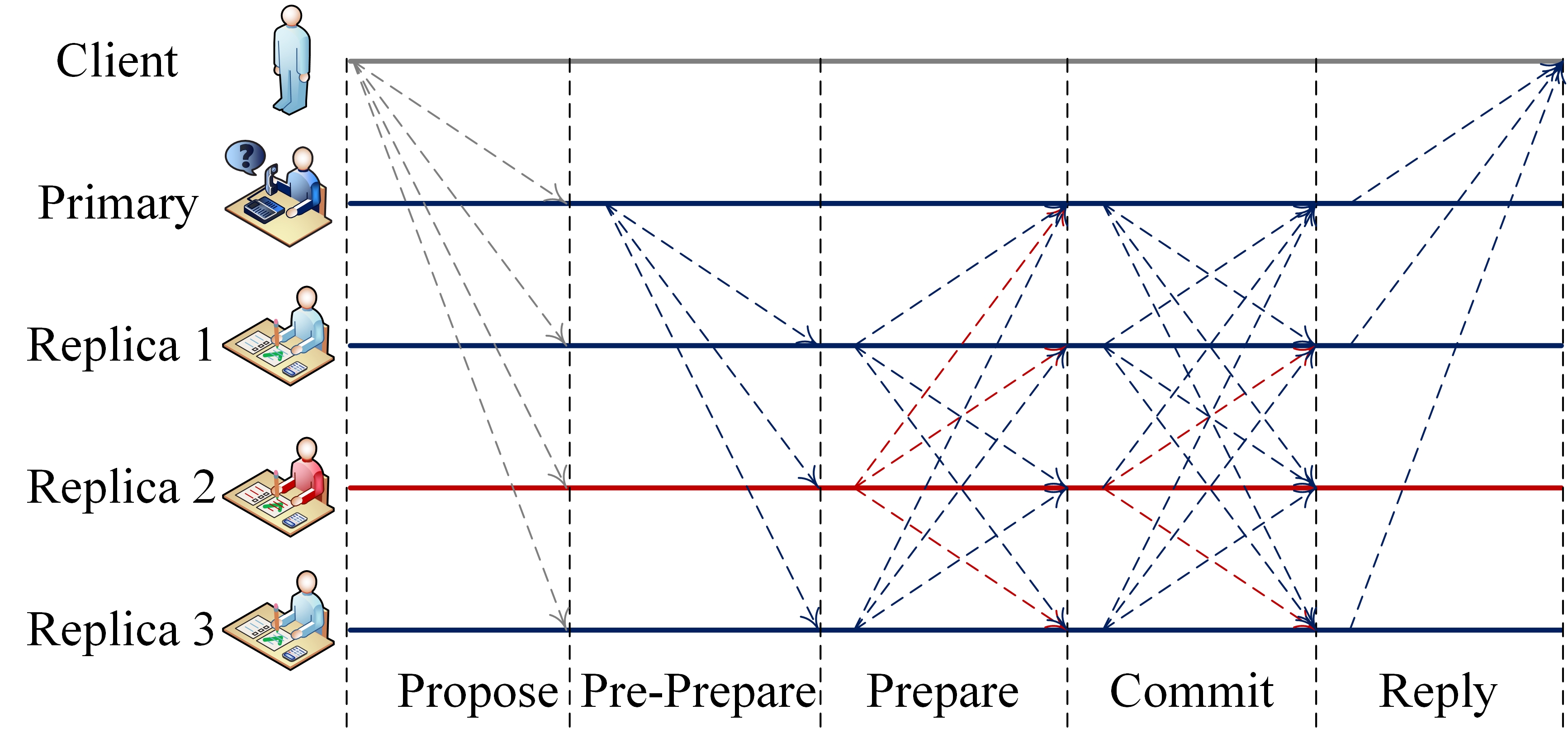}
	\caption{Process of PBFT algorithm}
	\label{fig1}
\end{figure}

PBFT utilizes a similar name to distinguish nodes as Paxos, i.e., primary and backup nodes. In a partially synchronous network, PBFT assumes the $u=3f+1$ model. 
Message authentication codes (MAC) \cite{B05} are used to achieve identity authentication between nodes in PBFT.
In the normal cases, PBFT relies on the operations to process proposals as shown in Fig.~\ref{fig1}.

First, in the \textsf{propose} phase, a client (user) uploads a proposal $ p $ to all nodes.
Second, in the \textsf{pre-prepare} phase, the primary node constructs a pre-prepare message $(\text{pre-prepare}, H(p),$ $s,v) $, where $ H(\cdot) $ is a one-way hash function, $s$ denotes the sequence number, and $ v $ represents the view number. The primary node sends the pre-prepare message to all replicas.
Third, in the \textsf{prepare} phase, every replica node confirms that for the same $(v,s)$, no conflicting preparation message has been received, then broadcasts the prepare message $(\text{prepare},H(p),s,v)$. 
Fourth, in the \textsf{commit} phase, after receiving $2f+1$ (including its own) valid prepare messages, a replica consider $p$ as prepared and broadcasts a commit message $(\text{commit},H(p),s,v)$. 
Fifth, in the \textsf{reply} phase, after receiving $2f+1$ (including its own) valid commit messages, a replica consider $p$ as committed, and returns the committed proposal with its signatures to the client \cite{YJZ+19}.

If a primary node behaves maliciously or does not respond, PBFT relies on a view-change mechanism to change a primary node. A checkpoint mechanism is designed to assist the view-change, where the maximum sequence number of all committed proposals is regarded as a stable checkpoint. The specific steps of a view-change mechanism are as follows.
First, in the \textsf{view-change message broadcast} phase, node $ i $ broadcasts a view-change message ${\rm vc}_i:(\text{view-change},v+1,S^*,C,U,i)$ where $v$ is the view number, $S^*$ stands for the current stable checkpoint number, $C$ denotes the set of $2f+1$ valid commit votes for $S^*$, and $ U $ is a set that contains the prepared messages whose serial number is greater than $ S^* $.
Second, in the \textsf{view-change acknowledgment} phase, a replica verifies the view-change message and constructs a corresponding view-change acknowledgment message ${\rm vca}_i:(\text{view-change-ack},v+1,i,j,H(\text{vc}_j)) $, where $ i $ is the current replica node, $ j $ is the node that sends the view-change message $\text{vc}_j$, and $H(\text{vc}_j) $ is the hash digest of the view-change message. The replica sends $\text {vca}_i $ directly to the new primary node of view $ v+1 $.
Third, in the \textsf{new-view broadcast} phase, for each view-change message, when the primary node collects $ 2f-1 $ view-change acknowledgment messages for $ \text{vc}_j $, then $ \text{vc}_j $ is valid and put into the set $ S $. The new primary node constructs a new-view message $ \text{nv}:(\text{new-view},v+1,S,U^*) $ where $ U^* $ includes the current stable checkpoint and a pre-prepare message with the smallest sequence number after the stable checkpoint. The node verifies the validity of the messages in $S$, updates its local states according to $U^*$, and enter view $v+1$. For more details, readers may refer to \cite{CL99}.

PBFT achieves the consistency and liveness properties of the state machine replication, and the communication complexity is $ O(n^3) $ in normal operations. In a view-change process, the communication complexity is $O(n^4)$. PBFT needs to be improved to make it better applied to sharding blockchains.


The HotStuff\cite{YMR+19} algorithm is proposed to improve PBFT, making the BFT algorithm and blockchain achieve a better combination. HotStuff adopts a partially synchronous network model, and the adversary model is $ u=3f+1 $. HotStuff mainly has three adaptions. First, it uses the pipeline technology to process proposals, that is, the message in a round contains a quorum certificate of the previous round and a new proposal. Second, it adopts the BLS threshold signature to aggregate $2f+1$ vote messages into a single signature, cutting down the communication complexity. Third, in each round, the leader who is responsible for collecting votes and sending the proposal will be changed. That is to say, the view-change occurs in each round. The essence is to integrate the view-change process into the normal operations by adding a voting round \cite{YJZ+19}.

\begin{figure}[h!]
	\centering
	\includegraphics[width=0.6\textwidth]{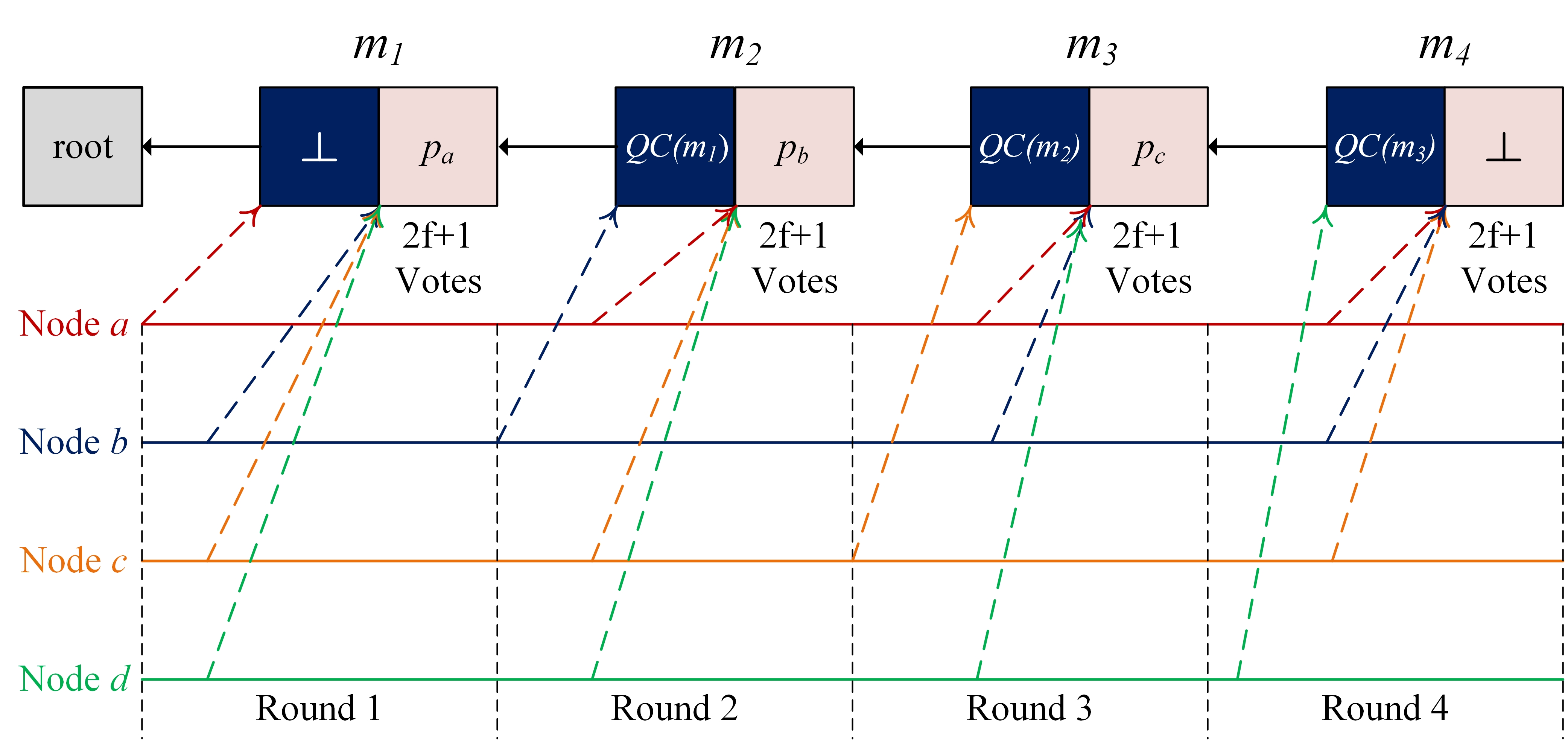}
	\caption{Process of HotStuff algorithm}
	\label{fig2}
\end{figure}

As shown in Fig.~\ref{fig2}, $p_a$ refers to the proposal of node $ a $ in the first round. The message of the first round is denoted as $m_1$. $m_1$ is broadcast by node $a$. Then the other nodes verify $m_1$ and vote for it. Node $b$ acts as a leader and collects valid votes. When the number of valid votes reaches $ 2f+1 $, node $b$ reconstructs a BLS threshold signature using the $2f+1$ valid votes and forms a quorum certificate $\text{QC}(m_1)$. Then node $b$ constructs a message $m_2$ by combining $ \text{QC}(m_1) $ and a new proposal $p_b$, and broadcasts $m_2$ to other nodes. At this time, node $a, c$ and $d$ act as replicas. After receiving $m_2$, they first verify the validity of $\text{QC}(m_1)$ and $p_b$, and then vote to $m_2$ if the verification is passed. The subsequent operations are similar. Note that the proposal is allowed to be $\perp$ to ensure the liveness property of the protocol as shown in round $4$, since it is necessary to consider the situation where there might be no transaction being uploaded. 
The linear view-change in HotStuff refers to the message sent by the leader since only a single threshold signature is in need during view-change instead of $2f+1$ votes in PBFT. 
For more details, readers may refer to \cite{yin2019hotstuff}.

In addition to PBFT and HotStuff, there are other Byzantine fault tolerant algorithms in partially synchronous networks, such as scalable Byzantine fault tolerance \cite{GGA+18} and PaLa \cite{CPS18}. 
They could be regarded as independent consensus algorithms or could be applied to sharding blockchains after some improvements.


\emph{(2) Combined with sharding blockchains:}
The research on the combination of Paxos and blockchains \cite{BW17,CAD18} is little.
In contrast, most instant sharding blockchains adopt the PBFT algorithm or some variants of it as the intra-shard consensus algorithm. ELASTICO \cite{LNZ+16} uses PBFT directly in every committee to process transactions. 
However, the committee of each shard in ELASTICO cannot finally complete the commitment of transactions. They have to send the signatures to the final committee, who will run PBFT to commit transactions. Therefore, ELASTICO cannot handle cross-shard transactions, and its transaction processing efficiency is low.

Omniledger designs Omnicon based on Byzcoin \cite{KJG+16}, which makes some adaptions of PBFT. We first introduce Byzcoin and then introduce the adaptions of Omnicon.

In Byzcoin, MAC is replaced by digital signatures, and they use a tree communication model and the CoSi \cite{STV+16} protocol, a scalable collective signing, to cut down message length. CoSi combines the well-known Schnorr multi-signatures \cite{Schnorr91} with communication trees, and generates a single signature on a message from a group of nodes. CoSi consists of four phases, namely, announcement, commitment, challenge, and response. Byzcoin combines CoSi with PBFT's voting mechanism. A leader's announcement is the same as that of the pre-prepare phase in PBFT. The commitment of other nodes implements the prepare votes. Then the leader collects enough votes and initiates a challenge phase, and the other nodes respond to it, which implements the commit phase in PBFT. Finally, a single signature from $2f+1$ votes is constructed by a leader. So the message length is reduced to $O(1)$ compared with $O(n)$ in PBFT. 

In Omnicon, the authors point out that the CoSi used in Byzcoin is susceptible to faults since the depth of the communication tree is much too large. So they change the message propagation mechanism. In each shard of Omnicon, nodes are divided into several groups based on some generated randomness, to decrease the depth of the communication tree. A shard leader requires signatures from several group leaders. Each group leader is responsible for collecting messages inside its corresponding group. If a group leader does not respond in a certain time, the shard leader will randomly choose another node in the group as a leader. If a shard leader is offline or behaves maliciously, Omnicon still relies on a view-change launched by all shard members to change the leader. 

ZILLIQA \cite{Zilliqa17} improves the efficiency of PBFT by using the EC-Schnorr multi-signature protocol \cite{Schnorr91}. The idea is similar to Byzcoin. In the process of signing, signers are required to maintain a bitmap which indicates who has signed the message. The bitmap is first built by a leader and could be used as an index for a verifier to verify the signature efficiently. 

Chainspace \cite{ASB+18} uses MOD-SMART implementation \cite{SousaB12} of PBFT as the intra-shard consensus. MOD-SMART makes adaptions to PBFT, where reliable broadcast is substituted for validated and provable consensus.

\paragraph{Asynchronous networks}
Next, we introduce the distributed consensus algorithms in asynchronous networks. 

Fischer, Lynch, and Paterson \cite{FischerLP85} propose the FLP impossible theorem. They point out that in a reliable asynchronous network that allows node failure, there is no deterministic consensus algorithm that could solve the consensus problem.

Honey Badger BFT \cite{MXC+16} is a well-known asynchronous BFT algorithm, so we take it as a representative and introduce its operations here. The adversary model of Honey Badger BFT is $u=3f+1$ where $u$ nodes try to reach agreement on $B$ transactions in a round. The asynchronous BFT algorithms mainly rely on reliable broadcast (RBC) and asynchronous binary agreement (ABA) as building blocks to realize consensus. The basic concepts of Honey Badger BFT are as follows.


First, in the transactions collection phase,
the $u$ participating nodes all collect transactions submitted by users.
Second, in the transaction threshold encryption phase, each node uses the threshold encryption function to encrypt $B/u$ of its transactions collected in every single round. 
Third, in the RBC broadcast phase, each node relies on the RBC to broadcast and collect messages. RBC contains two rounds, namely the echo and ready round. When a node receives a certain number of ready message on a value, it believes that the value arrives at all honest nodes. 
Fourth, in the ABA consensus phase, a leader is responsible for collecting all encrypted values sent by other nodes and initiating the ABA algorithm on them. In the ABA algorithm, nodes decide on whether the total transaction set is valid through multiple voting rounds. 
Fifth, in the transaction threshold decryption phase, if the encrypted transaction set is valid, then each node runs the threshold decryption algorithm. As long as the number of nodes who complete the decryption exceeds the predetermined threshold, i.e., $f+1$, the total transaction set could be decrypted and the transaction confirmation is completed. At the same time, the transaction censorship attack \cite{CKP+01,MXC+16} is prevented since the transactions appear in the form of ciphertext when a leader proposes the set. For more details, please refer to \cite{MXC+16}.


There are some other distributed consensus algorithms under the asynchronous network model, such as  \cite{Rabin83}  \cite{Ben-Or83}, asynchronous binary Byzantine agreement \cite{CKS05}, MinBFT \cite{VCB+13}, validated asynchronous Byzantine agreement \cite{AMS18}, BEAT \cite{DRZ18}, and Dumbo-MVBA \cite{LLT+20,GLT+20}.
These algorithms mainly rely on reliable broadcast and binary Byzantine agreement to reach a consensus within the committee.

As far as we know, there is no sharding blockchain that uses a distributed consensus algorithm in an asynchronous network as the intra-committee consensus. 
The reason might be that the confirmation of cross-shard transactions relies on the liveness of the consensus algorithm in each shard. When a sharding blockchain processes cross-shard transactions, multiple shards need to cooperate and respond within a certain time, while asynchronous distributed consensus algorithms generally sacrifice liveness to ensure security when a network partition occurs.

\subsubsection{Weak Consistency}
Eventual sharding blockchains still use PoW, PoS, or other weak consistency methods to generate blocks in each shard. There is no committee in each shard. If PoW is used to generate blocks, nodes in a shard need to find the pre-image of the hash function that meets the specific requirements. If PoS is adopted, a node needs to check whether it becomes a block producer through some mechanisms like VRF according to the stake it holds. 

Monoxide \cite{WW19} proposes chu-ko-nu mining to realize intra-shard consensus with PoW. Chu-ko-nu mining is mainly designed to defend against the 1\% attack where an adversary focuses his computational power on one single shard. In Monoxide, miners are required to mine on multiple blockchains using one hash function. The $m$ block headers at the end of $m$ different blockchains form a Merkle tree, where the root value of the Merkle tree and a $nonce$ are used as inputs of the hash function. In this way, the 1\% attack could be prevented since the inputs of the hash function used in mining are related to $m$ blockchains and change in real-time. However, since the $m$ blockchains are constantly extended and new blocks are generated, miners need to collect and verify all newly generated blocks of $m$ blockchains continuously to ensure their mining inputs are valid. As a result, miners have to collect and verify all transactions in the network, so in essence, Monoxide does not achieve scalability \cite{AKW19}.

Parallel Chains \cite{FGK+18} combines VRF and PoS to generate blocks in each shard. In each round, each node might become the block producer of one or more shards out of the $m$ shards. Each node uses VRF to generate $m$ outputs and corresponding proofs. If an output is lower than the specific value determined by the protocol, then the node becomes the block producer of the shard corresponding to the output. At this time, the node packages the transactions belonging to the shard to generate a block and broadcasts the block with the output and proof of VRF on the entire network.

\subsection{Problems and Future Directions}
\label{subsec:isc_pp}
With the emergence and continuous development of the blockchain technology, the classic state machine replication algorithms have been continuously researched and improved, since it matches well in the context of blockchain. However, there are still some problems hindering its further application. We summarize these issues from the perspective of instant sharding blockchains and eventual sharding blockchains as follows, and point out future research directions.

\subsubsection{Instant Sharding Blockchains}
The possible problems and research directions of instant sharding blockchains mainly include the following points.

\begin{itemize}[leftmargin=*]
	\item \textit{Reducing the communication complexity among shard members.} The distributed consensus algorithms inside a shard generally rely on multiple voting rounds to reach an agreement. 
	In the voting process, if each member broadcasts its own signature, collects and verifies the signatures of all other members, when the number of members increases, the broadcast and collection of signatures will cost huge communication bandwidth and time, thereby reducing the efficiency of the entire protocol.
	In particular, in a permissionless sharding blockchain, in order to ensure that there are sufficient honest nodes in a shard, the number of shard members needs to reach a certain security threshold. In this case, how to reduce the communication complexity inside a shard and improve the processing efficiency is a problem that needs to be solved.
	
	\item \textit{Malicious committee detection and recovery.}
	In an instant sharding blockchain, multiple committees run distributed consensus algorithms. Although most sharding blockchains are designed with a series of assumptions and analyses to ensure that each committee is honest with a high probability, it is still possible for a certain committee to be controlled by an adversary. In this case, how to detect malicious committees through other honest committees and design a specific mechanism to restore or replace malicious committees with honest committees is one of the future research directions.

	\item \textit{Efficient view-change mechanism inside a committee.}
	When a view-change occurs in a shard, a new leader is required to replace the old leader, and the views of all members in the committee are unified. In this process, how to reduce the communication complexity among members is a problem that needs to be solved. In addition, in a sharding blockchain, a leader not only serves as the message center inside a shard but also as a coordinator to transfer messages among different shards. Consequently, the leader has a higher overhead. How to share the burden of the leader and how to choose a new leader fairly and reasonably is one of the future research directions.
	
	\item \textit{Better combination with sharding blockchains.}
	In sharding blockchains, the intra-shard consensus is not just used to process transactions. In most cases, intra-shard consensus needs to handle cross-shard transactions. The processing of cross-shard transactions requires multiple shards to run multiple rounds of intra-shard consensus, and the target of the consensus is not just a simple transaction but might be an input of a transaction, to determine whether an input is available. In this way, there may be multiple types of inputs for the intra-shard consensus, and the rules for judging whether the inputs of different types are valid are also different. Therefore, the details of combining intra-shard consensus with the entire sharding blockchain protocol need to be designed more specifically, so as to process transactions in sharding blockchains more efficiently.
\end{itemize}

\subsubsection{Eventual Sharding Blockchains}
In eventual sharding blockchains, there are mainly two potential problems that need to be handled. 

\begin{itemize}[leftmargin=*]
	\item \textit{The 1\% attack problem.} 
	Taking PoW-based sharding blockchains as an example, an adversary could focus his computational power on a single shard to mine. If the mining process of each blockchain in each shard is independent, then for each blockchain, it is usually required that the computational power controlled by an adversary cannot exceed 51\% (if not consider selfish mining attacks). Assume that there are $m$ shards, that is, $m$ parallel blockchains, if the adversary concentrates its computational power on one of the shards, he only needs $51\%/m$ of the total computational power to fully control the blockchain.
	When $m$ is large enough, $51\%/m$ is approximately equal to $1\%$, so an adversary only needs $1\%$ of the computational power to control the shard.
	In this case, the blocks in this shard will all be generated by the adversary, which means that the adversary has full control over the current shard. The adversary could launch the double-spending attack, thereby destroying the security of the entire system. Therefore, in eventual sharding blockchains, how to prevent the 1\% attack while ensuring the computing, storage, and communication sharding (described in Section~\ref{subsec:background}) simultaneously is one of the future research directions.
	
	\item \textit{Complicated cross-shard transaction processing.}
	Due to the weak consistency property of eventual sharding blockchains, the blocks generated in a shard could not be confirmed as stable instantly. In general, a block has to reach a certain depth to be considered stable and valid. In other words, a certain number of blocks at the end of a blockchain must be truncated to confirm valid transactions. The number of blocks that is removed is related to the system security parameter such as $6$ in Bitcoin. As a result, cross-shard transaction processing is difficult in eventual sharding blockchains. We discuss this in detail in Section~\ref{sec:cross-shard}.
\end{itemize}

\section{Cross-Shard Transaction Processing}
\label{sec:cross-shard}
In this section, we analyze cross-shard transaction processing in sharding blockchains. In Section~\ref{subsec:cstp_bp}, basic concepts are given. In Section~\ref{subsec:cstp_es}, we classify existing approaches regarding cross-shard transaction processing into three categories, namely, two-phase commit (2PC) based, transaction split based, and relay transaction based solutions. For each type of processing method, its basic process is summarized and typical schemes are discussed. Section~\ref{subsec:cstp_pp} provides potential problems and future research directions. 

\subsection{Basic Concepts}
\label{subsec:cstp_bp}

In sharding blockchains, the probability for a transaction to be a cross-shard one is extremely high. The probability increases as the number of shards grows. As computed in RapidChain \cite{ZMR18}, when the number of shards is $16$, the proportion of cross-shard transactions is about $99.98\%$. So the method to process cross-shard transactions is of vital importance to the performance of a sharding blockchain system. 

In the process of cross-shard transaction processing, two problems need to be solved. 
First, the communication and processing methods among multiple shards need to be carefully designed. A transaction usually contains multiple inputs, which might be controlled by different shards. To confirm whether such a transaction is valid, multiple shards are required to cooperate and complete it together. If all inputs of the transaction are ready to be spent, and the sum of the transaction input value equals to that of the output value, then such a transaction is regarded as valid. On the contrary if one of the inputs has already been spent or does not exist, then the transaction is an invalid one and cannot be committed. 
Second, a mechanism is in need to prevent the double-spending attack, i.e., to prevent an input from being spent multiple times by different transactions. In a sharding blockchain, since different inputs are controlled by different shards, the double-spending attacks are different from those against ordinary blockchains, so special mechanisms need to be designed according to the actual situation.

\subsection{Existing Approaches}
\label{subsec:cstp_es}
The approaches to process cross-shard transactions could be divided according to whether there is a committee in each shard, i.e., whether the blockchain is an instant sharding blockchain or an eventual one. In instant sharding blockchains, the most common methods to process cross-shard transactions are two-phase commit based and transaction split solutions. Relay transaction based solutions are designed for eventual sharding blockchains.

\subsubsection{Two-Phase Commit Based Approaches}
Most cross-shard transaction processing approaches are designed on top of the two-phase commit protocol which contains a prepare phase and a commit phase. In a 2PC protocol, there is a coordinator who is responsible for collecting availability certificates of inputs and transmitting them among the related participating shards. 
To formalize the process of 2PC based solutions, we give the definition of an availability certificate. 

\begin{definition}[Availability Certificate]
	\label{def:ac}
	An availability certificate refers to a proof in the prepare phase of 2PC based cross-shard transaction processing methods that each shard provides for a transaction input, to prove that the input is available or unavailable.
\end{definition}

In the prepare phase, a coordinator collects certificates to prove that the inputs of a transaction are available from different shards. Such a proof is usually generated by shard members through running BFT to reach an agreement on if the inputs are available. So the proof might be some signatures or a single aggregated signature. Meanwhile, the inputs should be locked to prevent themselves from being spent by other transactions. 

Then in the commit phase, the coordinator sends all availability certificates of inputs to all related shards, including input and output shards. If all inputs are available, then the transaction is regarded as valid and committed in related shards. The inputs should be spent and outputs should be created. Else, if at least one input is unavailable (locked or already spent), the transaction is invalid. The previously locked inputs should be unlocked. 

According to the role of a coordinator, we divide current cross-shard processing methods into client-driven basic 2PC and shard-driven basic 2PC.

\begin{figure}[h!]
	\centering
	\includegraphics[width=0.6\textwidth]{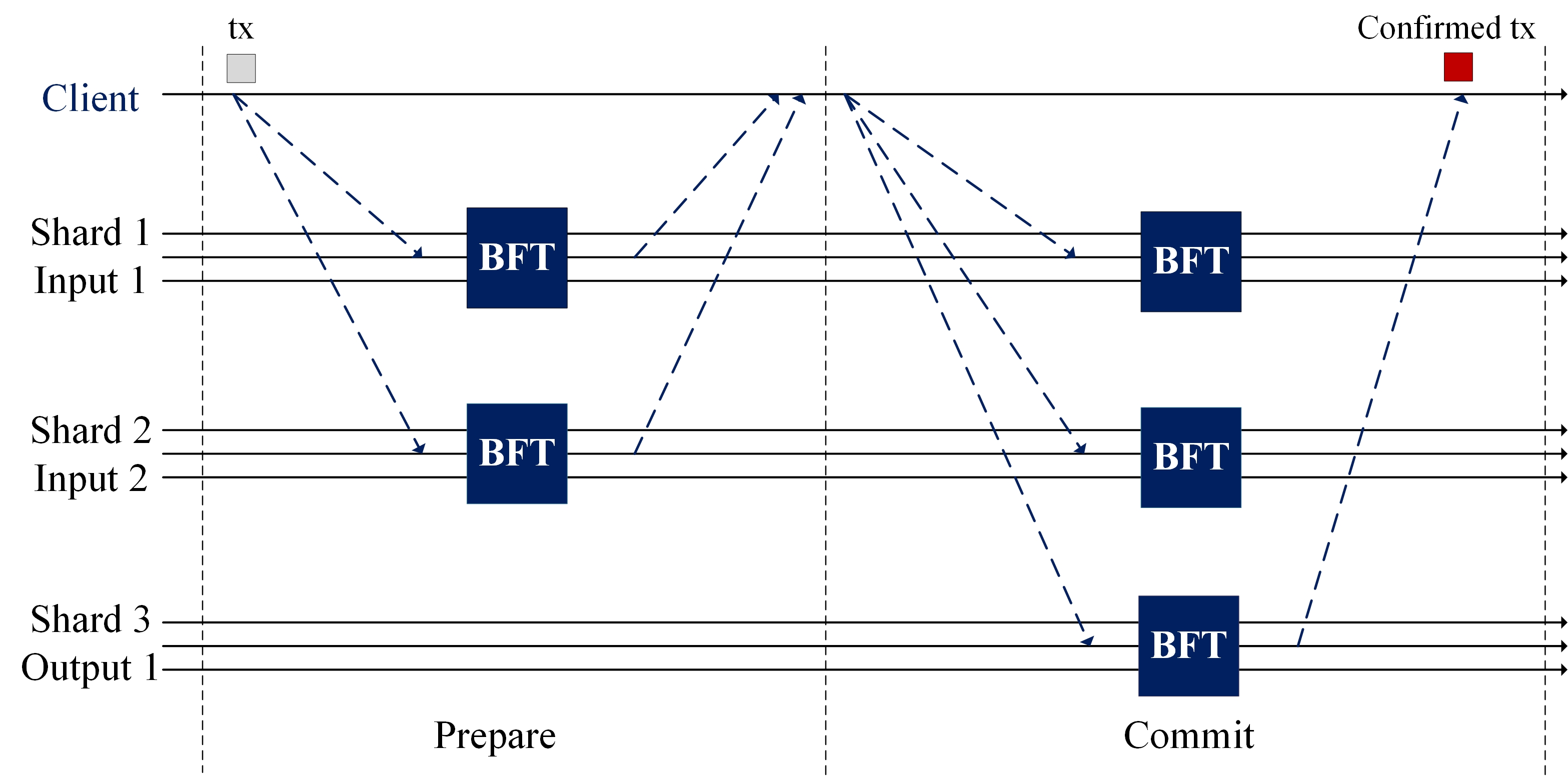}
	\caption{The flowchart of client-driven basic 2PC.}
	\label{fig:client_driven}
\end{figure}

\paragraph{Client-driven 2PC}
The basic procedure is shown in Fig.~\ref{fig:client_driven}. A client is responsible for collecting proofs in the prepare phase and transmitting them to related shards in the commit phase. 

Omniledger \cite{KJG+18} adopts the client-driven 2PC methods to process cross-shard transactions. In the prepare phase, an availability certificate is named as a proof-of-acceptance or a proof-of-rejection. A proof-of-acceptance is generated by a leader of an input shard committee, i.e., a leader signs on an input, to prove that the input is ready to be spent. 

\paragraph{Shard-driven 2PC}

In a shard-driven 2PC protocol, one or more shards play the role of coordinators. 
In the prepare phase, the availability certificates of inputs are generated by input shards. In the commit phase, a valid transaction will be accepted in all input and output shards. The confirmation information of the transaction will be sent to the client. Compared with client-driven 2PC, the burden on a client is released. The client just submits a transaction and waits for the response.

\begin{figure*}[h!]
	\centering
	\subfigure[All input shards as the coordinators.]{
		\centering
		\includegraphics[width=.46\linewidth]{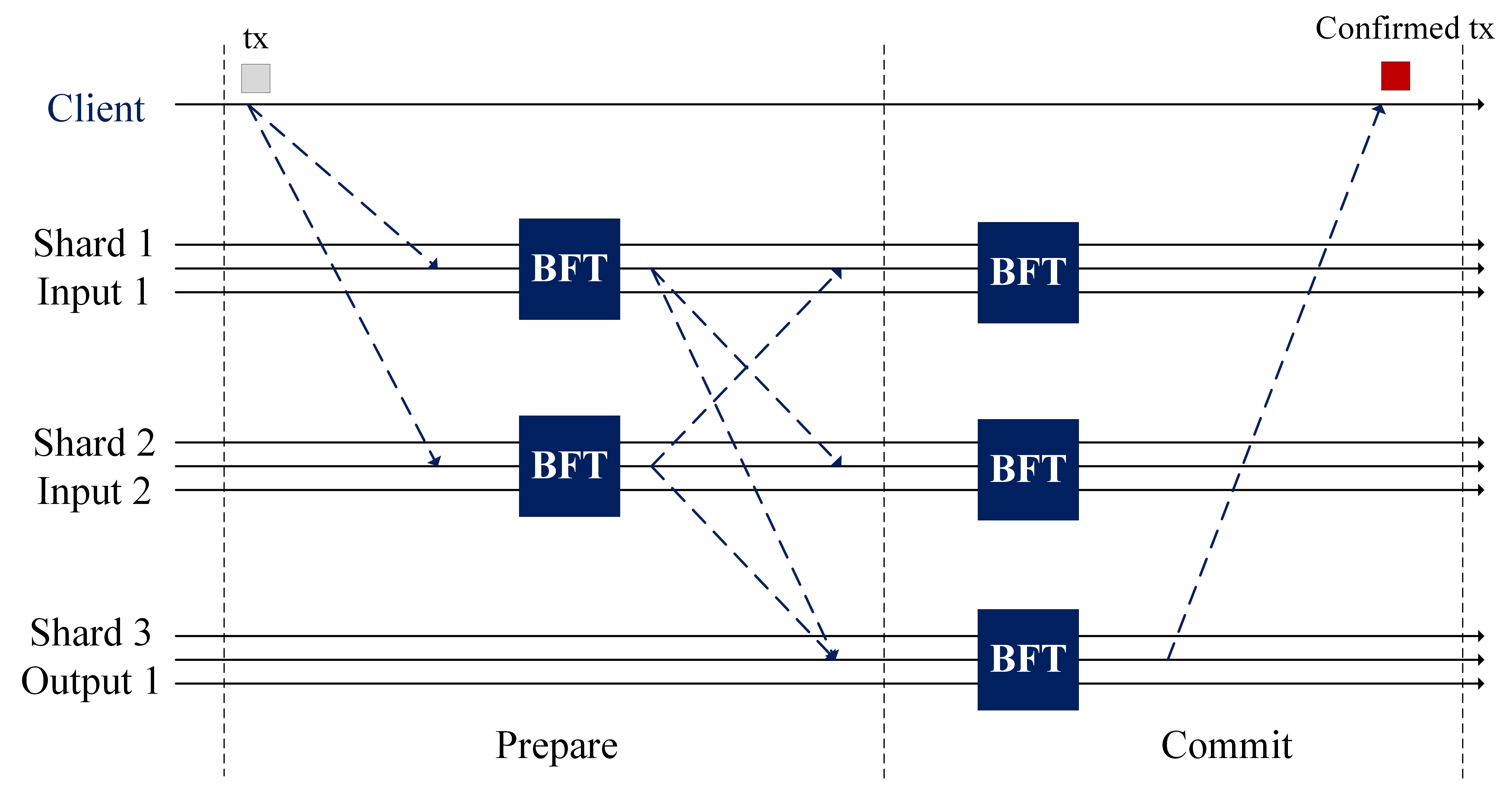}
		\label{fig:shard_driven}
	}
	\subfigure[An output shard as the coordinator.]{
		\centering
		\includegraphics[width=.46\linewidth]{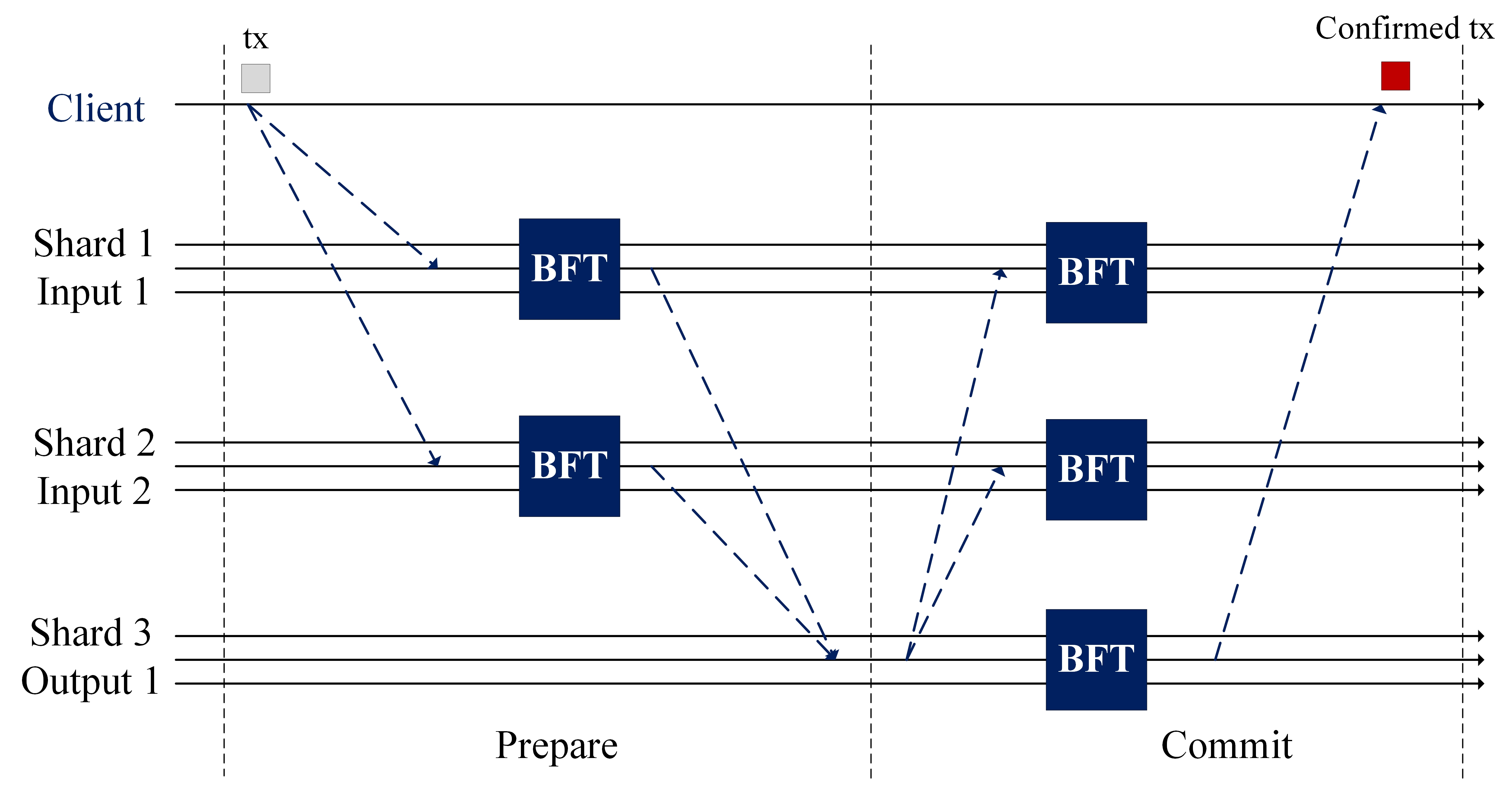}
		\label{fig:shard_driven-2}
	}
	\caption{The flowchart of shard-driven basic 2PC.}
	\label{fig:shard_driven_all}
\end{figure*}

The flowchart is shown in Fig.~\ref{fig:shard_driven_all}. Note that in Fig.~\ref{fig:shard_driven}, all input shards act as the coordinators, which means the availability certificates are transferred by the input shard directly. In Fig.~\ref{fig:shard_driven-2}, an output shard plays the role of a coordinator, collects the availability certificates, and forwards them to relative shards. We argue that in normal cases, the communication complexity of the two methods above is identical since the messages are simply collected and forwarded, without being aggregated, so the total number of messages remains unchanged.

Chainspace \cite{ASB+18}, RSTBP \cite{LLY+20} and FleetChain \cite{LLL+20} adopt the shard-driven 2PC methods to handle cross-shard transactions. The coordinators in Chainspace are the input shards. The availability certificates are produced by input shard committees running a BFT algorithm and broadcast among all participating shards, including input and output shards. 

\subsubsection{Transaction Split Based Approaches}
The transaction split based approach is proposed in RapidChain \cite{ZMR18} to handle cross-shard transactions by splitting a multi-input multi-output transaction into multiple single-input single-output transactions. A simple example is shown in Fig.~\ref{fig:transaction_split}.

For a transaction $\tx$ with two inputs $I_1$ and $I_2$ of shard $C_{in1}$ and $C_{in2}$, respectively, and one output $O$ belonging to shard $C_{out}$, a client submits $\tx$ directly to the committee members of $C_{out}$. Then $C_{out}$ splits $\tx$ into three following transactions, $\tx_1: I_1 \rightarrow I'_1$, $\tx_2: I_2 \rightarrow I'_2$ and $\tx_3: I'_1 + I'_2 \rightarrow O$ where $I'_1$ and $I'_2$ belong to $C_{out}$. If $\tx_1$ and $\tx_2$ are committed by $C_{in1}$ and $C_{in2}$ respectively, then $\tx_3$ will be executed successfully by $C_{out}$ and the cross-shard fund transfer process is completed. If there are more inputs for a cross-shard transaction, then each input corresponds to a new transaction.  

The transaction split based approach simplifies the processing of cross-shard transactions to some extent, while it introduces many new problems, which we will analyze in detail in the next section.

\subsubsection{Relay Transaction Based Approaches}
Relay transaction based approaches are usually adopted in eventual sharding blockchains. 
In eventual sharding blockchains, there is no BFT running in each committee, so transactions or availability certificates will not be confirmed immediately. As a result, 2PC based cross-shard transaction processing methods could not be used. 

In eventual sharding blockchains, a transaction is considered as committed after its block reaches a certain depth of the blockchain such as $6$ blocks in Bitcoin. The number of blocks (usually denoted by $\lambda$) is usually related to system security.
Consequently, in eventual sharding blockchains, the essence of processing cross-shard transactions is to ensure that the output shard does not treat the transaction as valid until the transaction in all related input shards is completely committed, i.e., reaches sufficient depth. 

The basic steps of relay transaction based solutions are as follows. First, the miners of the input shard collect the cross-shard transaction $\tx$, that is, an account $A$ wants to pay an account $B$ $y$ amount of fund. A miner verifies whether the balance of account $A$ is greater than $y$, and if it is, $\tx$ is regarded as a legal transaction. Second, a miner finds a PoW solution, constructs a block containing $\tx$, and broadcasts the block. Other miners verify the block, generate a corresponding relay transaction $\psi$, and send it to the miners of the corresponding output shard. Third, the miners of the output shard receive the relay transaction and verify the legitimacy of it, that is, $\psi$ has reached the block depth of $\lambda$ in the input shard. Fourth, a miner of the output shard finds a PoW solution, adds valid relay transactions including $\psi$ to the block, and broadcasts it. When the block containing the relay transaction $\psi$ reaches a $\lambda$ depth in the output shard, which is called a $\lambda$-confirmation, it is determined that the $y$ amount of money could be spent by the account $B$.

Monoxide \cite{WW19} adopts the relay transaction based solution to process cross-shard transactions. Meanwhile, the account model is utilized in their system. RChain \cite{CDE17} also utilizes a similar method to transfer value across shards. Buterin \cite{Buterin18} also proposes a cross-shard contract yanking method to deal with contracts that are related to multiple shards. The essence of the method is to use the relay transaction. 
Ostraka \cite{MME19} adopts a node differential environment model, i.e., each node's ability to process transactions is different. 

\subsection{Problems and Future Directions}
\label{subsec:cstp_pp}
Next, we analyze the possible problems of each processing approach for cross-shard transactions and point out future research directions.

\subsubsection{Two-Phase Commit Based Approaches}
We analyze potential problems in 2PC based approaches from the following two aspects, i.e., client-driven 2PC and shard-driven 2PC. 

\paragraph{Client-driven 2PC} Possible problems in client-driven 2PC based approaches are as follows.

\begin{itemize}[leftmargin=*]
	\item \textit{Malicious behaviors of the leader.} In the prepare phase, if an availability certificate, i.e., a proof of an input, is generated by a single leader like in Omniledger \cite{KJG+18} instead of a committee running BFT, then a malicious leader might provide false proofs or fail to respond. If an input is not available while a malicious leader still insists on signing its legality and providing proof-of-acceptance, other shards are likely to accept this proof. In this case, a double-spending attack is likely to be successful. The consistency and security of the entire blockchain protocol will be severely damaged. 
	
	\item \textit{Transaction input being locked.} Letting a client act as a coordinator could lead to an input being locked permanently if the client fails to send the corresponding proof to related input shards. This happens when a client is malicious or offline. Note that in a blockchain system, for a single transaction, its inputs are usually owned by the same entity. However, in some cases such as crowdfunding transactions, multiple inputs of a transaction might belong to different individuals. In this way, if a client does not provide the corresponding availability certificate, it could cause the transaction input to be locked and affect the liveness property of the system.
	
	\item  \textit{Increased burden on the client.} If a client is responsible for collecting and forwarding availability certificates, the client needs to record the shard state in the network and the IP addresses of participating nodes to communicate with the corresponding committee leader. Hence, the storage and communication overhead of a client is increased a lot, which is not desired. Especially for lightweight clients, such as those installed on mobile phones and smart homes, this overhead is impractical.
\end{itemize}

Therefore, client-driven 2PC approaches might have certain problems and are not widely used.

\paragraph{Shard-driven 2PC}
In shard-driven 2PC based existing approaches, there are also problems to be solved.

\begin{itemize}[leftmargin=*]
	\item \textit{Multiple calls of the BFT algorithm.} In most shard-driven 2PC approaches, to commit a single transaction, input and output shard committees need to run multiple times of BFT algorithm. For example, in Chainspace \cite{ASB+18}, in the prepare phase, every input shard needs to run one BFT algorithm, and in the commit phase, every input and output shard also have to invoke the BFT algorithm. Each BFT call requires communication among the members of the entire committee, which will cause more communication and computation overhead to each node. 
	So how to design a method to process multiple transactions in a single time is one of the future research directions.
	
	\item \textit{Possible attacks.} Shard-driven 2PC approaches might suffer from different types of attacks. For example, the replay attack is proposed in \cite{SBA+19} against cross-shard transactions. The essence of the attack is to use the previously generated transaction input availability certificate to disguise it as a proof of other transaction input. The way to prevent this kind of attack is simple: attach the relevant transaction ID information to the input availability certificate.
	Another type of attack is the transaction flooding attack where an adversary might manufacture a large number of transactions processed by a certain shard to cause the system liveness to be broken. Nguyen \textit{et al.} \cite{NND+19} propose OptChain which utilizes a novel transaction assignment method to distribute transactions to different shards. The method could be applied to common sharding blockchains to realize better performance. 
	In summary, many cross-shard transaction processing methods might be at risk of being attacked. In the future, it is necessary to analyze other possible attacks and design more secure processing methods.
	
	\item \textit{Malicious coordinators.} In 2PC based cross-shard transaction processing approaches, the role of the coordinator is very important. In shard-driven 2PC based solutions, the coordinator is generally a leader of the input or output shard. If a coordinator is malicious, it may delay the collection and forwarding process of the availability certificates, or even censor some of them. How to prevent the possible malicious behaviors of a coordinator is one of the directions of future research.
\end{itemize}

\subsubsection{Transaction Split Based Approaches}

\begin{figure}[h!]
	\centering
	\includegraphics[width=0.6\textwidth]{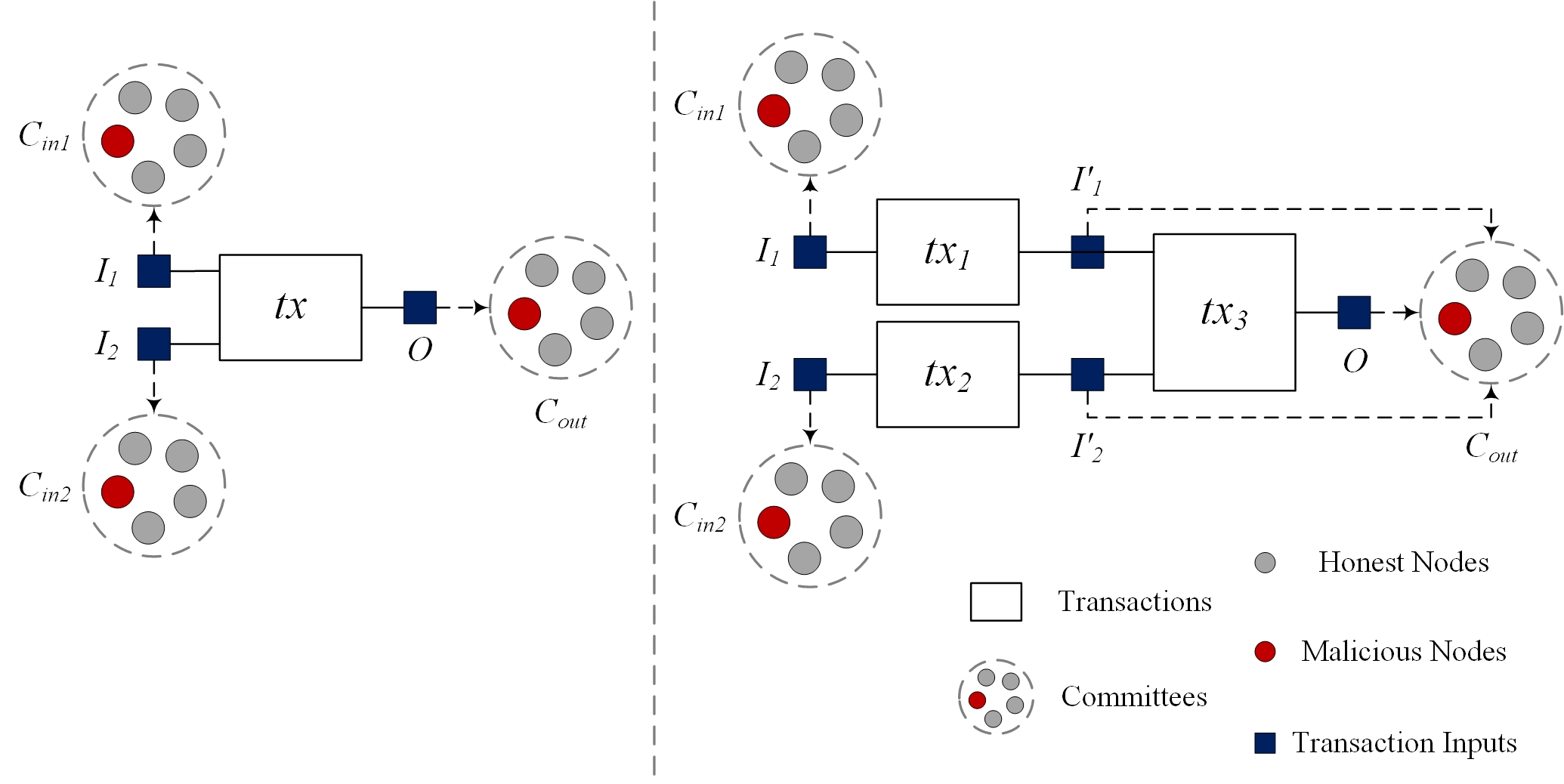}
	\caption{Transaction split based approaches.}
	\label{fig:transaction_split}
\end{figure}

In transaction split based approach shown in Fig.~\ref{fig:transaction_split}, e.g., RapidChain \cite{ZMR18}, there might exist several possible problems. 

\begin{itemize}[leftmargin=*]
	\item \textit{The way to generate a specific public key managed by $C_{out}$ is unclear.} The rule determining which shard is responsible for managing an input is not given. In the general case, it is based on public keys. In other words, the hash value of the last few digits of a public key address of a transaction input determines which shard should manage the input. However, in the process of transaction split, it is necessary to generate two or more public key addresses that belong to $C_{out}$, that is, $I'_1$ of $\tx_1$ and $I'_2$ of $\tx_2$ in the above instance. The method to generate such a public key address that satisfies the special requirement is not obvious, and the details of how to implement this process are not given.
	
	\item \textit{The new generated outputs might be illegally spent.} $I'_1$ and $I'_2$ are supposed to be generated by $C_{out}$, while the specific generation process is not given, that is, whether a leader of $C_{out}$ is responsible for generating $I'_1$ and $I'_2$, or whether it is generated by the entire committee $C_{out}$  running BFT. As we know, $I'_1$ and $I'_2$ are public key addresses, and behind them, there are corresponding private keys. How to ensure the private keys are only known to the owners is not analyzed. If a committee member or the leader of $C_{out}$ knows the private keys, $I'_1$ and $I'_2$ could be illegally spent. Also, if one of $\tx_1$ and $\tx_2$ is illegal, for example, $\tx_1$ successes and $\tx_2$ fails due to the unavailability of $I_2$, then the way to retrieve the private key of $I'_1$ for the client is not given. 
	
	\item \textit{The method to handle multi-output transactions is not given.} In practical applications, some transactions may include multiple outputs. For example, the outputs of a transaction may include change returned to the payer. In this case, the method for splitting a multi-input multi-output cross-shard transaction into multiple single-input single-output transactions is not given, and the whole process will be very complicated.

	\item \textit{The number of transactions is increased.} Transaction split leads to an increase in the number of transactions (at least three times as much as before), so the processing and storage overhead of the entire system is largely increased. 
\end{itemize}

In summary, the specific implementation details described above are not given, yet are critical to the system security. In order for the transaction split based solutions to be more widely and securely applied, the implementation details need to be further studied and designed.

\subsubsection{Relay Transaction Based Approaches}
There might exist some problems in relay transaction based approaches. 

\begin{itemize}[leftmargin=*]
	\item \textit{Transaction confirmation delay is long.} For a cross-shard transaction, it first needs to get $\lambda$-confirmation by the input shard. Then the corresponding relay transaction is required to obtain $\lambda$-confirmation by the output shard. After this, the cross-shard transaction is regarded as completed, then account $B$ could truly own the money and spend it. Therefore, the entire confirmation delay of a cross-shard transaction is equal to the time that at least $2\lambda$ blocks are committed.
	
	\item \textit{Multi-input transactions could not be easily processed by relay transaction based approaches.} Monoxide \cite{WW19} uses an account-based transaction model. Its cross-shard transaction processing method is actually a simplified version, which can only handle single-input single-output cross-shard transactions. In fact, whether it is an account-based model or a UTXO-based model, there might be multi-input cross-shard transactions. Taking the account-based model as an example, a user might have multiple different accounts, which are controlled by different shards. When the user initiates a transaction, he might use multiple accounts as inputs to complete a single transaction. This cannot be achieved in Monoxide \cite{WW19}. To process multi-input cross-shard transactions in an eventual sharding blockchain, a two-phase commit mechanism should be introduced, since some inputs might not be available. In an eventual sharding blockchain, the confirmation of transactions and availability certificates is not instant, and it is difficult to employ a lock-unlock mechanism for transaction inputs. In this way, an integral cross-shard transaction processing method will become extremely complicated in an eventual sharding blockchain, which is an open research direction in the future. 
	
	
	\item \textit{Once a fork appears in a blockchain, the security of the system will be severely damaged.} Eventual sharding blockchains are different from instant sharding blockchains. In each shard, the block confirmation is probabilistic. Even if a block gets $\lambda$-confirmation, the blockchain where the block is located still has a certain probability of being replaced by another blockchain, that is, a fork occurs. In this case, the cross-shard transactions that are considered valid previously might become invalid, while the output of the cross-shard transaction may have been spent by the owner. In this case, it is very difficult to roll back the system, and the security of the system is seriously damaged.
\end{itemize}

\section{Shard Reconfiguration}
\label{sec:shard_reconfiguration}
In this section, we introduce methods to conduct shard reconfiguration. In Section~\ref{subsec:sr_bp}, the reason, purpose, and basic steps of shard reconfiguration are explained. Section~\ref{subsec:sr_es} provides a taxonomy of existing approaches for shard reconfiguration. All approaches are divided into reconfiguration through random replacement and under specific rules. Basic steps as well as typical approachesd for each category are given. Section~\ref{subsec:sr_pp} points out the possible security and efficiency problems with regard to shard reconfiguration. 

\subsection{Basic Concepts}
\label{subsec:sr_bp}
We first explain the reason why a sharding blockchain needs to be reconfigured. Most blockchain systems require regular reconfiguration, including the update of shard members and state transmission between old and new shard members. This is because an adversary could launch a corruption attack by controlling a certain node after a certain period of time. If the members of a committee remain constant, the proportion of committee members controlled by an adversary may exceed a pre-defined safety threshold such as $1/3$ in PBFT \cite{CL99}. This will destroy the liveness and security of the entire system. Therefore, at regular intervals, the old committee members need to be replaced by new members to ensure that the number of nodes controlled by an adversary is always below the safety threshold.

There are three key problems to be considered in shard reconfiguration. First, it is necessary to ensure that the number of honest nodes in each committee in the new epoch after reconfiguration exceeds the safety threshold. Second, the system should be able to process transactions normally during reconfiguration. Third, the  corruption attacks will not succeed before the reconfiguration is completed.

The basic procedures of shard reconfiguration are as follows.
First, in some epoch $e$, the nodes that intend to participate in epoch $e+1$ are confirmed through some node selection method. Second, at the end of epoch $e$, the replacement arrangement of the old and new committee members are determined, that is, which part of the old members in each shard is replaced by which new nodes. Third, the new members of each shard communicate with the corresponding old members and get the shard UTXO data and historical transaction data. Fourth, the old members stop working, the new members start to process transactions as normal, and the whole protocol enters into epoch $e+1$.

Note that in most existing reconfiguration schemes, only some of the members in a committee are replaced. The reason is that the more nodes are substituted, the more time it takes for the new nodes to get the historical data, which might affect the efficiency of transaction processing. However, the liveness property of the whole system might be broken during the reconfiguration process.
\subsection{Existing Approaches}
\label{subsec:sr_es}
We divide existing approaches into reconfiguration through random replacement and under specific rules.

\subsubsection{Reconfiguration through Random Replacement}
Reconfiguration through random replacement means the selection of old members is a random procedure, i.e., each old member has an identical probability to be replaced. In general, reconfiguration through random replacement needs the following steps. 

First, in epoch $e$, a parameter $k$ that represents the number of members to be replaced is determined by the protocol. Then, after the node selection and allocation process, the new node list $\newnodes$ is confirmed. Third, for each shard, a seed $seed_{c}=H(c||\xi_e)$ is derived, where $c$ is the shard sequence number and $\xi_e$ is a randomness. Then for every shard, its seed is used as one of the inputs of a pseudorandomness generator \cite{Luby96} function $Perm(seed_c,n)$, to generate a permutation $\pi_{c}$. $\pi_{c}$ could be used as an indicator to select $k$ nodes out of $n$ old shard members. In this way, the $k$ old members to be replaced in each shard are selected. The new nodes could be assigned to $m$ shards in a similar way. A seed $seed=H(0||\xi_e)$ is first generated and a permutation $\pi_0$ could be obtained by computing $Perm(seed, mk)$ since there are $mk$ new nodes in the list $\newnodes$. At this time, the new committees for epoch $e+1$ are determined. Then newly joined members start to download historical UTXO and transaction data. Finally, new committees begin to work as normal and the protocol enters into epoch $e+1$.

Omniledger \cite{KJG+18} utilizes reconfiguration through random replacement where the reconfiguration parameter $k$ is set as $\log \frac{n}{m}$. SGX-Sharding \cite{DDC+18} also uses the same reconfiguration rule and parameter. The above two schemes rely on the epoch randomness generated by their protocols as the random seed to complete the random replacement process.

\subsubsection{Reconfiguration Under Specific Rules}
The update procedure of committee members could rely on other specific rules. We summarize these rules into the following two categories.

\paragraph{Chronological rule} This rule means that the node replacement is based on a chronological order, which is similar to the sliding window rule. Specifically, a new node replaces an old node that has been in the committee for the longest time. In some protocols such as Solida \cite{AMN+17}, every time a node is added, the committee performs a reconfiguration, and the newest node will replace the oldest one. Similarly, committee members could be divided into several parts according to the time when they join the committee. Each time a reconfiguration takes place, new nodes that find PoW solutions successfully replace the oldest $1/2$ (or $1/k$) of the existing committee members. This replacement rule is adopted in PaLa \cite{CPS18}, Byzcoin \cite{KJG+16}, etc.

\begin{figure}[h!]
	\centering
	\includegraphics[width=0.6\textwidth]{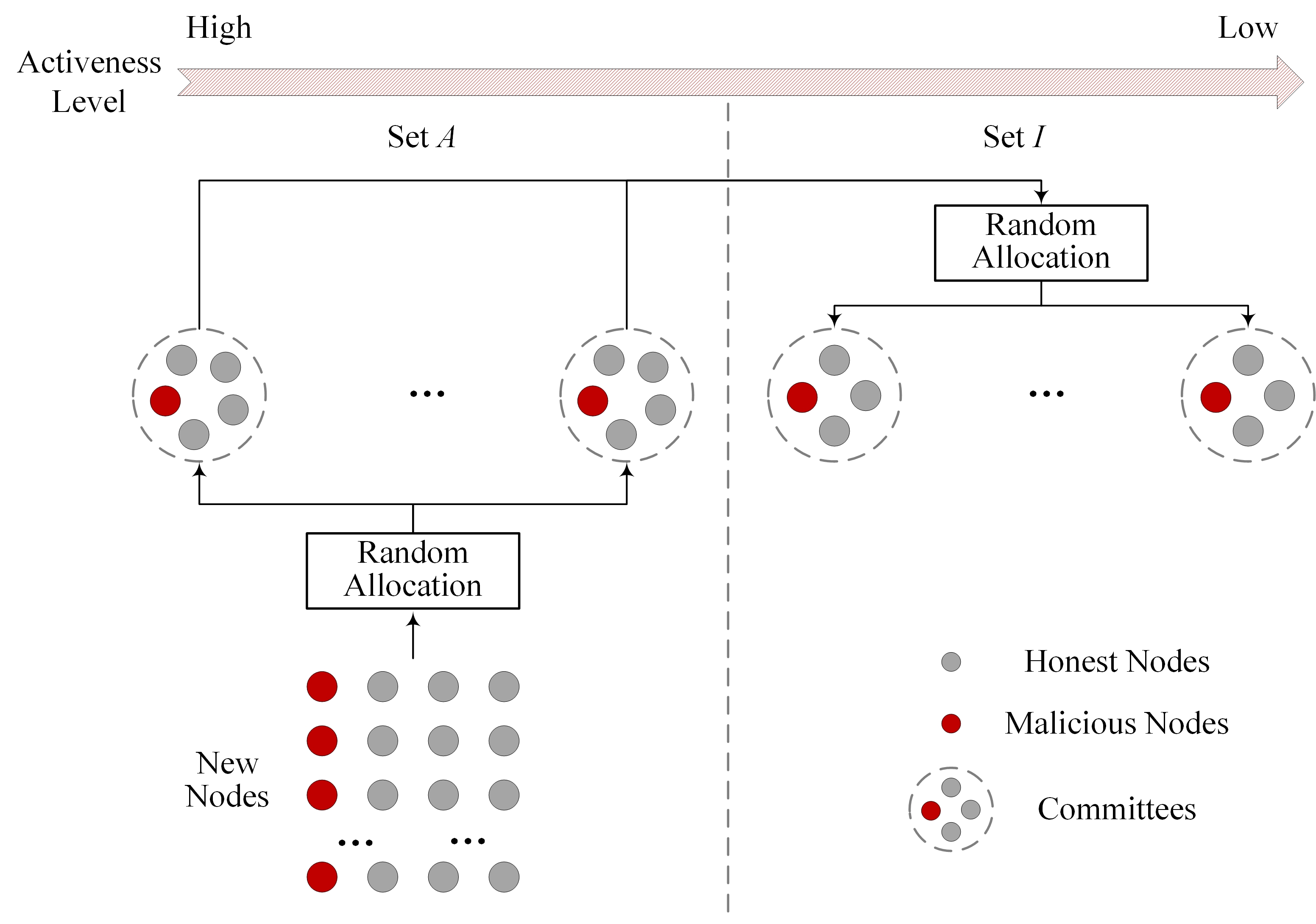}
	\caption{Reconfiguration under bounded Cuckoo rule.}
	\label{fig:reconfiguration_cuckoo}
\end{figure}

\paragraph{Bounded cuckoo rule} 
Reconfiguration under the bounded Cuckoo rule is shown in Fig.~\ref{fig:reconfiguration_cuckoo}. This rule dynamically adjusts the members of each committee based on the number of active members in each committee. At the end of epoch $e$, all committees are sorted according to the activeness level of all nodes, that is, the total number of transactions processed during the epoch. The $1/2$ committees with the highest activeness are put into a set $A$ (for ``active''), and the remaining committees are placed into a set $I$ (for ``inactive''). Then, the new nodes confirmed by the node selection mechanism are randomly assigned to a committee in the set $A$ according to the epoch randomness $\xi_e$. After all the new nodes are allocated, $k$ nodes are selected from each committee of set $A$, and these nodes will be kicked out of these committees and randomly assigned to committees in the set $I$. The reconfiguration procedure is done at this moment, and newly confirmed committees start to process transactions normally. RapidChain \cite{ZMR18} utilizes the bounded Cuckoo rule to complete committee reconfiguration.

\subsection{Problems and Future Directions}
\label{subsec:sr_pp}
We summarize potential problems that might occur in the existing approaches during the shard reconfiguration process. 

\subsubsection{Quantitative Analysis of the Corruption Parameter $\tau$}
$\tau$ is used to denote the time to complete a corruption process for an adversary, which is a crucial parameter in shard reconfiguration determining the system security. If an adversary could complete its corruption attack before the committees complete their reconfiguration, then the adversary might control the committee and destroy the system security. Generally speaking, the following condition applies for sharding blockchains that use PoW to select shard members: 
\begin{equation*}
	\tau > 2T_{epoch}
\end{equation*}
where $T_{epoch}$ denotes the time of an epoch. In sharding blockchains that use PoW to select members, nodes that want to participate in the protocol mine in some epoch $e$. Then nodes finding PoW solutions should submit their results to the reference committee or broadcast the results. So in the mining process in epoch $e$, the identities of new nodes are exposed to an adversary who can launch a corruption attack from this time on. When the protocol enters into epoch $e+1$, new nodes become committee members and work normally. So it is required that an adversary could not complete its corruption attack until epoch $e+1$ ends. Therefore, the total corruption parameter for an adversary should satisfy $\tau > 2T_{epoch}$. $T_{epoch}$ is related to the speed to find a certain number of PoW solutions and may vary in different sharding blockchains.
In existing sharding blockchain schemes, there is no discussion about the corruption time parameter. The analysis of it is one of the future research directions.

\subsubsection{Bootstrapping of New Joined Members} 
A new shard member should download historical data of the corresponding shard \cite{LSG+19}. This procedure is usually called bootstrapping in some literature. During the bootstrapping, there are two potential problems for new shard members. One is the security problem, where an adversary might produce some fake data to convince a new member. This kind of attack is easy to launch in a PoS-based blockchain since the cost of forging a fake chain is relatively low. The other problem is about the performance, that is, the downloading of a large amount of data takes a long time, making the system not work properly during reconfiguration. 
So how to ensure that honest committee members could confirm the correctness of the data received, and improve the efficiency of the entire reconfiguration process is worthy of in-depth study in the future. 

\subsubsection{Security Analysis of New Committees}
In reconfiguration solutions through random replacement, only a fraction of old members is replaced by new nodes. However, the analysis of random assignment in existing schemes \cite{KJG+18} assumes replacing all committee members instead of a certain proportion of members, so the analysis results are not accurate. Every time some members are replaced, it should be guaranteed that the honest members occupy more than a certain proportion of the entire committee. 
The security analysis of the reconfiguration process needs to be more rigorous, taking all shards into account.

In reconfiguration solutions under the chronological rule, an adversary could launch targeted corruption attacks on relatively new nodes. In this way, an adversary with an identical corruption attack ability could have a higher probability of controlling the committee. In other words, a stricter requirement is in need of an adversary's corruption parameter, and the security level of the system will become lower. 

As for the bounded Cuckoo rule, the committee members in set $I$ are constantly increasing, and there is no explanation of how to kick out old members in $I$. The increase of committee members will lead to excessive system overhead of the BFT algorithm. Furthermore, the scenario is not practical in general sharding blockchains. The reason is that, in order to prevent the transaction flood attack, the transactions in a sharding blockchain are randomly allocated to different shards. The transaction flood attack means that an adversary creates a large number of transactions supposed to be managed by a particular shard.
Therefore, the actual number of transactions processed by each shard is similar, and there is no obvious gap to distinguish them.

\subsubsection{Initial Setup of the Protocol}
In the initial phase of a sharding blockchain protocol, the method to safely initialize and create multiple genesis blocks should be designed. This process is usually called the protocol setup.
The protocol setup problem also exists in other general blockchains. Most existing blockchain protocols assume a trusted setup, where information such as the genesis block is set by a trusted third party. 
In sharding blockchains, the protocol setup is different from that of general blockchains since it is necessary to set the genesis block and genesis committee members for each shard. The details of these settings deserve a more in-depth study.
Besides, the way to initialize the protocol without relying on trusted setup \cite{AD15,GKL+18} such as using secure multi-party computing should be studied in deep.



\section{Motivation Mechanism}
\label{sec:motivation_mechanism}
In this section, we introduce motivation mechanisms that are important in sharding blockchain systems. Section~\ref{subsec:mm_bp} explains the meaning and purpose of a motivation mechanism. Section~\ref{subsec:mm_es} summarizes the related research from the following aspects, rewards for block producers and leader rewards, penalties for negative behaviors, and rewards based on reputation. Finally, we analyze the related problems and future research directions in Section~\ref{subsec:mm_pp}.

\subsection{Basic Concepts}
\label{subsec:mm_bp}
Blockchain systems need to design a motivation mechanism to encourage nodes to participate in the protocol. In general, the more nodes in the network, the safer the system. Besides, the node participating in the protocol needs to consume a certain amount of communication bandwidth and computational power. If the corresponding reward is not obtained, the node will lose the motivation to participate in the protocol. 
The difficulty in designing a motivation mechanism is to ensure that the rewards received by each node are fair and that malicious behaviors can be punished accordingly. In addition, when analyzing motivation mechanisms, it is usually necessary to assume that all nodes are rational, i.e., each node's behavior is to maximize its interests.

Motivation mechanisms consist of incentive and penalty parts.  Incentive mechanisms refer to the rewards for certain contributions of participating nodes in the blockchain. The rewards are generally in the form of tokens in the system. At the same time, certain nodes need to be punished by a penalty mechanism for negative sabotage or malicious behaviors. Generally speaking, in public blockchains, motivation mechanisms need to be carefully designed to encourage nodes to participate in the operations of the protocol. 

Since the motivation mechanism is not the focus of this paper, and there are very few studies on the incentive mechanism of sharding blockchains, we only briefly introduce the motivation mechanism.

\subsection{Existing Approaches}
\label{subsec:mm_es}

We classify motivation mechanisms into rewards for blockchain producers and leader, penalties for negative behaviors, and rewards based on reputation.

\subsubsection{Rewards for block producers and leaders}
There are multiple shards in a sharding blockchain, and each shard maintains its own blockchain. Each shard is constantly producing blocks, and the block producers should receive corresponding rewards. The block producer here might vary depending on the blockchain protocol. In an eventual sharding blockchain, such as Monoxide \cite{WW19} and Parallel Chains \cite{FGK+18}, a block producer is an individual node. In this case, a single node will get a complete block reward. In instant sharding blockchains, the committee in each shard runs an intra-shard consensus algorithm to generate blocks, so all members in the committee should get the corresponding block rewards. The rewards could be further distributed fairly according to the contributions of each member, i.e., the amount of communication and the number of transactions processed.

In instant sharding blockchains, a committee in a shard runs the BFT algorithm, which requires the collaboration of a leader. In addition, a leader is usually responsible for collecting, merging, and forwarding of messages within the committee, so its computation and communication costs are higher than the other committee members. Therefore, the incentive mechanism within the committee should be more elaborately designed to ensure the fairness of reward distribution.

Wang and Wu \cite{WW19lever} propose Lever as an incentive mechanism to distribute rewards among rational stakeholders in BFT consensus algorithms, and they further apply it to the sharding blockchain. 
Manshaei \textit{et al.} \cite{MJM+18} conduct an analysis of the motivation mechanisms in sharding blockchains based on the game theory. They propose an incentive-compatible reward mechanism to encourage shard members to participate in the protocol. 

\subsubsection{Penalties for negative behaviors} In a sharding blockchain, some negative behaviors need to be punished. Negative behaviors could be further divided into two categories. One is sabotage behaviors, which can also be understood as passive behaviors, such as committee members not voting on the leader's valid proposals, or a leader not proposing new blocks or transactions within a specified time. The other is malicious behaviors, such as a leader proposing two different proposals (blocks or transactions) in the same round, or the block proposed by a leader contains invalid transactions. Besides, committee members might vote for more than one proposal message in the same round. In general, the establishment of a penalty mechanism first requires nodes to submit a certain deposit before participating in the protocol, e.g., Casper FFG \cite{BG17}. When a malicious behavior is detected, the deposit may be deducted accordingly.

\subsubsection{Rewards based on reputation}
The reputation based motivation mechanism originates from the cooperative P2P scenarios, e.g., distributed storage systems \cite{KSG03} and is introduced to the blockchain area \cite{NGN+18}. Reputation usually refers to the unified evaluation and scoring of all nodes based on the past performances of participating nodes, including corresponding time, the number of transactions processed, and the number of malicious behaviors that are related to the nodes. In this process, different behaviors may come with different evaluation weights. Each participating node may eventually get a score, and the score will be updated as the system advances. When nodes perform well and participate actively, their rating scores will grow. On the contrary, if a node loses response or is detected to behave maliciously, then the node's score will decrease. The rewards in the system will be distributed according to each node's score. Nodes with higher reputation will get more rewards.
Bugday \textit{et al.} \cite{BOS19} apply learning methods to the establishment of node reputation in the blockchain, where the node reputation is dynamically changing.
Huang \textit{et al.} \cite{HWC+19} utilize high incentives to motivate nodes to behave themselves in the sharding blockchain. Similarly, the authors adopt a reputation-based evaluation system to measure the different processing capabilities of each node.
CycLedger \cite{ZLC+20} mainly uses the reputation to evaluate the mining ability of each miner. Block rewards are allocated according to the reputation value of each miner.

\subsection{Problems and Future Directions}
\label{subsec:mm_pp}

\subsubsection{Specific considerations for sharding blockchains}
The establishment of a motivation mechanism needs to take the differences between sharding blockchains and ordinary blockchains into account. 

First, according to the various intra-committee consensus algorithms, the distribution of rewards for committee members should be more specifically designed to realize fairness. In algorithms, especially those that adopt a stable leader such as PBFT \cite{CL99}, The leader is of vital importance to the stable operation of the system. In each round, the leader is responsible for broadcasting a proposal and collecting votes in the system.
Consequently, the leader will have a higher communication and computation overhead than other members, so it should obtain higher rewards. However, in this way, all members will hope to become a leader, and malicious nodes might take the opportunity to launch a view-change operation to replace the old leader. Moreover, an adversary might launch a DoS attack or network partition attack against the leader, causing its network to be paralyzed and replaced by a new leader. Relatively speaking, in the BFT algorithms that employ a continuously changing leader such as HotStuff \cite{YMR+19}, it is easier for the incentive mechanism to achieve fairness. 

Second, the influences caused by cross-shard transaction processing need to be considered. 
A coordinator is responsible for the forwarding of availability certificates, that is, to accomplish the communication among shards. 
The work of a coordinator needs to be carefully considered when designing a motivation mechanism. There may also be some malicious behaviors. For example, in a client-driven 2PC method, a client might not forward the transaction input availability certificate within a specified time. In sharding blockchains using shard-driven 2PC, an input shard leader might not forward availability certificates to other related shards. 

\subsubsection{Detailed analysis of the motivation mechanism}
As far as we know, no detailed theoretical analysis has been conducted on the motivation mechanism of sharding blockchains. The analysis requires a certain financial foundation, e.g., Nash equilibrium theory \cite{M99}. 
Therefore, for the sharding blockchains, a reasonable, comprehensive, and secure motivation mechanism and strict analysis of it are the future research directions.

\section{Related Work}
\label{sec:related_work}
In the following, we introduce the related work from different perspectives.

\subsubsection{Survey on Sharding Blockchain Systems}
\label{subsubsec:survey}
Wang \textit{et al.} \cite{WSN+19} give an overview of the research on sharding blockchain systems. However, their classification of the sharding blockchain components is abstract, and they do not provide a more in-depth analysis for each component. In contrast, in our paper, we provide a more complete characterization of components and their composition into a sharding blockchain system. Moreover,  
for each independent component, we provide a taxonomy of the existing approaches for that component. In addition, we deeply analyze possible problems and future research directions, which are of great importance to related researchers.

Yu \textit{et al.} \cite{YWYNZL20} take each existing scheme as a starting point and give basic details of each sharding blockchain scheme. However, they do not provide a macro vision and overall classification of sharding blockchain systems. Our paper provides a systematic taxonomy, and we also analyze related research on the security model and motivation mechanism of sharding blockchain systems that are not mentioned in the previous research.

\subsubsection{Modular Analysis of Sharding Blockchains}
\label{subsubsec:modula_analysis}
Avarikioti \textit{et al.} \cite{AKW19} provide a relatively formalized analysis of sharding blockchain systems, and propose some evaluation indicators, including communication, computation, storage complexity, and the scaling factor. Zamyatin \textit{et al.} \cite{ZAZ+19} analyze communication across distributed ledgers. Their analysis includes cross-shard communication between homogeneous blockchains and cross-chain swap such as sidechains between heterogeneous blockchains. 
Han \textit{et al.} \cite{HYZ20} analyze shard allocation protocols for sharding blockchain systems.
These studies mainly focus on a certain part of sharding blockchains, while our study combines a unified framework with thorough component analysis.

\subsubsection{Blockchain Consensus}
\label{subsubsec:blockchain_consensus}
Bano \textit{et al.} \cite{BSA+19,BanoAD17} study consensus mechanisms in the age of blockchain, including classical consensus, proof-of-X consensus, and hybrid consensus. Garay and Kiayias \cite{GarayK20} provide a consensus taxonomy in the blockchain era based on different network, setup, and computational assumptions. In their paper, consensus mechanisms are divided into consensus in the point-to-point setting, consensus in the peer-to-peer setting, and ledger consensus. Alsunaidi and Alhaidari \cite{AA19} conduct a comprehensive survey on popular blockchain consensus algorithms, focusing on their security and performance. Nguyen and Kim \cite{NK18} classify existing blockchain consensus algorithms into proof-based consensus and voting-based consensus. Xiao \textit{et al.} \cite{XZL+20} carry out a survey on blockchain consensus protocols and identify five core components, namely block proposal, block validation, information propagation, block finalization, and incentive mechanism. By contrast, our study conceptualizes functional components of sharding blockchain systems, going beyond consensus. 

\subsubsection{Blockchain Scalability}
\label{subsubsec:blockchain_scalability}
Some studies formalize specific properties to be followed by different solutions to the scalability of blockchain systems, but without aiming at a broad review as our study does.  
Xie \textit{et al.} \cite{XYH+19} study the scalability problem of blockchain from the perspectives of throughput, storage, and networking. Kim \textit{et al.} \cite{KKC18} analyze existing solutions to realize blockchain scale-out and classify them into different types, i.e., on-chain, off-chain, side-chain, child-chain, and inter-chain. Zhou \textit{et al.} \cite{ZHZ+20} summarize existing scaling schemes and provide potential research directions for solving the blockchain scalability problem.

\section{Conclusion}
\label{sec:conclusion}
This paper decomposes sharding blockchains into multiple components and analyzes the basic concepts, existing approaches, and potential problems of each component. On this basis, designing a new sharding blockchain system could be simplified into composing several different components. In this way, each component could be improved separately according to the current latest research, and the improved component could be integrated into a whole sharding blockchain system without affecting the security of other parts and the entire system. For each component, the possible problems and future research directions that are proposed in our paper are worthy of attention. We believe that our systematic and comprehensive research on sharding blockchains could give insights to future researchers.


\section*{Acknowledgment} 
\label{sec:acknowledgment}
Our deepest gratitude goes to the editors and reviewers for their careful work and meaningful suggestions that help improve this paper. 
This work is supported in part by the National Key R\&D Program of China (2017YFB1400702), in part by the National Natural Science Foundation of China (61972017, 61972018, 61972014, 72031001), in part by the National Cryptography Development Fund (MMJJ20180215), and in part by the Fundamental Research Funds for the Central Universities (YWF-20-BJ-J-1039).

\bibliographystyle{elsarticle-num}
\bibliography{bbsb}

\end{document}